\documentclass[english]{iopart}

\usepackage{graphicx}
\usepackage{url}
\usepackage{iopams} 

\expandafter\let\csname equation*\endcsname\relax
\expandafter\let\csname endequation*\endcsname\relax

\usepackage{amssymb}
\usepackage[latin1]{inputenc}
\usepackage{amsmath}
\usepackage{epsfig}
\newcounter{fig}
\begin{document}

\title[Differential algebra on lattice Green functions and Calabi-Yau operators]
{\Large Differential algebra on lattice Green functions and Calabi-Yau operators (unabridged version)}
\vskip .3cm 
{\bf 9th January 2014}

\author{S. Boukraa$||$, S. Hassani$^\S$, 
J-M. Maillard$^\pounds$, J-A. Weil$^\dag$}
\address{$||$  \ LPTHIRM and IAESB,
 Universit\'e de Blida, Algeria}
\address{\S  Centre de Recherche Nucl\'eaire d'Alger, 
2 Bd. Frantz Fanon, BP 399, 16000 Alger, Algeria}
\address{$^\pounds$ LPTMC, UMR 7600 CNRS, 
Universit\'e de Paris 6\footnote[1]{Sorbonne Universités 
(previously the UPMC was in Paris Universitas).}, Tour 23,
 5\`eme \'etage, case 121, 
 4 Place Jussieu, 75252 Paris Cedex 05, France} 
\address{$^\dag$ \  XLIM, Universit\'e de Limoges, 
123 avenue Albert Thomas,
87060 Limoges Cedex, France}

\begin{abstract}

We revisit miscellaneous linear differential operators mostly associated with lattice
Green functions in arbitrary dimensions, but also Calabi-Yau operators and order-seven 
operators corresponding to exceptional differential Galois groups. 
We show that these irreducible operators
are not only globally nilpotent, but are such that they are homomorphic 
to their (formal) adjoints. Considering these operators, or, sometimes, equivalent
operators, we show that they are also such that, either their 
symmetric square or their exterior square, have a rational solution.
This is a general result: an irreducible linear differential operator
homomorphic to its (formal) adjoint is necessarily such that either its 
symmetric square, or its exterior square has a rational solution,
and this situation corresponds to the occurrence of a special differential 
Galois group. We thus define the notion of being ``Special Geometry'' for 
a linear differential operator
if it is irreducible, globally nilpotent, and such that it is homomorphic 
to its (formal) adjoint. 
Since many Derived From Geometry $\, n$-fold 
integrals (``Periods'') occurring in physics, are seen 
to be diagonals of rational functions, we address
several examples of (minimal order) operators annihilating
diagonals of rational functions, and remark that they also seem to be,
systematically, associated with irreducible factors homomorphic 
to their adjoint.

\end{abstract}

\vskip .1cm

\vskip .3cm


\noindent {\bf AMS Classification scheme numbers}: 34M55, 
47E05, 81Qxx, 32G34, 34Lxx, 34Mxx, 14Kxx 

\vskip .3cm


\vskip .3cm

\section{Introduction}
\label{intro}

When one considers all the irreducible factors 
of the globally nilpotent linear differential operators
encountered in the study of $\, n$-folds 
integrals of the Ising class~\cite{Isingclass}
(or the one's displayed by other authors
in an enumerative combinatorics 
framework~\cite{Guttmann,GoodGuttmann}, or in a
Calabi-Yau framework~\cite{Batyrev,TablesCalabi,Almkvist}), 
one finds out that
 their differential Galois groups are not 
the $\,SL(N,\, \mathbb{C})$, or extensions of
 $\,SL(N,\, \mathbb{C})$, groups one could expect 
generically, but {\em selected} $\, SO(N)$,
 $\, Sp(N,\, \mathbb{C})$, $\, G_2$, ...
differential Galois groups~\cite{Katz}. 

\vskip .1cm

Along this line it is worth recalling that  
{\em globally nilpotent linear differential operators}
associated with generic $_nF_{n-1}$ {\em hypergeometric functions
with rational parameters}\footnote[5]{Their corresponding
linear differential operators are necessarily 
globally nilpotent~\cite{bo-bo-ha-ma-we-ze-09}.}, 
have $SL(N, \, \mathbb{C})$ (or extensions of $SL(N, \, \mathbb{C})$)
 differential Galois groups. For instance 
the $_3F_{2}$ hypergeometric function 
\begin{eqnarray}
\label{sl3}
 \hspace{-0.8in}&&\qquad \quad \quad 
 _3F_{2}([{{191} \over {479}}, {{359} \over {311}}, {{503} \over {89}}], 
[{{521} \over {151}},  {{401} \over {67}}],\,  x), 
\end{eqnarray}
has a $SL(3, \, \mathbb{C})$ differential 
Galois group\footnote[2]{One shows that there are no 
rational solutions of symmetric powers 
in degree 2,3,4,6,8,9,12,  using
an algorithm in M. van Hoeij et al.~\cite{WeilUlmer}}. 
In contrast, in the 
simplest examples, the emergence of ``selected''
 differential Galois groups can be seen very explicitely~\cite{Compoint},
and understood (from a physicist's viewpoint) 
as the emergence of some ``invariant''.   As far as the
$\, SO(3, \, \mathbb{C})$ group
is concerned\footnote[8]{This operator
 is actually homomorphic to its adjoint (see below) with
 non-trivial order-two intertwiners. }, let us consider 
the {\em non-Fuchsian}\footnote[1]{This operator has 
an irregular singularity at infinity. At $\, x \, = \, \infty$ 
the solutions behave like: $t \cdot \, 
(1 \, + 77/72  \, t^2 \,\, + \, \cdots)$, and 
$\exp(-2/t)/t \cdot \,(1 \, + \,13/36 \, t \, + \, \cdots) $ 
 where $\, t \, = \, \pm 1/\sqrt{x}$.}
 operator
 ($\theta \, = \, \, x \cdot d/dx$):
\begin{eqnarray} 
\label{nonfuchs}
\hspace{-0.8in}&&\qquad \quad \quad 
2\, \theta \cdot (3\, \theta -2) \cdot (3\, \theta -4) \, \, \,
 - 9\, x \cdot (2\, \theta +1), 
\end{eqnarray}
with the three $\, _1F_2$ hypergeometric solutions
\begin{eqnarray} 
\hspace{-0.8in}&& 
_1F_2\Bigl([{{1} \over {2}}],
 \, [-{{1} \over {3}}, \, {{1} \over {3}}], \, x\Bigr),
 \quad 
x^{2/3} \cdot \, _1F_2\Bigl([{{7} \over {6}}], 
\, [{{1} \over {3}}, \, {{5} \over {3}}], \, x\Bigr), 
\quad 
x^{4/3} \cdot \, _1F_2\Bigl([{{11} \over {6}}],
 \, [{{5} \over {3}}, \, {{7} \over {3}}], \, x\Bigr).
\nonumber 
\end{eqnarray}

If $\, f$ denotes a solution of this operator (in the above closed form 
or as a formal solution at the origin or at $\, \infty$),
one has the following {\em quadratic relation}
 $\, Q(f, \, f', \, f") \, = \, const.$,  where: 
\begin{eqnarray}
\hspace{-0.8in}&& Q(X_0, \, X_1, \, X_2) \,\,= \, \, \, 
9 \cdot (36\, x \, +5)\cdot \, x^2 \cdot X_2^2 \,
 -324 \cdot x^2 \cdot X_2 \cdot X_1\,\,-648 \,  x^2 \cdot X_2 \cdot X_0
\nonumber \\
\hspace{-0.8in}&&\qquad \quad 
+(81\, x\, -5) \cdot X_1^2 \, \,
+ 9 \cdot (36\, x \, -5) \cdot X_0 \cdot X_1 
\, + \, 9 \cdot (36\, x \, -5)  \cdot X_0^2. \nonumber 
\end{eqnarray}
The constant depends on the linear combination of solutions used.
For instance, with the first $\, _1F_2$ hypergeometric solution one has
$\, Q(f, \, f', \, f")\,= \, \, 225/4$, 
while with the two other $\, _1F_2$ solutions it reads 
$\, Q(f, \, f', \, f")\,= \, \, 0$.
In other words $\, Q$ is a {\em first integral}. 

\vskip .1cm

The emergence of such ``special'' differential Galois groups in
so many domains of theoretical physics is clearly something
{\em we need to understand better}.

\vskip .1cm

We have provided a large number of linear ODEs on various problems of
lattice statistical mechanics, in particular for the magnetic susceptibility
of the two-dimensional 
Ising model~\cite{bo-bo-ha-ma-we-ze-09,High,bernie2010,Khi6,ze-bo-ha-ma-05c,
Renorm,mccoy3,ze-bo-ha-ma-04,Short,Big,CalabiYauIsing}. 
These linear ODEs factorize into many factors of order ranging from one, to
12 (for $\, \chi^{(5)}$) and even 23 (for $\, \chi^{(6)}$).
As far as the factors of smallest orders (two, three and four) are concerned,
one can verify that {\em all these linear differential operators 
are homomorphic to their adjoint}.
Furthermore, one remarks experimentally, that 
their exterior square or symmetric square,
either have a {\em rational solution}, or are of an {\em order smaller 
than the order one would expect generically}. 
Quite often these differential operators are simply {\em conjugated to 
their adjoint}, i.e. the intertwiner between the operator and its adjoint,
is just an order-zero operator, namely a function.
In that case they can easily be recast into {\em self-adjoint} operators.
A large set of linear differential operators
 {\em conjugated to their adjoint}, can be found in
the very large list of Calabi-Yau order-four operators obtained 
by Almkvist et al.~\cite{TablesCalabi},
or displayed by Batyrev and van Straten~\cite{Batyrev},
 or some simple order-three operators
displayed in a paper by Golyshev~\cite{Golyshev,Bogner} (see also
Sanabria Malagon~\cite{Malagon}). 

\vskip .1cm

Throughout this paper we will see examples of {\em irreducible} operators
where these two differential algebra properties occur simultaneously.
On the one hand, these operators are {\em homomorphic
to their adjoint}, and on the other hand, their symmetric
or exterior square {\em have a rational}\footnote[1]{For hyperexponential 
solutions~\cite{SingUlm} (command expsols in DEtools), i.e. $\, N$-th 
root of rational solutions,
one must consider homomorphisms {\em up to algebraic extensions}.} {\em solution}.
These simultaneous properties correspond to {\em special differential
Galois groups}. In fact, these properties are equivalent\footnote[2]{
In a Tannakian formulation, one could say that the homomorphisms of
an operator $\, L_1$ with another  operator $\, L_2$
are isomorphic to the product
$\, Hom(L_1, \, L_2) \, \, \simeq  \, \, L_1 \, \otimes \, L_2^{\star}$,
giving, in the case of the homomorphisms of an operator
$\, L$ with its adjoint $\, L^{\star}$,
$\, Hom(L, \, L^{\star}) \, \, \simeq  \, \,$
$ L \, \otimes \, (L^{\star})^{\star} \,\, \simeq  \, L \,\otimes \, L$,
which is isomorphic to the direct sum
  $\, L \, \otimes \, L  \, \, \simeq $
$ \, \, \,  Ext^2(L) \, \oplus \, Sym^2(L)$.}.

\vskip .1cm

We will, in this paper, have a learn-by-example approach 
of all these concepts. In this respect, 
we will display, for pedagogical reasons, a set of enumerative combinatorics 
examples corresponding to miscellaneous 
{\em lattice Green functions}~\cite{Guttmann,GoodGuttmann,Thomas,GlasserGuttmann,Kou}
as well as  Calabi-Yau examples, together with order-seven 
operators~\cite{BognerGood,DettReit} associated 
with  differential Galois exceptional groups. We will show that 
these lattice Green operators, Calabi-Yau operators 
and order-seven operators associated with exceptional 
groups, are a perfect illustration of differential operators with 
{\em selected differential algebra structures}: they are homomorphic 
to their adjoint, and, either their symmetric, or exterior, powers
(most of the time squares) have a {\em rational solution}, or the 
previous symmetric, or exterior, powers of some {\em equivalent 
operators} have a rational solution. This situation corresponds
to the emergence of {\em selected differential Galois groups}, a situation we 
could call ``Special Geometry''. Among the Derived From Geometry $\, n$-fold 
integrals (``Periods'') occurring in physics, we have seen that they are
quite often {\em diagonals of rational functions}~\cite{Short,Big}. We 
will also address in this paper examples of (minimal order) operators annihilating
diagonals of rational functions, and will remark that they also seem to be
 associated with {\em irreducible factors homomorphic to their adjoint}.

\vskip .1cm 

\section{Adjoint of differential operators and invertible homomorphisms
of an operator with its adjoint}
\label{commentsadjoint}

In the next section, examples of linear differential operators 
corresponding to lattice Green functions on various lattices 
are displayed according to their order $\, N$ 
and their complexity. We focus on the differential algebra structures of 
these linear differential operators, in particular
with respect to an important ``duality'' with amounts
 to performing the {\em adjoint},
 or, more precisely (see 2.1 in~\cite{vdP}),
 the ``formal adjoint''  of the operator 
 ($D_x$ in the whole paper denotes the 
derivative $d/dx$):
\begin{eqnarray}
\label{adjointdef}
\hspace{-0.9in}&& L \, = \,\,   \sum_{n=0}^N \, a_n(x) \cdot D_x^n 
 \, \quad   \longrightarrow  \, \,\,  \,   \,  
adjoint(L)\,  =\,\,  (-1)^N \cdot \, \sum_{n=0}^N \, (-1)^n \cdot D_x^n \cdot a_n(x), 
\end{eqnarray}
that is:
\begin{eqnarray}
\label{adjointdef2}
\hspace{-0.9in}&&\sum_{n=0}^N \, a_n(x) \cdot {{d^n f(x)} \over {dx^n}}
 \, \,\, = \,\, 0 \,\, \, \, \,  \longrightarrow  \, \,\quad  
(-1)^N \cdot \, \sum_{n=0}^N \, (-1)^n  \cdot
 {{d^n (a_n(x) \cdot f(x))} \over {dx^n}} \, \,\, = \,\, 0.
\end{eqnarray}
Since exterior powers play a key role in this paper, it is worth 
recalling that the adjoint of an $\, N$-th order operator $\, L_N$ 
is nothing but the $\, (N-1)$-th exterior power\footnote[1]{In maple 
the exterior power is normalised to be a monic operator (the head 
polynomial is normalised to $\, 1$).} of this operator, 
up to a factor that is the Wronskian $\, W(L_N)$ of the operator:
\begin{eqnarray}
\label{Nminus1ext}
\hspace{-0.9in}&&\qquad \quad 
 W(L_N) \cdot \, adjoint(L_N)
 \,\,\, = \,\, \,\,
 Ext^{N-1}(L_N) \cdot \, a_N(x)  \cdot \, W(L_N).
\end{eqnarray}

\subsection{Homomorphisms of an operator with its adjoint}
\label{commentsadjointhomo}

Recall that two  operators $L$ and $\tilde{L}$, of the same order, 
 are called  homomorphic (see~\cite{Singer,vdP})
when there exist two operators $T$ and $S$ of {\em smaller order}
than  the one of $L$ and $\tilde{L}$,
such that\footnote[3]{The {\em intertwiner} $T$ is given by the
command \texttt{Homomorphisms}$(L, \,\tilde{L})$ 
of the DEtools package in Maple~\cite{Homomorphisms}.}:
\begin{eqnarray}
\label{recallintetwin}
 \hspace{-0.7in}&&\quad \quad \qquad \qquad
\tilde{L} \cdot \,T \,\,\, =\,\,\,\,\, S \cdot \, L.
\end{eqnarray}
The intertwiner $T$ maps the solutions of $L$ 
into the solutions of $\tilde{L}$. When $T$ and $L$ have 
{\em no common right-factor} (or equivalently when $S$ 
and $\, \tilde{L}$ has no common left factor), 
for example when $\, L$ is {\em irreducible},
 one can show that this map is bijective. When (\ref{recallintetwin}) 
holds, and $\, T$ and $\, L$ have 
{\em no common right-factor}, one says that 
$L$ and $\, \tilde{L}$ are {\em equivalent}. 
Thus, one also has {\em intertwiners} $\tilde{T}$, $\,\tilde{S}$
such that
\begin{eqnarray}
\label{otherintertwin}
\hspace{-0.7in}&&\quad \quad \qquad \qquad
 L \cdot \, \tilde{T} \,\,\,  =\,\,\,\, 
 \tilde{S} \cdot \,\tilde{L}, 
\end{eqnarray}

\vskip .1cm

We say that $ \, L$ is {\em self-adjoint} when
$\, L\, = \, \, adjoint(L)$. We say that $ \, L$ is
{\em conjugated with its adjoint} when there exists a
rational, or $\, N$-th root of rational,
function $\, f$ such that
 $\,L \cdot \, f \,= \,\, f \cdot \, adjoint(L)$, i.e. 
$L\cdot \, f$ is self-adjoint.
More generally,  a differential operator $\,L$ is
{\em homomorphic to its adjoint} (in the above sense)
when there exists an (intertwiner) operator $T$
(of {\em order less\footnote[8]{Note that the constraint on the order
 rules out the ``tautological'' intertwining relation, 
 satisfied by any operator, like  
 $\, L \cdot \, adjoint(T)  \, \, = \, \, \, T \cdot \, adjoint(L)$ 
 with $\, T \, = \, \, L$. }
 than that of} $L$)
such that\footnote[5]{
It is easy to show, in the case of an homomorphism of an 
operator $L$ with its adjoint, that the intertwiner on the 
right-hand-side of (\ref{homotoadj})
is necessarily equal to the adjoint of the intertwiner 
on the left-hand-side.
Actually, from the equivalence 
$\,L \cdot \, T\,= \, S \cdot \, adjoint(L)$, taking
adjoint on both sides gives
 $\,adjoint(T) \cdot \, adjoint(L) = \, L \cdot \, adjoint(S)$.
For irreducible $L$, the intertwiner is unique, 
so $S = \, adjoint(T)$.
} 
\begin{eqnarray}
\label{homotoadj}
 \hspace{-0.7in}&&\quad \quad \qquad \qquad
 L \cdot \,T \,\,  = \,\, \, adjoint(T) \cdot \, adjoint(L).
\end{eqnarray}
Again, this means that the operator $\, L \cdot \,T$ is self-adjoint.

\vskip .1cm

The typical situation we encounter in physics is such that the differential
operators are of a quite large order and factorize into many factors of various
orders (see the minimal order operators~\cite{High,Khi6} 
annihilating the $\, \chi^{(n)}$'s).
For these large order differential operators, we will 
systematically factorize the operator.
{\em The interesting concept amounts to seeing if 
 each irreducible factor in the factorization,
is homomorphic to its adjoint}.

\vskip .1cm

We end this section with two comments. 
For irreducible $L$, one deduces, from (\ref{recallintetwin})
 and (\ref{otherintertwin}), the equality
\begin{eqnarray}
\label{TTtilde}
 \hspace{-0.9in}&&\quad \quad \quad \qquad\qquad
 L \cdot \,\tilde{T} \cdot \,T \,\, = \,\,\, 
\tilde{S} \cdot \, S \cdot \, L,  
\end{eqnarray}
so the rest of the right division of 
$\tilde{T} \cdot \,T$ by $L$ is a constant. When $\tilde{L}$
 is the adjoint of $L$, we will see, in the sequel, that this relation 
on the intertwiners $T$ and $\tilde{T}$ makes a remarkable
 "decomposition" of $L$ emerge.
The second comment is on the homomorphisms of an operator 
with its adjoint in the {\em reducible} case. 
For two reducible differential operators, $\, L$ and $\, \tilde{L}$,
 of the {\em same order},
the relation (\ref{recallintetwin}) may hold
but may not be an equivalence of operators. 
% When the differential operator $L$ is reducible, there 
% is {\em no equivalence relation
% with the adjoint}.  
In the case of a reducible~\cite{Berman}
 operator having the unique factorization 
$\, L = \, L_n \cdot L_p$, with $ \, n \ne p$,
one can show that the homomorphism with the adjoint 
{\em just reduces} to a homomorphism of the {\em right factor} 
$ \, L_p$ with its adjoint.
% In the case of a reducible~\cite{Berman}  
%  operator with the unique factorization 
%  $L \, = \, L_n\cdot \, L_p$, one can 
% show that the homomorphism with the adjoint 
% {\em just reduces} to a homomorphism of the {\em right factor} 
% $L_p$ with its adjoint. 
% As far as the equivalent property to have a 
The corresponding rational solution for the
symmetric or exterior square is 
% concerned, one obtains 
precisely the rational solution induced by
 the right factor, since 
$\,Sym^2(L_p)$ (resp. $\, Ext^2(L_p)$) 
 is a right-factor of $\,Sym^2(L_n\cdot \, L_p)$ 
(resp. $\,Ext^2(L_n\cdot \, L_p)$).

\vskip .1cm

In the sequel, when studying homomorphism of an operator with its adjoint,
we will restrict to {\em irreducible} operators.

\vskip .1cm 

\section{Special ODEs from lattice statistical mechanics
 and enumerative combinatorics: lattice Green functions}
\label{specialODE}

We are going to display a set of miscellaneous examples 
of linear differential operators corresponding to {\em lattice Green 
functions} on various lattices. We will denote these lattice Green 
operators $\, G_n^{latt}$, where $\, n$ is the order of the 
operators\footnote[3]{Do not expect a simple match between the dimension 
of the lattice and this order $\, n$.}, and where  {\em latt} 
refers to the lattice one considers.

\subsection{Special lattice Green ODEs: body-centered cubic lattice and simple cubic lattice}
\label{warm20}

One of the simplest example of lattice Green function corresponds to
the order-three operator, for the body-centered cubic lattice,
given in equation (19) of~\cite{GoodGuttmann},
which reads
\begin{eqnarray}
\label{premierex}
\hspace{-0.9in}&&\quad \quad  \qquad  \qquad 
G_3^{bcc} \,\,  = \, \,\, \, \, 
 \theta^3 \, \, \,  -\, 64 \cdot x \cdot  (2\, \theta \, +\, 1)^3. 
\end{eqnarray}
This order-three operator has the $\, _3F_2$ hypergeometric solution 
\begin{eqnarray}
\label{solbcc}
\hspace{-0.9in}&&\quad \quad  \quad  
_3F_2\Bigl([{{1} \over {2}},\, {{1} \over {2}}, \, {{1} \over {2}}],
 \, [1, \, 1]; \, 512\, x\Bigr)
  \,  \,= \, \,  \,   \,  
\Bigl(\, _2F_1\Bigl([{{1} \over {4}},\, {{1} \over {4}}], \, [1]; 
\, 512\, x\Bigr)\Bigr)^2,
\end{eqnarray}
where the $\, _2F_1$ hypergeometric function is a series with 
{\em integer coefficients}:
\begin{eqnarray}
\hspace{-0.95in}&&\,  
_2F_1\Bigl([{{1} \over {4}},\, {{1} \over {4}}], \, [1]; \, 512\, x\Bigr)
 \,  = \, \, \,   1\,\,  +32\, x\, +6400\, x^2\, +1843200\, x^3
\, +623001600\, x^4\,  + \, \, \cdots 
\nonumber 
\end{eqnarray}
The operator (\ref{premierex}) is conjugated to its adjoint: 
$\,\,\, x \cdot \, adjoint(G_3^{bcc})  \,  = \, \, \, G_3^{bcc}\cdot \, x$.
The symmetric square of $\, G_3^{bcc} $ is of {\em order five}
(instead of the order six one could expect generically 
for order-three operators).

\vskip .2cm 

The most well-known example of lattice Green function
has been obtained~\cite{Joycecube} for the 
{\em simple cubic lattice}. The lattice Green function
corresponds to the order-three operator 
(see equation (19) in~\cite{GoodGuttmann})
\begin{eqnarray}
\label{Gsc}
\hspace{-0.9in}&&\quad \quad \quad G_3^{sc} \,\, \,  = \, \, \, \, \, 
\, \theta^3 \, \, \,  -\, 2\, x \cdot 
(10\, \theta^2 \, +10 \,\theta \, +3)  \cdot (2\, \theta \, +\, 1)  
\,  \, \,  
 \\
\hspace{-0.9in}&&\quad \quad \quad \qquad \qquad  \quad \,  \, 
+18 \, x^2 \cdot   (2\,\theta \, +\, 3)
 \, (2\,\theta \, +\, 2) \, (2\, \theta \, +\, 1),
\nonumber 
\end{eqnarray}
This order-three operator  (\ref{Gsc}), when divided by 
$\, x$ on the left, is {\em exactly self-adjoint}.
The symmetric square of $\, G_3^{sc}$ is of order five (instead of the 
generic order six).

The solution of (\ref{Gsc}), which corresponds to a series
expansion with {\em integer coefficients}, is the 
Hadamard product of  $(1-4\, x)^{-1/2}$ with a Heun function, 
and is also the square of another Heun function which can also
be written in terms of $\, _2F_1$ hypergeometric functions 
with {\em two possible algebraic pullbacks}:
\begin{eqnarray}
\label{selected}
\hspace{-0.9in}&&\,\, \,\, \,  
  HeunG\Bigl(9,\,  {{3} \over {4}}, \, {{1} \over {4}},\,
  {{3} \over {4}},\,  1,\,  {{1} \over {2}}; \, \,  36 \,x\Bigr)^2
\, \,   = \, \, \, 
HeunG\Bigl({{1} \over {9}},\,  {{1} \over {12}},\,  
{{1} \over {4}}, \, {{3} \over {4}},\,  1,\,  {{1} \over {2}};
 \, \,  4\,x\Bigr)^2
\nonumber \\
\hspace{-0.9in}&& \, \, \, \,    = \, \, 
 (1-4\, x)^{-1/2}\, \star \, HeunG(1/9, 1/3, 1, 1, 1, 1; \, x)
\, \, = \, \, \, C_{\pm}^{1/2} \, \cdot  \,
 _2F_1\Bigl([{{1} \over {6}}, \, {{1} \over {3}}],
\,  [1]; \, P_{\pm}  \Bigr)^2 
\nonumber \\ 
\hspace{-0.9in}&& \, \,  \, \,    = \, \, \,  \, 
1\,\,  +6\, x\,\,  + 90\, x^2\,\,  + 1860\, x^3 \, 
 +44730\, x^4\, + 1172556\, x^5\, \,  \,+ \, \, \cdots
\end{eqnarray}
where the algebraic pull-backs $\, P_{\pm}$ and algebraic 
prefactors $\, C_{\pm}$ read:
\begin{eqnarray}
\hspace{-0.9in}&&\quad   P_{\pm}\, \,  =  \, \, \,\, 
 54 \cdot x \cdot \Bigl(1 -27\,x\, +108\,x^2
\,\,\, \pm (1\, -\, 9\, x) \cdot ((1-36 \,x)
\cdot (1-4\,x))^{1/2}\Bigr),
\nonumber \\
\hspace{-0.9in}&&\quad  C_{\pm}\, \,  =  \, \, \,
 -18\, x\,+ {{5} \over {2}} \,\,\, 
\pm{{3} \over {2}} \cdot ((1-36\,x)\cdot (1-4\,x))^{1/2}.
\end{eqnarray}
The fact that these selected Heun functions (\ref{selected})
correspond to {\em modular forms}~\cite{CalabiYauIsing} 
can be seen on the relation between the two algebraic pullbacks,  
$\, y \, = \, \,  P_{+}$ and $\, z \, = \, \,  P_{-}$,
namely the genus-zero {\em modular curve}\footnote[1]{
Which is {\em exactly a rational modular curve} already found
for the order-three operator $\, F_3$ in~\cite{CalabiYauIsing}.}:
\begin{eqnarray}
\label {encore}
\hspace{-0.9in}&&\quad \quad  \quad 
4\cdot \, y^3 \,z^3 \,\,\, -12 \,\,y^2 \,z^2 \cdot (z+y)\,\,
+3 \, \,y \, z\cdot(4\, y^2\,4\, z^2 \, -127\,\,y\, z)\,
\nonumber \\
\hspace{-0.9in}&&\quad \quad  \quad \quad \qquad \quad \quad 
-4 \cdot (y \, +z) \cdot (y^2\, +\, z^2\, +83 \, \,y \, z)
\,\, +432  \cdot \,y \, z \, \,\, \, = \,\, \, \,\, 0. 
\end{eqnarray}
\vskip .1cm
{\bf Remark:} If one compares two Heun functions
with the same singular points and the same critical exponents, 
which just differ by their {\em accessory parameter}, namely 
$\, HeunG(9,3/4,1/4,3/4,1,1/2, 36 \,x)$ 
and\footnote[5]{In~\cite{Joycecube} Joyce adopted the Heun function 
notation used by Snow~\cite{Snow}, which corresponds to a 
{\em change of sign in the accessory parameter} $\, q$ 
in the Heun function 
$\, HeunG(a,q,\alpha,\beta, \gamma, \delta, x)$. 
Therefore $\, HeunG(9,3/4,1/4,3/4,1,1/2,*)$ is denoted 
$\, F(9,-3/4,1/4,3/4,1,1/2,*)$ in~\cite{Joycecube}. Unfortunately 
this old notation, different from the one used, for instance, in Maple, 
may contribute to some confusion in the literature. } 
 $\, HeunG(9,-3/4,1/4,3/4,1,1/2, 36 \,x)$, 
one sees that the first one corresponds to a {\em modular form}
and to series with {\em integer coefficients}, while
the second one 
{\em is not even a globally bounded series}~\cite{Short,Big}. These
 two Heun functions
$\, HeunG(9, \pm 3/4,1/4,3/4,1,1/2, 36 \,x)$ 
 are solutions of  order-two linear differential operators
\begin{eqnarray}
\label{firstHeun}
\hspace{-0.9in}&&\quad \quad  \quad 
H_2^{(\pm)} \,\, = \, \,\,
 \theta^2 \, \,  \, -x \cdot (40\, \theta^2+20\, \theta\, \pm 3)
 \, \, \,  
+9 \cdot x^2  \cdot (4\, \theta\, +3) \cdot (4\, \theta\, +1), 
\end{eqnarray}
which are, both, conjugated to their adjoint:
\begin{eqnarray}
\hspace{-0.95in}&&\,
f(x) \cdot \, adjoint(H_2^{(\pm)})  \,\, = \, \, \, H_2^{(\pm)}    \cdot \, f(x)  
\quad  \hbox{with:} \quad   
f(x) \, = \, \, x \cdot \, ((1-36\,x)\cdot (1-4\,x))^{1/2}.
\nonumber 
\end{eqnarray}

\vskip .1cm 

\subsection{Special lattice Green ODEs: face-centered cubic lattice}
\label{warm21}

A third order linear differential operator corresponds
 to the {\em lattice Green function} of the 
{\em face-centered cubic} lattice
(see equation (19) in Guttmann's 
paper~\cite{GoodGuttmann}):
\begin{eqnarray}
\label{G3bcc}
\hspace{-0.9in}&&\quad \quad 
G_3^{fcc} \,\,  = \, \,\,  \, \theta^3 \,\,\, 
- \, 2\,  x \cdot \theta \cdot (\theta +1) \cdot (2\, \theta +1) \, 
 - \, 16\,  x^2 \cdot  (\theta +1) \cdot (5\, \theta^2\, +10\, \theta +6)
 \nonumber \\
\hspace{-0.9in}&& \qquad \quad  \qquad \qquad \, 
-96\,  x^3 \cdot  (\theta +1) \cdot (\theta +2) \cdot (2\, \theta +3),
\end{eqnarray}
where $\, \theta$ is the homogeneous derivative: 
$\, \theta \, = \, \, x \cdot \, d/dx$. 
This operator, once divided by $\, x$, is {\em exactly self-adjoint}: 
$\,\,\, 1/x \cdot \, G_3^{fcc} \, = \, adjoint(1/x \cdot \, G_3^{fcc})$. 
The symmetric square of $\, G_3^{fcc}$ is  of order {\em five} 
(instead of the order six one could expect for generic 
order-three operators).

\vskip .1cm 

Let us introduce, instead of $\, G_3^{fcc}$, the 
equivalent operator $\, \tilde{G}_3^{fcc}$ such
 that\footnote[1]{The operator $\, \tilde{G}_3^{fcc}$ can be obtained 
from the rightdivision of the $\, LCLM(G_3^{fcc}, \, D_x)$ by $\, D_x$,
the  operator $\, S_1^{fcc}$ being obtained  from the rightdivision 
of the $\, LCLM(G_3^{fcc}, \, D_x)$ by  $\, G_3^{fcc}$.}
\begin{eqnarray}
\label{G1TTbcc}
\hspace{-0.9in}&&\quad \quad \quad \quad  \qquad  \qquad 
 S_1^{fcc} \cdot  \, G_3^{fcc}  \, \,\,\, =  \,  \, \,\,\, \, 
   \tilde{G}_3^{fcc} \cdot  \,\,\, D_x.
\end{eqnarray}
where the order-one intertwiner $\, S_1^{fcc}$ reads up to a factor
\begin{eqnarray}
\label{G1Dbcc}
\hspace{-0.9in}&&\quad \quad \quad \quad  \qquad  \qquad
D_x \,  \,  \, 
- {{d \ln(\rho(x)) } \over {dx }},
\end{eqnarray}
where the Wronskian $\, \rho(x)$ is a rational function: 
\begin{eqnarray}
\label{rhoG1TTbcc}
\hspace{-0.9in}&&\quad \quad \quad \quad \quad 
\rho(x) \, \, = \, \, \, \, {{ 6\, x \, + \, 1} \over {
x \cdot \, (4\, x\,+1)^2 \cdot \,(12\,x \,-1) }}.
\end{eqnarray}
We find that the {\em symmetric square} of the equivalent 
operator $\, \tilde{G}_3^{fcc}$ has a {\em rational solution} $\, r(x)$:
\begin{eqnarray}
\label{r(x)}
\hspace{-0.9in}&&\quad \quad \quad \quad \qquad \qquad
r(x) \,\,  = \, \,\, 
{{1} \over {x^2 \cdot \, (4\, x\, +1)^2\,  (12\, x\, -1) }}.
\end{eqnarray}
More precisely, the {\em symmetric square} of the equivalent 
operator $\, \tilde{G}_3^{fcc}$ is the 
{\em direct sum}\footnote[3]{A consequence of the fact that, 
in this case, the differential Galois group is {\em reductive}
so its representations are {\em semi-simple}.} 
of an order-one operator
and an order-five operator:
\begin{eqnarray}
\label{symmsquareDD}
\hspace{-0.9in}&& \quad \,  
Sym^2(\tilde{G}_3^{fcc}) \,\,  = \, \,\,  M_1 \, \oplus \, M_5 
\qquad \hbox{where:} \qquad \quad 
 M_1 \,\,  = \, \,\, D_x \, - \, {{ d \ln(r(x))} \over {dx }}.
\end{eqnarray}

The Wronskian of $\, G_3^{fcc}$  is the square root
of a rational function. The 
differential Galois group is not the generic 
$\, SL(3, \,\mathbb{C})$ one could expect for a generic
order-three operator, but is equal to the orthogonal group 
$ \,O(3, \,\mathbb{C})$:  the rational solution (\ref{r(x)}) 
of $\, Sym^2(\tilde{G}_3^{fcc})$, comes from an invariant 
of degree $\, 2$ for the differential Galois group.

In fact the operator (\ref{G3bcc}) 
happens to be the {\em symmetric square} of an order-two 
operator\footnote[5]{Conjugated to its adjoint by the 
function $\, (1-12\,x)^{1/2} \cdot \, x$}: 
\begin{eqnarray}
\hspace{-0.9in}&&\qquad \quad  \qquad
\theta^2 \,\,\, - \, 2\,  x \cdot \theta \cdot (4\, \theta +1) 
\,\, -24 \, x^2 \cdot \, (\theta +1)  \,  (2\, \theta +1).
\end{eqnarray}
From that last remark, one immediately deduces that the differential 
Galois group must be the differential Galois group of an order-two 
operator, generically $ \,SL(2, \,\mathbb{C})$. Indeed,
 $ \,O(3, \,\mathbb{C})$ is a symmetric square of 
$ \,SL(2, \,\mathbb{C})$ (see~\cite{SingerFano}). It is 
shown in~\cite{SingerFano} that a third order operator has a
symmetric square of order five (instead of six) if, and only if, it is
the symmetric square of a second order operator.
 
\vskip .1cm 

\subsection{Special lattice Green ODEs:  diamond lattice}
\label{warm25}

Another example can be found 
in Guttmann and Prellberg~\cite{GoodGuttmann,Thomas},
 and corresponds to
an order-three operator which has the following $\, _3F_2$
solution
\begin{eqnarray}
{{1} \over {(4\, -\, x^2)^3}} \, \cdot \, \, \, 
_3F_2\Bigl([{{1 } \over {3}}, \, {{1 } \over {2}}, \,{{2 } \over {3}}],\, 
[1, \, 1], \, \, {{27 \, x^4 } \over { (4\, -\, x^2)^3}}   \Bigr), 
\end{eqnarray}
associated with the {\em Green function of the diamond lattice}. This 
 order-three linear differential operator reads:
\begin{eqnarray}
\label{Dia}
\hspace{-0.9in}&&\quad \quad   G_3^{diam}  \, \,  \,  \,= \,\,\,  \, \, \, \,  
64 \cdot \theta^3 \, \,\, - \,16 \,  x^2 \cdot 
(7 \,\theta^3 \, +27 \,\theta^2 \,+42 \,\theta \, +24) 
\, \nonumber \\
\hspace{-0.9in}&&\quad \quad \quad  \quad \quad  \quad \quad
+ \,12 \,  x^4 \cdot  
 (5 \,\theta^3 \, +42 \,\theta^2 \,+124 \,\theta \, +128) \,
  \\
\hspace{-0.9in}&&\quad \quad  \quad \quad \quad \quad \quad \quad \quad
- \, \,  x^6 \cdot  
 (13 \,\theta^3 \, +171 \,\theta^2 \,+762 \,\theta \, +1152)
\,\,   \, + \, \,  x^8 \cdot (\theta \, +6)^3, 
 \nonumber
\end{eqnarray}
which can be seen to be conjugated to its adjoint:
\begin{eqnarray}
\hspace{-0.9in}&&\qquad \quad \quad 
{{(x^2-4)^2} \over {x }} \cdot  G_3^{diam}  
\,\,\, \, \, \, = \,\,  \,\,\, \, \, \,
adjoint(G_3^{diam}) \cdot {{(x^2-4)^2 } \over {x }}.
 \nonumber
\end{eqnarray}
The symmetric square of $\, G_3^{diam}$ is  of order {\em five} 
(instead of the order six one could expect for generic 
order-three operators).

Let us introduce, instead of $\, G_3^{diam} $, the 
equivalent operator $\, \tilde{G}_3^{diam}$ 
\begin{eqnarray}
\label{G1TTDiam}
\hspace{-0.9in}&&\quad \quad \quad \quad  \quad \quad \quad 
 S_1^{diam} \cdot  \, G_3^{diam}   \, \,\,\, =  \, \,\,\, \,
    \tilde{G}_3^{diam} \cdot  \,\,\, D_x.
\end{eqnarray}
where the order-one intertwiner $\, S_1^{diam}$ reads
\begin{eqnarray}
\label{G1Diam}
S_1^{diam}  \,\,\, =  \, \,\,\, 
 -\, {{ (x^2-4)^2 } \over { x }}  \cdot \,r(x) \cdot \, 
\Bigl(D_x \,  \, 
- {{ 1 } \over {\rho(x) }} \cdot \,
 {{d \rho(x) } \over {dx }}    \Bigr),
\end{eqnarray}
where $\, r(x)$ and $\, \rho(x)$ are rational functions: 
\begin{eqnarray}
\label{RG1TTDiam}
\hspace{-0.9in}&&\quad \quad \quad \quad \,
r(x) \,\, =  \, \,\,
 {{1} \over {(x-2)^5 \, (x-1) \, (x+1) \, (x+2)^5 \cdot \, x^2}},
 \qquad \quad \, \, \hbox{and:}
\end{eqnarray}
\begin{eqnarray}
\label{rhoG1TTDiam}
\hspace{-0.9in}&&\quad \quad \quad \quad \, 
\rho(x) \, \, = \, \, \, 
(3\, x^3\, -8\, x+4)\, (3\, x^3\, -8\, x\, -4) \cdot \, x^2.
\end{eqnarray}
Again, the {\em symmetric square}
 of the equivalent operator $\, \tilde{G}_3^{diam}$ 
has the {\em rational solution} $\, r(x)$.

The Wronskian of (\ref{Dia}) is the square root
of a rational function. The 
differential Galois group is again  $ \,O(3, \,\mathbb{C})$.

As in the previous example, $\, G_3^{diam}$  is 
the {\em symmetric square} of an order-two operator: 
\begin{eqnarray}
\label{Dia2}
\hspace{-0.9in}&&\qquad \qquad \quad 
 16 \cdot \theta^2 \,\, - 12 \, x^2 \cdot \, (2 \,\theta^2\, +3 \,\theta \, +2)
\nonumber \\
\hspace{-0.9in}&&\qquad \qquad \qquad \qquad 
 \, +3 \, x^4 \cdot \,  (3 \,\theta^2\, +11 \,\theta \, +12)
 \,\,  -\, x^6 \cdot \, (\theta \, +3)^2.
\end{eqnarray}

\vskip .1cm 

\subsection{Order-three operators conjugated to their adjoint}
\label{Joyce}
In fact, these  previous results, for the bcc, sc, fcc, diamond lattices,
 can be seen as the consequence of the
 following general result on {\em order-three} linear differential 
operators (without any loss of generality we restrict
to monic operators) 
\begin{eqnarray}
\hspace{-0.9in}&&\qquad \quad  \qquad
L_3 \, \, = \, \,\,  \, D_x^3 \,\,  \, + \, a_2(x) \cdot \, D_x^2 
\,  \,+ \, a_1(x) \cdot \, D_x\,  \,+ \, a_0(x).  
\end{eqnarray}
Any (monic) order-three operator which is conjugated to its adjoint, 
namely \\
$\, L_3 \cdot \, f(x) \, = \, \, \, f(x) \cdot \, adjoint(L_3)$, 
is the symmetric square of an order-two operator
\begin{eqnarray}
\label{L2sousjacent}
\hspace{-0.9in}&&\,\,
L_2 \,\, = \, \,\, 
D_x^2 \, + \, b_1(x) \cdot \, D_x\, + \, b_0(x), 
 \quad \quad  \, \hbox{with:} \,\, \quad \, \, \,
 b_1(x) \,  = \,
 -{{1} \over {2}}  \, {{1} \over {f(x) }} \, {{d f(x) } \over {dx }}, 
\end{eqnarray}
\begin{eqnarray}
\hspace{-0.9in}&& \quad \hbox{where:}\quad \quad \quad
b_0(x) \, \, \, = \,\, \, \, \, {{a_1(x)} \over {4}}  \, \,\, 
 +{{1} \over {8}}\,  {{1} \over {f(x) }} \cdot \, {{d^2 f(x) } \over {dx^2}} 
\, \,  \, -{{1} \over {4}} \cdot \, \Bigl({{1} \over {f(x) }} \cdot \, 
{{df(x) } \over {dx}} \Bigr)^2. 
\nonumber   
\end{eqnarray}
Note that one necessarily has $\, a_2(x) \, = \, \, 3 \, b_1(x)$.
The Wronskian of $\, L_3$ is necessarily equal to $\, f(x)^{3/2}$, and  
the order-two operator (\ref{L2sousjacent}) is conjugated 
to its adjoint by a function: 
\begin{eqnarray}
\hspace{-0.9in}&&\qquad \qquad \quad 
f(x)^{1/2} \cdot \, adjoint(L_2) 
\,\, \,\,  = \, \,\,\,  \,  L_2  \cdot \, f(x)^{1/2}.
\end{eqnarray}
The symmetric square of such an order-three operator $\, L_3$, conjugated 
to its adjoint, is  of order {\em five} (in contrast to order six 
for symmetric squares of generic order-three operators).

\vskip .1cm 

\subsection{Special lattice Green ODEs: 4D
 face-centered cubic lattice}
\label{warm22}

A slightly more involved example, corresponding to
the {\em four-dimensional
 face-centered cubic lattice Green function},
 can be found in paragraph 2.5 
of Guttmann's paper~\cite{GoodGuttmann} (it is also ODE 
number 366 in the list of Almkvist
et al.~\cite{TablesCalabi}). This order-four linear 
differential operator  
\begin{eqnarray}
\label{FC}
\hspace{-0.9in}&&\qquad  G_4^{4Dfcc} \,\,\,\, = \, \,\,\, \, \,
\theta^4 \,\,\, 
+ \,  x \cdot (39 \cdot \theta^4 -30\cdot \theta^3
   -19\cdot \theta^2 -4 \, \theta)
 \nonumber \\
\hspace{-0.9in}&&\qquad \quad \quad  \quad \quad 
\,+ \, 2\,  x^2 \cdot (16 \cdot \theta^4 -1070 \cdot \theta^3
   -1057 \cdot \theta^2 -676 \, \theta\, -192) 
\nonumber \\
\hspace{-0.9in}&&\qquad \quad  \quad \quad  \quad \,
- \, 36\,  x^3 \cdot (171 \cdot \theta^3 
   +566 \cdot \theta^2 +600 \, \theta\, +316)
\cdot (3\, \theta\, +2)
\nonumber \\
\hspace{-0.9in}&&\qquad \quad \quad \quad \quad 
 \,- \,2^5 \, 3^3 \,  x^4 \cdot 
(384 \cdot \theta^4 +1542 \cdot \theta^3
   +2635\cdot \theta^2 +2173 \, \theta\, +702)
\nonumber \\
\hspace{-0.9in}&&\qquad \quad  \quad \quad \quad \,
 - \,2^6 \, 3^3 \,  x^5 \cdot (1393 \cdot \theta^3 
   +5571 \cdot \theta^2 +8378 \, \theta\, +4584)
\cdot (\theta\, +1)
\nonumber \\
\hspace{-0.9in}&&\qquad \quad \quad  \quad \quad 
 \,- \,2^{10} \, 3^5 \,  x^6 \cdot (31 \cdot \theta^2
 +105 \, \theta\, +98)
\cdot (\theta\, +1) \cdot (\theta\, +2) \\
\hspace{-0.9in}&&\qquad \quad  \quad \quad  \quad \,
- \,2^{12} \, 3^7 \,  x^7 \cdot 
(\theta\, +1) \cdot (\theta\, +2)^2 \cdot (\theta\, +3)
\nonumber \\
\hspace{-0.9in}&&\quad \,  \, = \, \, \, \,  \,
x^4 \cdot \, \,(1+3\,x)\,(1+4\,x)\,  (1+8\,x)
(1+12\,x)\,(1+18\,x)^2 \,(1-24\,x) \cdot D_x^4 
\, \,\,\, + \, \, \cdots
 \nonumber 
\end{eqnarray}
can be seen to be conjugated to its adjoint by a function $\,f^{4Dfcc}$:
\begin{eqnarray}
\hspace{-0.95in}&&   
G_4^{4Dfcc}  \cdot f^{4Dfcc} \,\,\, = \, \,\,
f^{4Dfcc} \cdot adjoint(G_4^{4Dfcc}), 
 \quad \hbox{with:}\, \quad \, 
 f^{4Dfcc} \, = \, \, x \cdot (1\, +18 \, x)^3.
\nonumber
\end{eqnarray}
The {\em exterior square}  of operator (\ref{FC}) 
is an irreducible {\em order-five} operator 
(not order-six as could be expected):
one easily checks that 
the ``Calabi-Yau condition'' (see~\cite{Almkvist} 
and (\ref{CalabiCond}) below)
is actually satisfied for operator (\ref{FC}).
If one considers
an operator $\, \tilde{G}_4^{4Dfcc}$, non-trivially
 homomorphic~\cite{Singer,vdP} to $\, G_4^{4Dfcc}$,
its {\em exterior square} is, now, an operator of (the generic)
order six, and it has a {\em rational solution}. For instance, 
if we consider the operator $\, \tilde{G}_4^{4Dfcc}$ 
equivalent to $G_4^{4Dfcc}$ 
\begin{eqnarray}
\label{G1FC}
\hspace{-0.9in}&&\quad \quad \quad 
  \quad  \quad  \quad \quad 
 S_1^{4Dfcc} \cdot  \, G_4^{4Dfcc}   \,\,\, =  \, \,\,\, \, 
    \tilde{G}_4^{4Dfcc} \cdot  \,\,\, D_x.
\end{eqnarray}
where 
\begin{eqnarray}
\label{G1FCs}
\hspace{-0.9in}&&\quad \quad \quad  \, 
S_1^{4Dfcc}  \,  \,\,\, =  \, \,\,\, \, \,
-\, {{ r(x)} \over { (1\, + \, 18 \, x)^3 \cdot \, x}} \cdot \, 
\Bigl(D_x \, - \, {{ d\ln(\rho(x))} \over {dx}} \Bigr), \, \, 
\qquad \hbox{ with} \\
\label{rRG1FC}
\hspace{-0.9in}&&\quad \quad \,  r(x)
\,\, =  \, \,\,\,\, 
 {\frac {18\,x+1}{{x}^{3} \cdot \,  
\, (3\,x+1) \,  (4\,x+1) \, 
  \left( 8\,x+1 \right) \, (12\,x+1) \, (24\,x-1) }},
\qquad \quad \hbox{ and}
 \nonumber \\
\label{rhoRG1FC}
\hspace{-0.9in}&&\quad \quad \,
\rho(x) \, \,\, = \, \, \,\, \,
(1119744 \,x^5 \, +508032 \,x^4 \, +82512 \,x^3 
\, +6318 \,x^2 \, +237 \,x \, +4) \cdot \, x,
\nonumber 
\end{eqnarray}
we find that the {\em exterior square} of $\, \tilde{G}_4^{4Dfcc}$ 
has the {\em rational solution} $\, r(x)$.

This is a situation we will encounter many times:
switching to an equivalent operator ``desingularizes'' 
the drop of order of the exterior (resp. symmetric)
 $\, n$-th power of an operator, into a situation of 
emergence of a rational solution for that $\, n$-th power. 

The Wronskian of $\, G_4^{4Dfcc}$  is a rational function. As the 
exterior square of $\, \tilde{G}_4^{4Dfcc}$ 
has a {\em rational solution}, the 
differential Galois group is included in the {\em symplectic group}
  $\, Sp(4, \,\mathbb{C})$. Moreover, its symmetric square
 being irreducible, theorems A.5 et A.7
of Beukers et al.~\cite{BBH} show that the differential Galois group 
 is exactly $\, Sp(4, \,\mathbb{C})$.

\vskip .1cm 

\subsection{Another version of the order four operator
for 4D fcc lattice}
\label{Order-four}

Another version\footnote[5]{Corresponding to a change of 
variable: $\, F(x)    \, \rightarrow \, F(x/(1-18\,x))/(1-18\,x)$.} of 
 $\,G_4^{4Dfcc}$, can be found in the unpublished paper 
of D. Broadhurst (see equation (68) in~\cite{Broad})
and corresponds to the order-four operator
\begin{eqnarray}
\label{Broad}
\hspace{-0.9in}&&\quad \quad G_4^{4D} \, \, = \, \, \,  \, \,
(2\, \theta)^4 \,\, \,
 + \,  \sum_{j=1}^{6} \, (-1)^j \cdot \,P_j(2\, \theta \, +j)
 \, \,\,\,\,  = \, \, \,\,  \,    \, 
h_4 \cdot D_x^4 \,\,\, + \, \, \cdots, 
\end{eqnarray}
with 
\begin{eqnarray}
\hspace{-0.95in}&&P_1(u) \,  = \,  \,
  105\, u^4 +166\, u^2\,  +17,  
  \quad \quad \quad \quad \, \, 
P_2(u) \,   = \, \, 2 \cdot (2095 \, u^4 +2912\, u^2\,  +432),
  \nonumber \\
\hspace{-0.95in}&&P_3(u) \,  = \, \, 
 72 \cdot \,(1155\, u^4 \, -892\, u^2\,  +577), 
 \,   \, \,    \, 
 P_4(u) \,  = \, \, 
 864 \cdot \, (1011 \, u^4 \, -5059\, u^2\,  +4900), 
  \nonumber \\
\hspace{-0.95in}&&P_5(u) \,  \, = \, \, \,  \, 
75600 \cdot \, (u^2 \, -\,9) \, (61\, u^2 \, -\,145), 
\,   \, \,  \,   P_6(u) \,  = \, \,
 9525600 \cdot \, (u^2 \, -\,4) \, (u^2 \, -\,16),
\nonumber 
\end{eqnarray}
and where the head polynomial $\, h_4$ reads: 
\begin{eqnarray}
\hspace{-0.9in}&&\quad 
 h_4 \,  \, \,= \, \, \,  \, \,
16\, x^4 \cdot (1-6\,x)\,(1-10\,x) 
\,(1-14\,x)  \,(1-15\,x) \,(1-18\,x)\,(1-42\,x).
 \nonumber 
\end{eqnarray}
Operator $\, G_4^{4D}$ is a globally nilpotent operator, 
as can be seen on the Jordan 
form reduction of its
 $\, p$-curvature~\cite{bo-bo-ha-ma-we-ze-09,Deitweiler}.

This irreducible order-four operator $\, G_4^{4D}$ 
is conjugated to its adjoint 
(or exactly self-adjoint with a different normalization):
\begin{eqnarray}
\hspace{-0.6in}&&\qquad \quad \quad \quad 
G_4^{4D} \cdot \, x \, \, \,  \, = \, \, \,  \, \,  \, 
x \, \cdot \, adjoint(G_4^{4D}). 
\end{eqnarray}
Its {\em exterior square} is an irreducible 
{\em order-five} linear differential 
operator (not order-six as could be expected):
 the ``order-five Calabi-Yau condition'' 
(see (\ref{CalabiCond}) below)
is  satisfied for (\ref{Broad}).
Similarly to  (\ref{G1FC}) and (\ref{G1TT}), 
one can switch, by operator equivalence, to 
an operator $\, \tilde{G}_4^{4D}$ such that its {\em exterior square}
has a {\em rational solution}: 
\begin{eqnarray}
\label{G1TT68}
\hspace{-0.9in}&&\quad \quad \quad  \quad \quad \quad \quad \quad 
 S_1^{4Dfcc} \cdot  \, G_4^{4D}  \,\,\,\, =  \, \,\,\, \,\,
   \tilde{G}_4^{4D} \cdot  \,\,\, D_x.
\end{eqnarray}
where the order-one intertwiner $\, S_1^{4D}$ reads
\begin{eqnarray}
\label{G1Diam}
\hspace{-0.5in}&&\quad \quad \quad  \quad \quad 
S_1^{4D}  \,\,\, =  \, \,\,\, 
 -\, {{  r(x)} \over { x }}  \cdot 
\Bigl(D_x \,  \, 
- {{d \ln(\rho(x)) } \over {dx }}   \Bigr),
\end{eqnarray}
with
\begin{eqnarray}
\label{rhoG1TT68}
\hspace{-0.9in}&&\quad  \, 
\rho(x) \, \, = \, \, \, 
(63504000\,x^5\, -17388000\,x^4\,+1644948\,x^3\,
-64578\,x^2\,+950\,x\,-3) \cdot  \, x, 
\nonumber
\end{eqnarray}
and 
where $\, r(x)$ is the rational function:
\begin{eqnarray}
\label{rG1TT68}
\hspace{-0.95in}&&\quad \,  
r(x) \,   = \, \,\,  
{\frac {1}{{x}^{3} \cdot \, \, (6\,x-1)  \, (10\,x-1)\,
  \, (14\,x-1)  \, (15\,x-1) \,
  \, (18\,x-1)  \, (42\,x-1) }}.
\end{eqnarray}

As above, the {\em exterior square} of 
the equivalent operator $\, \tilde{G}_4^{4D} $
has (obviously) the {\em rational solution} $\, r(x)$:
the differential Galois group of $\, G_4^{4D} $ is  not 
$\, SL(4, \,\mathbb{C})$
but is included in the {\em symplectic group}  $ \,Sp(4, \,\mathbb{C})$.

One can get, directly, a rational solution if one switches
to the {\em linear differential system} associated 
with  $\, G_4^{4D}$, calculates\footnote[1]{In order to do these 
calculations download the three Maple Tools files
TensorConstructions.m and IntegrableConnections.m 
in the web page~\cite{Secret}. Using DEtools, you will need to use, on
the order-six operator $\,G_4^{4D}$ written in a monic way,
 the command  companion-system, then the command exterior-power-system( ,2),
 and, finally, the command 
RationalSolutions([],[x]).
} 
the {\em exterior square of that system} 
and seeks for the rational solution 
of that exterior square system.

One gets, that way, the rational solution 
of the exterior square  of the {\em companion system} of  $\,G_4^{4D}$:
\begin{eqnarray}
\hspace{-0.7in}&&\, \qquad \quad  
\Big[0, \,  \,  \,\, 0, \, \,  \, \, r(x),  \,  \,  \,
\, -\, r(x),  \,  \,  \, \,
 {x}^{2} \cdot \, q_5 \cdot \, r(x)^2, \, 
 -\, x \cdot \, q_6 \cdot \, r(x)^2\Bigr],   
\end{eqnarray}
where:
\begin{eqnarray}
\hspace{-0.95in}&&q_5 \,  = \, \, 
\, 85730400\,{x}^{6}\, -36892800\,{x}^{5}\,+6114528\,{x}^{4}\,-498960\,{x}^{3}\,
+20950\,{x}^{2}\,-420\,x\,+3, 
\nonumber \\
\hspace{-0.95in}&&q_6 \,  = \, \,  190512000\,{x}^{6}
-72198000\,{x}^{5}+10262808\,{x}^{4}-690804\,{x}^{3}+22406\,{x}^{2}-304\,x+1.
\nonumber 
\end{eqnarray}

At first sight, for a physicist, performing calculations 
on a linear differential operator, like  $\, \tilde{G}_4^{4D} $, 
and finding that its exterior square has the 
{\em rational solution} (\ref{rG1TT68}), seems simpler, and 
more natural, than introducing the {\em companion system}. 
As far as practical calculations are concerned, the calculations on the 
linear differential systems turn out to be {\em drastically more efficient},
and allow to handle symmetric and exterior powers constructions 
on larger examples.

\vskip .1cm

\vskip .1cm 

\subsection{Special lattice Green ODEs: 5D staircase polygons}
\label{warm24}

Another example of Guttmann and Prellberg~\cite{GoodGuttmann,Thomas},
corresponding to the generating
function of the {\em five-dimensional staircase polygons},
 is the order-four operator 
\begin{eqnarray}
\label{Prell}
\hspace{-0.9in}&&\quad  G_4^{5D} \, \,  \,  \,= \, \, \, \, \,  \,  \,
\theta^4 \, \,\,  \,  - \, x \cdot 
(35\, \theta^4 \, +70 \,\theta^3 \, +63 \,\theta^2 \,+28 \,\theta \, +5) 
\\
\hspace{-0.9in}&&\,  \quad \quad  \quad \qquad 
+ \, x^2 \cdot (259\, \theta^2 \, +518 \,\theta \, +\, 285) 
\cdot  (\theta \, +\, 1)^2\, 
- \,225\,  x^3 \cdot   (\theta \, +\, 1)^2 
\cdot   (\theta \, +\, 2)^2
\nonumber \\
\hspace{-0.9in}&&\qquad  \quad \quad \quad \quad  \qquad 
= \, \, \, \,\,   \,\,  
x^4 \cdot (1\,+35\,x\, +259\, x^2\, -225 \,x^3) \cdot D_x^4
 \,\,\,  \, \, + \, \, \cdots
  \nonumber 
\end{eqnarray}
which can be seen to be conjugated  to its adjoint:
\begin{eqnarray}
\hspace{-0.9in}&&\qquad \qquad  \quad \quad  
G_4^{5D} \cdot x 
\,\,\, \, = \,\,  \,\,\, \, \,
x \cdot adjoint(G_4^{5D}).
 \nonumber
\end{eqnarray}
The exterior square operator of the  order-four operator (\ref{Prell}) 
is an irreducible {\em order-five} operator (not order-six as could be expected):
 the ``order-five Calabi-Yau condition'' (see 
(\ref{CalabiCond}) below) is satisfied for (\ref{Prell}).
Let us introduce, instead of $\, G_4^{5D}$, the equivalent operator 
$\, \tilde{G}_4^{5D}$ corresponding to the intertwining relation
\begin{eqnarray}
\label{G1TT}
\hspace{-0.9in}&&\quad \quad \quad \quad \quad  \quad \quad 
 S_1^{5D} \cdot  \, G_4^{5D}  \, \,\,\, =  \, \,\,\,  \, \, 
   \tilde{G}_4^{5D} \cdot  \,\,\, D_x,
\end{eqnarray}
where the order-one intertwiner $\, S_1^{5D} $ reads
\begin{eqnarray}
\label{G1}
S_1^{5D}  \,\,\, =  \, \,\,\,\,   -\, {{r(x)} \over {x }}  \cdot 
\Bigl(D_x \,  \, 
- {{d \ln((60\, x\, +1)\, (3\,x\, -1) \, x) } \over {dx }}   \Bigr),
\end{eqnarray}
and where $\, r(x)$ is the rational function:
\begin{eqnarray}
\label{RG1TT}
\hspace{-0.9in}&&\quad \quad \quad \quad \quad \quad 
r(x) \,\,\, =  \, \,\,\, \, 
 {{1} \over {(225\, x^3  \, -259 \, x^2 \, -35 \, x \, -1) \cdot \, x^3}}.
\end{eqnarray}
We find, again, that the exterior square of the equivalent operator 
$\, \tilde{G}_4^{5D}$ 
has the {\em rational solution} $\, r(x)$. The Wronskian 
of $\, G_4^{5D}$ is a rational function. The 
differential Galois group is, again (see A.5 et A.7 in 
appendix A of~\cite{BBH}), the {\em symplectic group}
  $\, Sp(4, \,\mathbb{C})$.

\vskip .1cm

\vskip .1cm

\subsection{Order-six operator by Broadhurst 
and Koutschan: the lattice Green function of  
the five-dimensional fcc lattice\newline}
\label{Order-six}

A more involved example of order-six, 
can be found in Koutschan's paper~\cite{Kou} 
and in an unpublished paper 
of D. Broadhurst (see equation (74) in~\cite{Broad})
and corresponds to a {\em five-dimensional fcc lattice}
\begin{eqnarray}
\label{Broad6}
\hspace{-0.9in}&&\quad \quad 
G_6^{5Dfcc} \,  \,\, = \, \,  \, \,  \, \,
3^4 \cdot \theta^5 \cdot (\theta\, -1) \,  \,\, 
+ \,  \sum_{j=1}^{12} \, x^j \cdot \,Q_j(\theta)
 \, \, \, \,  = \, \, \,  \, \,  
h_6 \cdot D_x^6 \, \,  \,  + \, \, \cdots , 
\end{eqnarray}
where the polynomials $\, Q_j$ are degree-six polynomials
with integer coefficients,
and where the head polynomial $\, h_6$ reads: 
\begin{eqnarray}
\label{h6lambda}
\hspace{-0.9in}&&\quad \quad h_6\, \, = \, \, \,  \, \,
 x^6 \cdot \, \lambda(x) \,  \cdot p_6,
\nonumber  \\
\hspace{-0.9in}&&\hbox{with:} \quad \quad \quad
\lambda(x)  \, = \, \, \,  \, 
(1-4\,x) \,(1-8\,x)\, (1+16\,x)\, (1-16\,x)
\, (1-48\,x)\, (3-16\,x), 
\nonumber  \\
\hspace{-0.9in}&&  \hbox{and:} \quad \quad \quad
  p_6\, \,  \, = \, \, \,  \,  \, \, 916586496\, x^6\,\,  
-571981824\, x^5 \, \, +67242496\, x^4\, \, 
-8372096\, x^3
\nonumber \\
\hspace{-0.9in}&&\qquad \qquad \qquad \qquad \quad \quad 
 +315096\, x^2\,\, \,  -6840\, x\, \,  +27. 
\end{eqnarray}

This order-six linear differential operator has,
 at the origin $\, x\, = \, \, 0$,
{\em two independent analytic solutions} (no logarithms, 
it is {\em not} MUM\footnote[1]{MUM means maximally unipotent 
monodromy~\cite{GoodGuttmann,CalabiYauIsing,Almkvist1}.}~\cite{Big}). 
 One can  build, from these two solutions, a one-parameter 
family of analytic solutions:  
\begin{eqnarray}
\hspace{-0.9in}&& \, \, \, \,\,\,   1 \,\,\,   +8 \cdot x \cdot c
\, \, \,\,\,
  + {{8} \over {3}} \cdot (41 \cdot c \, -2) \cdot x^2 \,\,  \, \,
  + {{32} \over {27}} \cdot (1933 \cdot c -286) \cdot x^3 
 \nonumber \\
\hspace{-0.9in}&&   \, \,  \, \,  \quad \,\,\,
 + {{8} \over {27}} \cdot (183136 \cdot c -25537) \cdot x^4\,\,\, 
\, + {{256} \over {2025}} \cdot (12082067 \cdot c -1788992) \cdot x^5
 \, \,  \, \, +  \, \cdots \nonumber 
\end{eqnarray}
which, for $\, c\, = \, 1$, (and only this value) becomes a series 
with {\em integer}\footnote[2]{The integrality of these coefficients 
has been checked with $\, 2000$ coefficients, and  
the coefficients  
$\, c_{c \cdot 10000} \cdot x^{c \cdot 10000}$ coefficients,
for $\, c\, = \, 1, \, 2, \, 3, \, 4$, have also been seen to be 
integers.
}
 coefficients:
\begin{eqnarray}
\hspace{-0.8in}&&1 \,\,\, +8 \cdot x 
\,\, \,+104 \cdot x^2 \,\,\, +1952 \cdot x^3
\, \,+46696 \cdot x^4\, \,\,
 +1301248 \cdot x^5 \,\, \, +40047584 \cdot x^6 
 \nonumber \\
\hspace{-0.8in}&& \quad  \quad \quad 
\,+1319992192 \cdot x^7 \,\,\, \,+45737941096 \cdot x^8 \, \,\,
+1646328483008 \cdot x^9 \,\,\,\,
\, + \, \cdots \nonumber 
\end{eqnarray}

The question of the integrality of such D-finite series,
emerging from physics,
is addressed in previous papers~\cite{Short,Big}.

\vskip .1cm

{\bf Remark:} The other unique independent no-log series 
starting with $\, x$ reads:
\begin{eqnarray}
\hspace{-0.9in}&&z_0(x) \,\,  \, = \, \, \, \,\,\, 
 x \,\,\,  \,  + {{41 }  \over { 3}} \cdot x^2\,\, \, \, 
+ {{7732 }  \over {27 }} \cdot x^3\,\, \, 
+ {{183136 }  \over {27 }}  \cdot x^4\,\,\,   
+ {{386626144 }  \over {2025 }} \cdot x^5
\nonumber \\
\hspace{-0.9in}&& \quad  \quad  \qquad \quad
\,+ {{ 106836145888}  \over {18225 }} \cdot x^6\, 
\,+ {{172725353100416 }  \over {893025 }} \cdot x^7
\,\,\, + \,\,  \cdots 
\nonumber 
\end{eqnarray}
It is {\em not a globally bounded} series~\cite{Short,Big}, 
i.e. it is {\em not} a series that can be 
recast into a series with integer coefficients after a rescaling 
of the variable.
%  by an integer. 

\vskip .1cm 

This order-six linear differential 
operator is
 {\em globally nilpotent}~\cite{bo-bo-ha-ma-we-ze-09,Deitweiler}, 
the Jordan reduction of its
 $\, p$-curvature~\cite{bo-bo-ha-ma-we-ze-09,Deitweiler} reading: 
\begin{eqnarray}
\hspace{-0.8in}&&\qquad \qquad \quad \quad 
J_6 \, \, = \, \, \, \,\,
\left[ \begin {array}{cccccc}
0&0&0&0&0&0 \\ \noalign{\medskip}
0&0&1&0&0&0 \\ \noalign{\medskip}
0&0&0&1&0&0 \\ \noalign{\medskip}
0&0&0&0&1&0 \\ \noalign{\medskip}
0&0&0&0&0&1 \\ \noalign{\medskip}
0&0&0&0&0&0
\end {array} \right].
\end{eqnarray}

We found that the order-six operator $\, G_6^{5Dfcc}$
is {\em non-trivially homomorphic to its adjoint}, with 
a simple {\em order-one} intertwiner
\begin{eqnarray}
\label{Homo1}
\hspace{-0.9in}&&\qquad \quad  \quad G_6^{5Dfcc} \cdot T_1^{5Dfcc} 
\, \, \, = \, \, \,  \,\,  
 adjoint(T_1^{5Dfcc}) \cdot \, adjoint(G_6^{5Dfcc}),
\end{eqnarray}
with:
\begin{eqnarray}
\hspace{-0.9in}&&\qquad \quad T_1^{5Dfcc} 
\, \, \, = \, \,\,  \,  \, \, x^2 \cdot p_2 \cdot p_6 \cdot
 \Bigl(D_x  \, - \, \, 
{{1} \over {2}} \cdot {{ d \ln(R(x)) } \over {d x }}  \Bigr),
 \qquad  \quad \quad \hbox{where}
 \nonumber \\
\label{p2}
\hspace{-0.9in}&&\qquad \quad 
R(x) \, = \, \, \,  {{p_2^5 } \over {x^4 \cdot p_6^4 }}  
\quad \quad \quad  \,  \hbox{with} \quad  \quad \quad  \, \, 
 p_2 \, \, = \, \, \,  \, \, 1152\, x^2\, -56\, x\, -3.
\end{eqnarray}
Introducing 
\begin{eqnarray}
\label{rho}
\hspace{-0.9in}&&\qquad \quad \quad \quad \qquad 
\rho(x) \, \, \, = \, \,\,  \,  \, \,
 {\frac {{p_2}^{6}}{{p_6}^3 \cdot \, x^2}},
\end{eqnarray}
the previous order-one intertwiner $\, T_1^{5Dfcc}$, 
can be seen as the product  
of the rational function $\, \rho(x)$,
 and of a {\em self-adjoint} order-one operator $\, Y^{s}_1$:
\begin{eqnarray}
\label{sY1}
\hspace{-0.9in}&&\quad \quad \, 
 T_1^{5Dfcc} \, = \,  \, \rho(x) \cdot Y^{s}_1, 
\quad  \quad \quad
Y^{s}_1  \, \,  \, = \, \, \, \ \,  {{1} \over {R(x)}}
\cdot  \Bigl(D_x  \, - \, \, 
{{1} \over {2}} \cdot {{ d \ln(R(x)) } \over {d x }}  \Bigr).
\end{eqnarray}

\vskip .1cm 

The other intertwining relation is a bit more involved
 since the intertwiner is an 
{\em order-five} linear differential operator $\, S_5^{5Dfcc}$
\begin{eqnarray}
\label{Homo5}
\hspace{-0.9in}&&\qquad \quad  adjoint(S_5^{5Dfcc}) \cdot G_6^{5Dfcc} 
\, \, \, \,  = \, \, \,  \, \, adjoint(G_6^{5Dfcc}) \cdot S_5^{5Dfcc}, 
\end{eqnarray}
where
\begin{eqnarray}
\hspace{-0.9in}&&\quad \qquad \quad 
S_5^{5Dfcc} \, \, \, \, = \, \, \,  \, \,  \,
{{x^2 \cdot \, \lambda(x)  \cdot \, p_2^5 } \over { p_6^3}}  \cdot \Bigl( 
D_x^5 \, \,  \,
- \, {{1} \over {2}} \cdot  {{d \ln(\mu(x))} \over {d x}} 
 \cdot D_x^4 \,  \, + \,\,  \cdots 
\Bigr)  \nonumber 
\end{eqnarray}
with $\, \lambda(x)$ as above in (\ref{h6lambda}), and:
\begin{eqnarray}
\hspace{-0.7in}&&\, \, \quad \quad \quad \quad 
 \mu(x)\,\, \, = \, \,\, \,  \, \,
 - \, {{  p_2^5} \over { \lambda(x)^5 \cdot \, x^{20}}}.\qquad   
\nonumber 
\end{eqnarray}

Quite remarkably, introducing the {\em same} function
 $\, \rho(x)$ as for $\, T_1^{5Dfcc}$ (see (\ref{rho})),
the previous order-five intertwiner $\, S_5^{5Dfcc}$, can be seen as the product  
$\, S_5^{5Dfcc} \, = \,  \, \rho(x) \cdot Y^{s}_5 $,
of the rational function $\, \rho(x)$ (see (\ref{rho})) 
and of a {\em self-adjoint} order-five operator
\begin{eqnarray}
\label{sY5}
\hspace{-0.9in}&&\qquad \quad \quad \quad 
Y^{s}_5  \, \,  \, = \, \, \, \ \,  {{ x^4 \cdot \, \lambda(x)} \over {p_2}}
\cdot    \Bigl( 
D_x^5 \, \,  \,
- \, {{1} \over {2}} \cdot  {{d \ln(\mu(x))} \over {d x}} 
 \cdot D_x^4 \,  \, + \,\,  \cdots 
\Bigr).
\end{eqnarray}

The self-adjoint order-five  irreducible operator $\, Y_5^{s}$ has a solution 
analytic at $\, x \, = \, 0$, (the other solution have log terms),
 which has the following expansion
\begin{eqnarray}
\hspace{-0.9in}&&\quad \,\, \, \,  
1\, \,\, +8\,x\, \, +102\,{x}^{2}\, \, +{\frac {487192}{243}}\,{x}^{3}\, \, 
+{\frac {86597215}{1944}}\,{x}^{4}\, \,
 +{\frac {22841991292}{16875}}\,{x}^{5}\,
\nonumber  \\
\hspace{-0.9in}&&\quad  \qquad  \quad 
 \, +{\frac {1874527149707741}{49207500}}\,{x}^{6}\, \, 
+{\frac {40302470144568331141}{29536801875}}\,{x}^{7}\,
\, \,\, + \,\,  \cdots 
\end{eqnarray}
This solution-series is {\em not globally bounded}~\cite{Short,Big}.
The examination of the formal series solutions at $\, x\, = \, 0$
corresponds to a MUM structure. 

The self-adjoint order-five irreducible operator $\, Y_5^{s}$ 
is such that its {\em symmetric square 
is of order 14 instead of the order 15 expected generically}
 (its exterior square 
is of order 10 as it should, with no rational solution).

The Wronskian of this order-six linear differential 
operator $\, G_6^{5Dfcc}$ is the square root of a 
rational function: 
\begin{eqnarray}
\hspace{-0.6in}&&\quad \qquad 
 W(G_6^{5Dfcc}) \,\, \,= \, \,\,\, \,
\Bigl( {{p_6^2} \over {x^{28} \cdot \lambda(x)^7 }} \Bigr)^{1/2}.
 \nonumber 
\end{eqnarray}

The previous homomorphisms of the order-six operator $\, G_6^{5Dfcc}$ with
its adjoint, namely (\ref{Homo1}) and (\ref{Homo5}), can be simply rewritten
 in terms of the self-adjoint operators $\, Y^{s}_1$ and $\, Y^{s}_5$:
\begin{eqnarray}
\label{Homo1Homo5}
\hspace{-0.9in}&&\qquad \quad  \quad  
G_6^{5Dfcc}\cdot  \rho(x) \cdot \,  Y^{s}_1 
\, \, \, \,  = \, \, \, \,  \,\,  
  Y^{s}_1  \cdot \,   \rho(x)   \cdot \, adjoint(G_6^{5Dfcc}),  \\
\label{Homo1Homo5bis}
\hspace{-0.9in}&&\qquad  \quad \quad  
 Y^{s}_5 \cdot \, \rho(x) \cdot \, G_6^{5Dfcc} 
\, \, \, \,  = \, \, \,  \, \,  \, 
 adjoint(G_6^{5Dfcc}) \cdot \rho(x) \cdot \,  Y^{s}_5. 
\end{eqnarray}
From these two intertwining relations it is 
straightforward\footnote[3]{Using the identity 
 $\, adjoint(\Omega \, +\, f(x)) \, = \, \, adjoint(\Omega) +f(x)$
valid for 
{\em any even order operator} $\, \Omega$, and
 for {\em any} function $\, f(x)$.} to see
that an operator of the form 
\begin{eqnarray}
\label{formOmega}
\hspace{-0.9in}&&\qquad \qquad \quad  
\Omega_6 \, \, \, \,= \, \, \,  \,\,\, \, 
  Y^{s}_1 \cdot \, \rho(x) \cdot \, Y^{s}_5 
 \, \,\,\,  + \,\, \, {{\alpha } \over {\rho(x)}}, 
\end{eqnarray}
satisfies the {\em same} intertwining relations 
(\ref{Homo1Homo5}) and (\ref{Homo1Homo5bis}),
as $\, G_6^{5Dfcc}$.
It is, in fact, a straightforward calculation to see that 
the order-six operator $\, G_6^{5Dfcc}$ is 
{\em actually of the form} (\ref{formOmega}) with
 $\, \alpha \, = \, \, -192$: 
\begin{eqnarray}
\label{finalresult}
\hspace{-0.9in}&&\qquad \qquad \quad  
G_6^{5Dfcc}\, \, \, \,= \, \, \,  \,\,\, \, 
  Y^{s}_1 \cdot \, \rho(x) \cdot \, Y^{s}_5 
 \, \,\, \, - \,\, \, {{192 } \over {\rho(x)}}. 
\end{eqnarray}
Recalling section \ref{commentsadjoint}, and, more precisely, 
the fact that the right division of 
$\tilde{T} \cdot \,T$ by $L$ is a constant (see (\ref{TTtilde})),
one can rewrite (\ref{finalresult}) as:
\begin{eqnarray}
\label{finalresultrewrit}
\hspace{-0.9in}&&\qquad \qquad \quad  
\rho(x) \cdot \,   Y^{s}_1 \cdot \, \rho(x) \cdot \, Y^{s}_5 
 \, \, \, \,= \, \, \,  \,\, \,
 192  \,\, \,  +  \,\rho(x) \cdot \, G_6^{5Dfcc}.   
\end{eqnarray}
In other words, the two intertwiners $\,\rho(x) \cdot \,   Y^{s}_1$
and  $\, \rho(x) \cdot \, Y^{s}_5$ are {\em inverse of each other
modulo the operator} $\, \rho(x) \cdot \, G_6^{5Dfcc}$. 

The operator $\, Y^{s}_5$ is 
{\em not globally nilpotent}~\cite{bo-bo-ha-ma-we-ze-09}, 
however we see the emergence of some structure: the 
Jordan form of the $\, p$-curvature of this self-adjoint 
operator $\, Y^{s}_5$ reads
\begin{eqnarray}
\label{pcurv5}
\hspace{-0.95in}&& \quad \qquad  \qquad  \qquad \quad 
\left[ \begin {array}{ccccc}
 0&0&0&0&0\\ 
\noalign{\medskip}0&0&1&0&0\\ 
\noalign{\medskip}0&0&0&1&0\\ 
\noalign{\medskip}0&0&0&0&1\\ 
\noalign{\medskip}0&R_1^{p}&0&R_2^{p}&0
\end {array} \right],  
\end{eqnarray}
its characteristic polynomial reading 
$\,\,\, x \cdot \, (x^4 \, -R_2^p \cdot \, x^2 \, -R_1^p)$. 

The {\em exterior square} $\, Ext^2(G_6^{5Dfcc})$
is an order-fifteen linear differential operator 
which {\em does not have a rational solution}
(or a hyperexponentional solution, see chapter 4 
of~\cite{vdP} and~\cite{Barkatou}), thus {\em excluding a symplectic 
structure} with a $\, Sp(6, \, \mathbb{C})$
 differential Galois group.

In contrast, its {\em symmetric square}  $\, Sym^2(G_6^{5Dfcc})$,
 which does not have a rational solution, is of 
{\em order $\,20$ instead of the generic order $\, 21$}. In fact,
 the {\em associated differential system}
  {\em does have a rational solution}. Again, we see that
the operator representation is not ``intrinsic'' enough,
differential systems are more  ``intrinsic'' from a practical
viewpoint.  The emergence (for the system) of a rational solution 
for the symmetric square means that the 
differential Galois group is included\footnote[5]{In fact,
 an argument of  Katz~\cite{Katz} enables, in principle, to see 
whether the differential Galois group is included in  
$\, O(6, \, \mathbb{C})$ or actually equal to 
$\, O(6, \, \mathbb{C})$. This argument is difficult to work out here.
} 
in the orthogonal group $\, O(6, \, \mathbb{C})$.

From that viewpoint, the 
order-six operator $\, G_6^{5Dfcc}$ seems to contradict an ``experimental'' 
principle\footnote[2]{See Katz's 
book~\cite{Katz} and most of the explicit 
examples known in the literature.} 
that orthogonal groups occur from odd order operators,
and symplectic  groups occur from even order operators. In fact 
the exceptional character of this even order operator 
comes from this decomposition (\ref{finalresult}) in terms 
of {\em odd order} intertwiners (see (\ref{Homo1Homo5}) 
and (\ref{Homo1Homo5bis})).

The log structure of the solutions is exactly the same as the one
of a symmetric square of an order-three operator, $\, Sym^2(L_3)$, 
which might suggest that the differential Galois group would be
the differential Galois group of a MUM order-three operator (generically 
$\, SL(3, \, \mathbb{C})$).

\vskip .1cm

\subsubsection{System representation of  $\, G_6^{5Dfcc}$ \newline}
\label{revisitBrod6}

The following calculations are performed using the
 ``system representation'' as in the previous section.
One gets\footnote[3]{In order to do these calculations 
on the linear differential systems, 
 download the Maple Tools files
TensorConstructions.m and IntegrableConnections.m 
in the web page~\cite{Secret}. Using DEtools, you will need to use, on
the order-six operator $\, G_6^{5Dfcc}$, the command 
 companion-system,then the command 
symmetric-power-system(,2)) and finally the command 
RationalSolutions([],[x]).
} the {\em rational solution} of the {\em symmetric square}
 of the companion system:
\begin{eqnarray}
\label{ratsystBroad6}
\hspace{-0.95in}&&  
[c_1, \, \,\,  \,  \, c_2, \, \,\, \,  \,  c_3, \, \, \, \, 
\cdots \, ,  \,\, \, c_{21}] 
\,\,\,\, =   \\
\hspace{-0.95in}&&  \, \, \, \, = \, \, \, \, 
\Big[0, \,  \,  0, \, \,   0, \, \,  0, \, \,  
  {{2 \cdot \, Q_5} \over {\delta}},  \,  \,  \, 
 -{{20 \, x^3 \cdot \, Q_6} \over {\delta^2}}, 
 \,  \,  \,0, \, \, \, 0,  
 -{{2 \cdot \, Q_5} \over {\delta}},  \,  \, 
{{12 \, x^3 \, Q_{6}} \over {\delta^2}},  \,  \, 
 -{{2 \, x^6 \, Q_{11}} \over {\delta^3}},  \,  \, 
 {{ Q_5} \over {\delta}}, 
\nonumber \\
\hspace{-0.95in}&&  \quad \quad  \quad \,  \,  \, 
 -{{4 \, x^7 \cdot \, Q_{6}} \over {\delta^2}}, \, \, 
{{2 \, x^6 \, Q_{14}} \over {\delta^3}},  \,  \,
 {{6 \, x^9 \, Q_{15}} \over {\delta^4}},  \,  \, 
 {{ x^6 \, Q_{16}} \over {\delta^3}},  \,  \,
  \,  \,  {{ -2 \, x^9 \, Q_{17}} \over {\delta^4}},  \,  \, 
 {{ 8 \, x^{12} \, Q_{18}} \over {\delta^5}},  \,  \,
  {{  x^{12} \, Q_{19}} \over {\delta^5}}, 
 \nonumber \\
\hspace{-0.95in}&& \qquad    \quad  \qquad  \,  \, \,
 \,  \,  
{{  -8 \, x^{15} \, Q_{20}} \over {\delta^6}}, 
 \,  \,  \,  \,
  {{  4 \, x^{18} \, Q_{21}} \over {\delta^6}}
\Bigr],\nonumber
\end{eqnarray}
where 
\begin{eqnarray}
\hspace{-0.95in}&&\qquad \quad  c_1  \,= \, \,c_2 \, = \, \,
c_3  \,= \, \, c_4 \, = \, \,c_7 = \, \, c_8 \,  = \,  \,0, 
\qquad 
c_5  \,=  \,\,c_9  \,=  \, \,c_{12}, 
\end{eqnarray}
and where, recalling $\, p_2$ in (\ref{p2}) 
and $\, \lambda$ in (\ref{h6lambda})
\begin{eqnarray}
\hspace{-0.95in}&&\qquad \quad  
\delta \, \, = \, \, \, \, -x^4 \cdot \, \lambda(x), 
\qquad  \quad \quad  \,  \quad 
 Q_5 \, \, = \, \,\, \,  p_2, 
\nonumber 
\end{eqnarray}
\begin{eqnarray}
\hspace{-0.95in}&&\qquad \quad
Q_6 \, \, = \, \,\,  \, \, 14495514624\,{x}^{8}
-8191475712\,{x}^{7}+1552941056\,{x}^{6}-94273536\,{x}^{5}
\nonumber \\
\hspace{-0.95in}&& \qquad \quad \qquad    \quad \quad
 -3440640  \, x^4\, \,  +498624\,{x}^{3} \, 
-3632\,{x}^{2} \, -609\,x \, +9, 
\nonumber 
\end{eqnarray}
the other $\, Q_n$'s being much larger polynomials.

\vskip .1cm 

If one wants to stick with an operator description,
similarly to (\ref{G1FC}) or (\ref{G1TT}), one can switch, by operator
equivalence\footnote[1]{In Maple, choosing
 an intertwiner $\, R$, e.g. $\, R = \, D_x^2$,
 this amounts to performing 
$S_2$ = rightdivision(LCLM($G_6^{5Dfcc}$,$ \, R$),$\, R$). 
Note that the LCLM with $\, D_x$ instead of $\, D_x^2$, still yields 
a symmetric square of order 20 instead of 21.}, to an operator 
such that its symmetric square is of the generic order 21 and 
has a  {\em rational} solution. 

The denominator of the monic order-twenty operator
 $\, Sym^2(G_6^{5Dfcc})$ is of the form
 $\, x^{16} \cdot \lambda(x)^5 \cdot  p_{278}\, $,
where $\, p_{278}$ is a polynomial of degree 278 in $\, x$.

Let us introduce, for $\, n \, \ge \, 2$, 
 an equivalent operator
$\, G_6^{(n)}$, corresponding to an intertwining by $\, D_x^n$
\begin{eqnarray}
\label{G1Broad6}
\hspace{-0.6in}&&\qquad \quad  \quad  \quad 
 S_2^{(n)} \cdot \, G_6^{5Dfcc} \, \, \,  \, = \, \, \,  \, \,  \, 
G_6^{(n)} \cdot \, D_x^n, 
\end{eqnarray}
For $\, n\, = \, 2$, the {\em symmetric square} of 
$\,G_6^{(n)} $ has the {\em rational solution}
\begin{eqnarray}
 \label{RatBroad6}
 \hspace{-0.95in}&& \,   {{p_2} \over { x^4 \cdot \, \lambda(x)}} \, = \, \,  
 {\frac {1152\,{x}^{2}-56\,x-3}{{x}^{4} \cdot \, (16\,x+1)  \, (8\,x-1) 
 \, (4\,x-1)  \, (16\,x-1)  \, (48\,x-1)  \, (16\,x-3) }},
 \end{eqnarray}
which is nothing but $\, c_5/2$ in (\ref{ratsystBroad6}). For the 
 symmetric square of the other $\, \tilde{G}_6^{(n)}$'s one finds,
respectively for $\, n\, = \, 3$,  $\, n\, = \, 4$ and  $\, n\, = \, 5$,  
the rational solutions $\, c_ {16}$, $\, c_ {19}$ and   $\, c_ {21}$ in
 (\ref{ratsystBroad6}). More generally the rational solution 
reads:
\begin{eqnarray}
\hspace{-0.6in}&&\qquad \qquad \quad 
{{P_{12\, n \, -22}(x) \cdot \, x^{8\, n\, -12} } \over {
 x^{2\, n} \cdot \, \delta^{2\, n \, -3}}}, 
\end{eqnarray}
where $\, P_m(x)$ is a polynomial of degree $\, m$ in $\, x$.

Getting (or even only checking) the
 rational solution (\ref{RatBroad6})
for the symmetric square of the
equivalent operator (\ref{G1Broad6}), paradoxically, 
corresponds to massive calculations compared to obtaining the rational 
solution on the symmetric square of the 
companion system (see (\ref{ratsystBroad6})).

\vskip .1cm 

\subsection{Koutschan's order-eight operator: the lattice Green function of  
the six-dimensional fcc lattice}
\label{ordereight}

A slightly more spectacular\footnote[2]{It is a 
quite large~\cite{Kouweb} order-eight linear differential operator 
of 52 Megabytes.} example of order-eight, $\, G_8^{6Dfcc}$,
has been found by Koutschan~\cite{Kou} for a {\em six-dimensional 
face-centered cubic lattice}. Its  {\em exterior square
is of order 27}. The irreducibility of this order-eight operator
is hard to check\footnote[3]{One can, however, check that 
this operator has no rational solutions.}.
One finds again, at the origin $\, x=\, 0$,
 that there are two independent analytical solutions (no logarithms).
Since the order-eight operator $\, G_8^{6Dfcc}$ has
 {\em two analytical solutions}, it {\em cannot be} MUM~\cite{Big}
at $\, x \, = \, 0$.

A linear combination of these solutions
 is globally bounded~\cite{Short,Big}.
It is such that, 
after rescaling, it can be recast into 
a series with {\em integer} coefficients~\cite{Short,Big}:
\begin{eqnarray}
\hspace{-0.90in}&&1\,\,\, +\,60\,x^2\,\,\,+\,960\,x^3\,\,+\,30780\,x^4
\,+\,996480\,x^5\,\,\,+\,36560400\,\,x^6\,\, +\,1430553600\,\,x^7
 \nonumber \\
\hspace{-0.90in}&&\, \quad \, \, \, 
 +\,59089923900\,x^8 \,\,+\,2543035488000\,x^9
\,\,+\, 113129280527760\,x^{10} \,\, \,+\,\, \cdots 
\end{eqnarray}

In order to fix the normalization, 
this order-eight operator will be 
analyzed in a monic
form: $\, G_8^{6Dfcc} \, = \, \, \, D_x^8 \, + \, \, \cdots \, $
This order-eight operator is (non-trivially) 
{\em homomorphic to its adjoint}, 
one intertwiner being of order {\em six},
the other one being of order  {\em two} 
\begin{eqnarray}
\label{interKou8}
\hspace{-0.7in}&& \qquad \quad  \quad  adjoint(S_6^{6Dfcc})  \cdot  G_8^{6Dfcc} 
 \, \, \,   = \, \,  \,   \,      adjoint(G_8^{6Dfcc})  \cdot  S_6^{6Dfcc}, 
\\
\label{interKou8bis}
\hspace{-0.7in}&&\qquad \quad \quad  G_8^{6Dfcc} \cdot T_2^{6Dfcc}
 \, \, \,   = \, \,  \,   \, adjoint(T_2^{6Dfcc}) \cdot adjoint(G_8^{6Dfcc}). 
\end{eqnarray}
where, again, noticeably, the two intertwiners are, after the 
{\em same} rescaling {\em self-adjoint operators}. Let us introduce
\begin{eqnarray}
\hspace{-0.90in}&& \qquad \quad \qquad \qquad 
a(x) \, \,  \,  = \, \,  \, \,
  {{ x^6 \cdot \, p_{5}^4} \over {p_{25}}} \cdot \, \lambda(x),
\end{eqnarray}
 where the polynomial $\, p_{5}$ reads
\begin{eqnarray}
\label{p5}
\hspace{-0.9in}&&\qquad  \, p_{5} \,\,\, = \, \,\, \,
56\,{x}^{5}\,\, +625\,{x}^{4}\,-1251\,{x}^{3}\,
 -24840\,{x}^{2}\, -65556\,x\, -38880, 
\end{eqnarray}
where $\, \lambda(x)$  reads:
\begin{eqnarray}
\hspace{-0.90in}&& \quad \quad \lambda(x) \,\, = \, \, \,\, 
(x-1)  \, (x-3)  \, (x+24)
  \, (2\,x \, +15)  \, (7\,x+60)  \, (2\,x \, +3)  \, (4\,x \, +15) 
\nonumber \\
\hspace{-0.90in}&& \qquad \qquad \quad  \qquad \quad \times 
 \, (x\, +9)  \, (x\, +5)  \, (x\, +4)  \, (x\, +15)^4,
\end{eqnarray}
and where the polynomial $\, p_{25}$ is a quite large polynomial of degree 25.

The intertwiners $\, T_2^{6Dfcc}$ and $\, S_6^{6Dfcc}$ are, respectively, of 
the form $\, T_2^{6Dfcc} \, = \, \, a(x) \cdot \, Y_2^{s}$ and 
$\, S_6^{6Dfcc} \, = \, \, a(x) \cdot \, Y_6^{s}$, 
where $\, Y_2^{s}$ and  $\, Y_6^{s}$ are two irreducible {\em self-adjoint}
 order-two and order-six operators 
\begin{eqnarray}
\label{selfinterKou8}
\hspace{-0.95in}&&\quad  \quad  \qquad \quad  
 Y_2^{s}  \, \, \,= \, \, \,  \,\,\, {{1} \over {W_2(x)}}
\cdot \Bigl(D_x^2 \,\, 
- \, {{d \ln(W_2(x))} \over {dx}} \cdot \, D_x \, \, + \, \, \cdots  \Bigr), 
\end{eqnarray}
and:
\begin{eqnarray}
\label{selfinterKou8bis}
\hspace{-0.95in}&& \quad  \quad  \qquad \quad  
 Y_6^{s}  \, \, \,= \, \, \,  \,\,\,  {{1} \over {W_6(x)^{1/3}}}
\cdot \Bigl(D_x^6 \,\, - \, {{d \ln(W_6(x))} \over {dx}}
 \cdot \, D_x^5 \, \, + \, \, \cdots  \Bigr). 
\end{eqnarray}
Their corresponding two Wronskians $\,W_2(x)$ and $\,W_6(x)$  read
 respectively:
\begin{eqnarray}
\label{wrskW2W6}
\hspace{-0.95in}&& \quad \quad \quad  \,\,
 W_2(x)\, \,= \, \, \,
 {{x^{11} \cdot \, \lambda(x)^2 \cdot \, p_5^3 } \over {
(x+15)^3 \cdot \,  p_{25}}}, 
\qquad  \quad \,  W_6(x)^{1/3}\, \,= \, \, \,
 {{ (x+15)^3 \cdot \, p_5 } \over {\lambda(x) \cdot \, x^{5}  }}.
\end{eqnarray}

These self-adjoint operators are 
{\em not globally nilpotent}~\cite{bo-bo-ha-ma-we-ze-09}, 
however we see, again, the emergence of some 
structure: the (Jordan form of the) 
$\, p$-curvature of these two self-adjoint operators read
\begin{eqnarray}
\label{pcurv}
\hspace{-0.95in}&& \quad \qquad \qquad   \quad 
\left[ \begin {array}{cc}
 0&1 \\ 
\noalign{\medskip}  S_1^{p}&0
\end {array} \right], \qquad  \quad 
\left[ \begin {array}{cccccc}
 0&1&0&0&0&0\\ 
\noalign{\medskip}0&0&1&0&0&0\\ 
\noalign{\medskip}0&0&0&1&0&0
\\ \noalign{\medskip}0&0&0&0&1&0\\ 
\noalign{\medskip}0&0&0&0&0&1\\ 
\noalign{\medskip} R_1^{p}&0& R_2^{p}&0&R_3^{p}&0
\end {array} \right] 
\end{eqnarray}
where $\, S_1$ is a different rational function for each prime $\, p$, 
for instance for $\, p\, = \, \, 11$
\begin{eqnarray}
S_{1} \, \, \, = \, \, \,  \,  
\,{\frac { 4 \cdot \, \left( 5+x \right) ^{2}}{ \left( x+7 \right)
 \cdot \, x \cdot \,\left( {x}^{2}+7\,x+5 \right)\cdot \, 
  (x+4) \cdot \,  (x+2) }}, 
\end{eqnarray}
and similarly for the $\, R_n$'s of the $\, 6 \times 6$ 
matrix in (\ref{pcurv}).   

\vskip .1cm

The intertwining relations (\ref{interKou8}) give, 
in terms of the self-adjoint operators (\ref{selfinterKou8})
and (\ref{selfinterKou8bis}):
\begin{eqnarray}
\label{interKou8ter}
\hspace{-0.9in}&&\qquad \quad \quad \quad 
Y_6^{s}  \cdot \, a(x) \cdot \, G_8^{6Dfcc} 
 \, \, \,\, \,  = \, \,  \,   \, \,   
   adjoint(G_8^{6Dfcc})  \cdot a(x) \cdot  \,  Y_6^{s}, 
\nonumber \\
\hspace{-0.9in}&&\qquad \quad \quad \quad   
 G_8^{6Dfcc} \cdot a(x) \cdot  \,  Y_2^{s}
 \, \, \, \,\,  = \, \,  \,   \,  \, 
 Y_2^{s} \cdot \, a(x)  \cdot \,  adjoint(G_8^{6Dfcc}),
\end{eqnarray}
which yield  
\begin{eqnarray}
\label{commute}
\hspace{-0.95in}&&\quad  \quad
{\cal K}_8 \cdot  \, {\cal M}_8  \, \,   = \, \,  \,  
{\cal M}_8  \cdot  \, {\cal K}_8
 \qquad \qquad \qquad \hbox{and} \\
\hspace{-0.95in}&&\quad  \quad
adjoint({\cal M}_8)  \cdot  \, adjoint({\cal K}_8)
  \, \,   = \, \,  \, 
adjoint({\cal K}_8)  \cdot  \,adjoint({\cal M}_8),
 \qquad \quad  \quad \hbox{where:} 
\nonumber \\
\label{defcommute}
 \hspace{-0.95in}&&\quad \quad 
{\cal K}_8  \, \,   = \, \,  \,  a(x) \cdot  \, G_8^{6Dfcc} 
\, \,  \qquad \hbox{and:} \qquad \, \, 
{\cal M}_8  \, \,   = \, \,  \,  
 a(x) \cdot  \,  Y_2^{s}  \cdot  \, a(x) \cdot  \, Y_6^{s}. 
\end{eqnarray}
A commutation relation of linear differential operators, 
like (\ref{commute}), is a drastic constraint on the 
operators. As $\, {\cal K}_8$ is {\em irreducible},
the commutation (\ref{commute}) 
forces $\, {\cal K}_8$ to be of the form
 $\,\,  \alpha \cdot \,  {\cal M}_8 \, + \, \beta$, where $\, \alpha$
and $\, \beta$ are constants. 
We may thus guess, from  
the intertwining relations (\ref{interKou8ter}), a decomposition 
of the order-eight operator $\, G_8^{6Dfcc}$, similar to the 
one we had for $\, G_6^{5Dfcc}$, of the form 
\begin{eqnarray}
\label{decompKou8}
\hspace{-0.9in}&&\qquad \qquad \quad  
G_8^{6Dfcc}\, \, \, \,= \, \, \,  \,\,\, \, 
  Y^{s}_{2} \cdot \, a(x) \cdot \, Y^{s}_{6} 
 \, \,\,\,  + \,\, \, {{\alpha } \over {a(x)}},  
\end{eqnarray}
where $\,  Y^{s}_{2}$ and  $\, Y^{s}_{6}$
are two self-adjoint operators of {\em even} order
(instead of odd order for $\, G_6^{5Dfcc}$).
This is, indeed, the case. The operator $\, G_8^{6Dfcc}$
has the noticeable decomposition:
\begin{eqnarray}
\label{finalresult8}
\hspace{-0.95in}&& \qquad \qquad \quad  
G_8^{6Dfcc}\, \, \, \,= \, \, \,  \,\,\, \, 
  Y^{s}_{2} \cdot \, a(x) \cdot \, Y^{s}_{6} 
 \, \,\,\,  + \,\, \,
 \,  {{87480} \over {a(x)}}.  
\end{eqnarray}
Again, and similarly to what has been done for $\, G_6^{5Dfcc}$
(see (\ref{finalresultrewrit})), 
one can rewrite (\ref{finalresult8}) as:
\begin{eqnarray}
\label{finalresult8rewrit}
\hspace{-0.95in}&& \qquad \qquad \quad  
  a(x) \cdot \, Y^{s}_{2} \cdot \, a(x) \cdot \, Y^{s}_{6} 
\, \,\,= \, \, \,  \, -87480 \, \,\, \,  +  a(x) \cdot \,G_8^{6Dfcc}, 
\end{eqnarray}
which means that the two intertwiners $\, a(x) \cdot \, Y^{s}_{2}$
and $\, a(x) \cdot \, Y^{s}_{6}$ are inverse of each other 
modulo the operator $\, a(x) \cdot \,G_8^{6Dfcc}$. From 
(\ref{finalresult8rewrit}) one sees that a solution 
of $\, G_8^{6Dfcc}$ is an eigenfunction of
$\, a(x) \cdot \, Y^{s}_{2} \cdot \, a(x) \cdot \, Y^{s}_{6}$ 
with the eigenvalue $\, -87480$.

\vskip .1cm 

The examination of the formal series solutions, at $\, x\, = \, 0$, 
of the self-adjoint order-six $\, Y_6^{s}$ operator 
% corresponds to
shows a MUM structure. 
The $\, Y_6^{s}$ operator has one analytic solution 
at $\, x \, = \, 0$, (the other solutions have log terms),
 which has the following expansion:
\begin{eqnarray}
\label{solY6}
\hspace{-0.9in}&& Sol(Y_6^{s}) 
\, \,\, = \, \, \, \, \,
 1 \, \, \,  +{\frac {197}{11520}}\,{x}^{2}\, \, 
+{\frac {8559443}{1889568000}}\,{x}^{3}\, \, 
+{\frac {381585241573}{154793410560000}}\,{x}^{4}\,
\nonumber  \\
\hspace{-0.9in}&&\quad    \,  \,\,
+{\frac {35207145815207429}{27209779200000000000}}\,{x}^{5} \, 
\,+{\frac {150944307721060740999089}{182807922003148800000000000}}\,{x}^{6}
 \, \, \,\, + \, \, \cdots 
\end{eqnarray}
This solution-series (\ref{solY6}), again, is {\em not\footnote[1]{This is 
also the case for the self-adjoint order-two $\, Y_2^{s}$ operator. Its solution 
analytic at $\, x \, = \, 0$ is {\em not} globally bounded~\cite{Short,Big}.} 
globally bounded}~\cite{Short,Big}. One deduces immediately,
from decomposition (\ref{finalresult8}), an interesting eigenvalue result:
 the order-eight operator $\, a(x) \cdot \, G_8^{6Dfcc}$ has the {\em not globally 
bounded} eigenfunction (\ref{solY6}), corresponding to
 the {\em integer eigenvalue} $\,87480$.

\vskip .1cm 

The self-adjoint order-five  irreducible operator $\, Y_6^{s}$ is such that its 
{\em exterior square 
is of order 14 instead of the order 15 expected generically}
 (its symmetric square 
is of order 21 as it should, with no rational solution).

\vskip .1cm 

The symmetric square of $\, G_8^{6Dfcc}$ is of the (generic) order 36. However
the {\em exterior  square} of $\, G_8^{6Dfcc}$ is of order 27 
{\em instead of the (generic) order} 28.

\vskip .1cm 

{\bf Remark:} The adjoint of $\, G_8^{6Dfcc}$ has the 
following decomposition, straightforwardly deduced from (\ref{finalresult8}):
\begin{eqnarray}
\label{adjfinalresult8}
\hspace{-0.95in}&& \qquad \qquad \quad  
adjoint(G_8^{6Dfcc}) \, \, \, \,= \, \, \,  \,\,\, \, 
  Y^{s}_{6}   \cdot \, a(x) \cdot \,  Y^{s}_{2}
 \, \,\,\,  + \,\, \,
 \,  {{87480} \over {a(x)}}.  
\end{eqnarray}
So we can expect the Wronskian of $\, Y^{s}_{2}$ to be 
a (rational) solution of its exterior square.
% We have verified that {\em this is actually the case}.
We have verified that {\em this is indeed the case}.
\vskip .1cm 

Similarly to the previous order-six operator $\, G_6^{5Dfcc}$,
one could try to switch to equivalent operators (see (\ref{G1Broad6})),
calculate the exterior  square of these equivalent operators,
and try to find the corresponding rational solution
 (see (\ref{RatBroad6})).
These, at first sight, straightforward calculations are, 
in fact, too ``massive''. The way to get the rational solution
is, in fact, to switch to {\em differential systems}
 (see (\ref{ratsystBroad6})).

\vskip .1cm 

\subsubsection{System representation of  $\, G_8^{6Dfcc}$ \newline}
\label{revisitBrod8}

In fact, even after switching to a differential system
 using the same tools~\cite{Secret}
that we used for obtaining (\ref{ratsystBroad6}), 
we found that the resulting calculation exceeded 
our computational capacity. These calculations mostly amount to 
finding a transformation that reduces 
the system to a system with simple poles. 
We need a second ``trick'' to be able to achieve 
these calculations and get the rational solution 
of the differential system.
An easy way is to rewrite the system\footnote[2]{In the Maple  
TensorConstruction tools found at \cite{Secret}, the 
command  \texttt{Theta\_companion\_system}(L)
returns two matrices  $\frac{1}{p(x)}A_\theta$ and $\,P_{\theta}$ such that,
 for $Y=\,(y,y',\ldots, y^{(n-1)})^T$, we have $\,Y=\, P_{\theta} Y_\theta$
and $Y'_{\theta}=\,\frac{1}{p(x)}  \,A_\theta  \, Y_\theta$, where 
$\,A_\theta$ has no finite poles and $p(x)$ is squarefree, it has only simple roots.
This gives the correspondence between the original companion system 
and the $\theta$-companion system.}
 in terms of the {\em homogeneous derivative}
$\, \theta \, = \, \, x \cdot \, D_x$. Switching to this companion
 system\footnote[3]{If one is reluctant to switch 
to companion systems in $\, \theta$,
another way to achieve these calculations is to perform 
a reduction on the matrix of the corresponding system (moser-reduce
in Maple) {\em before} calculating the symmetric powers 
of the system (an operation that 
preserves the order of the poles).} in
 $\, \theta$, one automatically has simple poles 
for the system.

With all these tricks and tools, we finally found 
that the linear {\em differential system} for the {\em exterior square} 
of the order-eight $\, G_8^{6Dfcc}$  operator is of {\em order 28 and has a 
rational solution}. Note that the intertwiners (\ref{interKou8}) have 
been found from this rational solution: seeking straightforwardly for 
the intertwiners (\ref{interKou8}) 
% yields  
needs too massive calculations !

The rational solution reads
\begin{eqnarray}
\label{ratsystBroad8}
\hspace{-0.95in}&&  
[c_1, \, \,\,  \,  \, c_2, \, \,\, \,  \,  c_3, \, \, \, \, 
\cdots \, ,  \,\, \, c_{28}] 
\,\,\,\, =   \\
\hspace{-0.95in}&&  \quad 
\, \, = \, \, \, \, \Big[0, \,  \,  0, \, \,   0, \, \,  0, \, \,  
  {{P_5} \over {\delta}},  \,  \,  \,  
 {{x^4 \, P_6} \over {\delta^2}},  \,  \,  \, 
  {{x^8 \, P_7} \over {\delta^3}},  \,  \,  \, 
0, \, \, 0, \, \,   {{P_{10}} \over {\delta}},  \,  \,  \,  
 {{x^4 \, P_{11}} \over {\delta^2}},  \,  \,  \, 
 {{x^8 \, P_{12}} \over {\delta^3}},  \,  \,  \, 
 {{x^{12} \, P_{13}} \over {\delta^4}}, 
\nonumber  \\
\hspace{-0.95in}&&  \qquad \quad \quad 
 \,  \,  \, 
 {{P_{14}} \over {\delta}},  \,  \,  \, 
{{x^4 \, P_{15}} \over {\delta^2}}, 
 \,  \,  \,{{x^8 \, P_{16}} \over {\delta^3}},  
 \,  \,  \,
{{x^{12} \, P_{17}} \over {\delta^4}},  \,  \,  \,
{{x^{16} \, P_{18}} \over {\delta^5}},  \,  \,  \,
{{x^3 \, P_{19}} \over {\delta^2}}, 
 \,  \,  \,{{x^7 \, P_{20}} \over {\delta^3}},  \,  \,  \,
{{x^{11} \, P_{21}} \over {\delta^4}},
\nonumber  \\
\hspace{-0.95in}&&  \qquad \quad \quad \quad  \,  \,  \,  \,  \, 
 \,{{x^{15} \, P_{22}} \over {\delta^5}},  \,  \,  \,
{{x^6 \, P_{23}} \over {\delta^3}}, 
{{x^{10} \, P_{24}} \over {\delta^4}},  \,  \,  \,
{{x^{14} \, P_{25}} \over {\delta^5}},  \,  \,  \,
{{x^9 \, P_{26}} \over {\delta^4}},  \,  \,  \,
{{x^{13} \, P_{27}} \over {\delta^5}},  \,  \,  \,
{{x^{12} \, P_{28}} \over {\delta^5}}
\Big],
\nonumber
\end{eqnarray}
where
\begin{eqnarray}
\hspace{-0.95in}&& \qquad \quad 
 \delta \,\,\,\, =  \,\,\,\, 
(x-1)  \, (x-3)  \, (x+24)  \, (2\,x \, +15)  \, (7\,x \, +60) 
 \, (2\,x\, +3)  \, \nonumber \\
\hspace{-0.95in}&& \qquad  \qquad 
\quad \qquad \times (x\, +15)  \, (4\,x\, +15)  \, (x\, +9) 
 \, (x\, +5)  \, (x\, +4) \cdot \, x^5, 
\end{eqnarray}
and where the polynomial $\, P_n$ in (\ref{ratsystBroad8})
are too large to be displayed here. The polynomials 
$\, P_{28}, \, P_{27}, \, P_{25}, \,\, P_{22},\, P_{18} $  are
of degree 49, polynomials 
$\, P_{26}, \, P_{24}, \,  P_{21}, \,\, P_{17}$  are
of degree 38, $P_{13}$ is of degree 37, polynomials 
$\,  \, P_{23}, \, P_{20}, \, P_{16},\, P_{12}, \,P_{7} \, $  are
of degree 27, polynomials 
$\,  \, \, P_{19}, \, P_{15},\, P_{11}, \,P_{6} \, $  are
of degree 16, and $\, P_{14}, \, P_{10}, \,\, P_{5}$  are
of degree 5. Furthermore we have some equalities like 
$\,P_{14}= \, -P_{10} \, = \, P_{5}$, 
$\, P_{11} \, = \, \, -2 \cdot \, P_{15} \, = \, \, -2/3 \cdot \, P_{6}$   

Having this rational solution at our disposal, we can, {\em now}, 
find  the rational solutions for the exterior square of the 
equivalent operators: 
\begin{eqnarray}
\label{G1Broad8}
\hspace{-0.6in}&&\qquad \quad  \quad  \quad 
 G_2^{(n)} \cdot \, G_8^{6Dfcc} 
 \, \, \,  \, = \, \, \,  \, \,  \, 
    G_8^{(n)} \cdot \, D_x^n. 
\end{eqnarray}

Recalling  (\ref{p5}), the rational solution for $\, n\, = \, \, 2$
reads $\, p_5/\delta$. The differential Galois group of $\, G_8^{6Dfcc}$ 
is included in (and probably equal to) $\, Sp(8,\, C)$.

\vskip .1cm 

{\bf Remark:} The same calculations can be performed 
on all the linear differential operators 
we have encountered in lattice statistical 
mechanics~\cite{bo-bo-ha-ma-we-ze-09,High,bernie2010,Khi6,
ze-bo-ha-ma-05c,Renorm,mccoy3,ze-bo-ha-ma-04,Big,CalabiYauIsing}:
all the examples we have tested give operators whose irreducible 
factors are actually equivalent to their adjoint. 

\vskip .1cm

\subsection{Generalization of the decomposition for higher order operators }
\label{generali}

The remarkable decompositions (\ref{finalresult})
 and (\ref{finalresult8}),
encountered with $\, G_6^{5Dfcc}$ and $\, G_8^{6Dfcc}$
can easily be generalized.
In fact, one can systematically introduce the
 {\em even} order operators 
\begin{eqnarray}
\label{genr1}
 \hspace{-0.95in}&& \quad \quad \quad\quad   \qquad 
 M_{2p}^{(n, \, 2p\, -n)} \, \,\, = \, \, \,  \, \, 
L_{2p \, -n} \cdot \, a(x) \cdot \, L_n 
\, \,\, + \, \, \, {{ \lambda} \over { a(x)}}, 
\end{eqnarray}
or, after rescaling\footnote[2]{Do note that the rescaled operators
(\ref{genr2}) are such that the functions annihilated by $\, L_n$ are
automatically {\em eigenfunctions} of $\, {\tilde M}_{2p}^{(n, \, 2p\, -n)}$
with {\em eigenvalue} $\, \lambda$.},
\begin{eqnarray}
\label{genr2}
 \hspace{-0.95in}&& \quad \quad \quad \quad  \qquad 
 {\tilde M}_{2p}^{(n, \, 2p\, -n)} \, \,\, = \, \, \,  \, \, 
a(x) \cdot \, L_{2p \, -n} \cdot \, a(x) \cdot \, L_n 
\, \,\, + \, \, \,  \lambda, 
\end{eqnarray}
where the $\, L_m$'s  are {\em self-adjoint operators} of order $\, m$.
They are, naturally, homomorphic to their adjoint, 
with intertwiners corresponding to these
decompositions (\ref{genr1}) and (\ref{genr2}):
\begin{eqnarray}
\label{genr1inter}
 \hspace{-0.95in}&& \quad \quad \quad   \quad 
    a(x) \cdot \, L_n  \cdot \,  M_{2p}^{(n, \, 2p\, -n)} 
\, \,\, = \, \, \,  \, \,  
 adjoint(M_{2p}^{(n, \, 2p\, -n)}) \cdot \, a(x) \cdot \, L_n,  
\nonumber \\
\hspace{-0.95in}&& \quad \quad \quad \quad 
   M_{2p}^{(n, \, 2p\, -n)}  \cdot \,a(x)  \cdot \, L_{2p \, -n} 
\, \,\, = \, \, \,  \, \,  
L_{2p \, -n} \cdot \, a(x) \cdot \,  adjoint(M_{2p}^{(n, \, 2p\, -n)}),  
\nonumber 
\end{eqnarray}
\begin{eqnarray}
\label{genr2inter}
 \hspace{-0.95in}&&  \quad   \quad 
    L_n  \cdot \,  {\tilde M}_{2p}^{(n, \, 2p\, -n)} \, \,\, = \, \, \,  \, \,  
 adjoint( {\tilde M}_{2p}^{(n, \, 2p\, -n)})  \cdot \, L_n,  
\nonumber \\
\hspace{-0.95in}&&  \quad \quad  
  {\tilde M}_{2p}^{(n, \, 2p\, -n)}  \cdot \, a(x) 
 \cdot \, L_{2p \, -n} \cdot \, a(x)
\, \,\, = \, \, \,  \, \,  
 a(x) \cdot \, L_{2p \, -n} \cdot \, a(x)
 \cdot \,  adjoint( {\tilde M}_{2p}^{(n, \, 2p\, -n)}).  
\nonumber 
\end{eqnarray}
Experimentally we have seen (for instance with our two previous 
lattice Green functions examples of order six and eight, 
see (\ref{Homo1Homo5}) and (\ref{Homo1Homo5bis}) for (\ref{finalresult}), 
and (\ref{interKou8}) and (\ref{interKou8bis}) for (\ref{finalresult8})), 
that this corresponds to {\em two different 
types} of operators: the operators with {\em even} $\, n$,
% the ones where $\, n$ is {\em even}, 
for which the {\em exterior square} of 
an equivalent operator (or of the corresponding 
differential system) will have a rational 
solution (yielding a symplectic differential Galois group),
and the operators with {\em odd} $\, n$,
% the ones where $\, n$ is {\em odd} 
for which the
{\em symmetric square} of the corresponding differential system 
will have a rational solution (yielding an 
orthogonal differential Galois group).

\vskip .2cm

\section{Focus on order-four differential operators: Calabi-Yau conditions}
\label{CalabiYau}

It has been underlined by A.J. Guttmann that these lattice Green functions
are (most of the time) solutions of Calabi-Yau ODEs, or higher order Calabi-Yau
 ODEs~\cite{Guttmann,GoodGuttmann}. The definition of 
Calabi-Yau ODEs, and some large lists of Calabi-Yau ODEs,
can be found in~\cite{Batyrev,TablesCalabi,Almkvist1,Almkvist2}.
Calabi-Yau ODEs are defined by several constraints, some 
are natural like being MUM, others (like some cyclotomic constraints)
are essentially introduced, in a classification perspective
like~\cite{TablesCalabi} to provide hopefully exhaustive
lists of Calabi-Yau ODEs, some are related to the concept
of ``{\em modularity}'', requiring the {\em integrality} of various series
like the nome or the Yukawa coupling. Therefore, in the definition
of Calabi-Yau ODEs, there is some ``mix'' between analytic and differential
constraints, and constraints of a more arithmetic\footnote[1]{In order 
to disentangle these various constraints see~\cite{Short,Big}. }, 
or algebraic geometry character. Among all these constraints defining 
the Calabi-Yau ODEs, the most important one is the so-called 
``{\em Calabi-Yau condition}''. 
Let us consider a (monic) order-four
 linear differential operator: 
\begin{eqnarray}
\label{Omega4}
\hspace{-0.9in}&&\quad \quad \Omega_4 
\, \,  \,  \,   \, = \, \, \, \, \,  \, 
D_x^4 \,\, \,\,  +  \,\, a_3(x) \cdot  D_x^3 \, \,\, 
\, +  \,\, a_2(x) \cdot  D_x^2  \,  \,\, 
 +  \,\, a_1(x) \cdot  D_x   \,\,  \, +  \,\, a_0(x). 
\end{eqnarray}
The exterior square of  (\ref{Omega4}), $\,Ext^2(\Omega_4)$,   
reads, up to an overall factor:
\begin{eqnarray}
\hspace{-0.3in}&&\qquad C_6(x) \cdot Ext^2(\Omega_4) 
\, \, \, \,    \,= \,\, \, \, \,   \, \,\, 
 \sum_{n=0}^{6} \, C_n(x) \cdot \,  D_x^n,
\end{eqnarray}
where the  $\, C_i(x)$'s are 
polynomial expressions of
 $\, a_3(x)$, $\, a_2(x)$,  $\, a_1(x)$, $\, a_0(x)$
and of their derivatives (up to the third derivative). 

The vanishing  condition $ \, C_6(x) \, = \, \, 0$, 
which reads
\begin{eqnarray}
\label{CalabiCond}
 \hspace{-0.9in}&& \quad   \quad \quad \quad \quad 
8 \, a_1(x) \,\, \, + \,\,a_3(x)^3 \, \, \, \, 
-4 \cdot a_3(x) \cdot a_2(x) \,\, \, 
 +6 \cdot a_3(x) \cdot {{ d a_3(x)} \over {dx}}\, 
 \nonumber \\
 \hspace{-0.9in}&& \qquad \qquad \quad  \quad \quad \quad \quad \quad 
 \,\,   -8 \cdot {{ d a_2(x)} \over {dx}}\,\, \, \, 
+ \, 4 \cdot {{ d^2 a_3(x)} \over {dx^2}}\,  \, \, \, \,=
 \, \,  \, \,  \, 0,  
\end{eqnarray} 
is satisfied if, and only if, the exterior square is of {\em order five}, 
instead of the order six one expects generically. It is 
called  ``{\em Calabi-Yau condition}'' by 
some authors~\cite{Almkvist} and
is one of the conditions for the ODE to be a {\em Picard-Fuchs equation}
of a family of Calabi-Yau manifolds (see (11) in~\cite{fast}). 
This Calabi-Yau condition (\ref{CalabiCond}) is actually {\em preserved by 
pullbacks, but not by operator equivalence}. Note that 
this Calabi-Yau condition (\ref{CalabiCond}) is 
{\em preserved by the adjoint transformation} (see \ref{weakpreserved}).

Do note 
that such condition is actually independent 
of $\, a_0(x)$ in  (\ref{Omega4}). Also note that 
all the order-four operators $\, M_4$ that 
can be written as the sum of the symmetric-cube of 
an order-two operator\footnote[5]{If the order-two operator
$\, M_2$ is chosen to be globally nilpotent its Wronskian is an $\, N$-th root 
of a rational function, and $\, M_4\, = \, Sym^3(M_2) \,  + \, f(x)$
 is conjugated to its adjoint
up to the cube of this Wronskian. }, and a function, 
 $\, M_4 \, = \, Sym^3(M_2) \,  + \,  \, f(x)$, automatically verify 
the Calabi-Yau condition (\ref{CalabiCond}). This gives a practical way
to quickly provide examples of order-four operators satisfying 
the Calabi-Yau condition (\ref{CalabiCond}).

Of course similar Calabi-Yau conditions can be introduced 
for higher order operators, imposing,
for order-$N$ operators, that their {\em exterior squares} 
are of order {\em less than} the generic $\, N \cdot \, (N-1)/2$
order. These  higher order {\em Calabi-Yau conditions} actually 
correspond to consider 
{\em self-adjoint operators} (see \ref{self}, see also~\cite{Bogner}).  

Furthermore similar ``Calabi-Yau conditions'' can be introduced 
for {\em symmetric squares instead of exterior squares},
imposing, for order-$N$ operators, that  their symmetric squares 
are of order {\em less than} the generic $\, N \cdot \, (N+1)/2$
order. For an order-three operator written in a monic form
\begin{eqnarray}
\label{Omega3}
\hspace{-0.9in}&&\quad \quad \quad \quad \quad 
 \Omega_3 
\, \,  \,  \,   \, = \, \, \, \, \,\,  \,   \, 
D_x^3 \, \,\, 
\, +  \,\, a_2(x) \cdot  D_x^2  \,  \,\, 
 +  \,\, a_1(x) \cdot  D_x   \,\,  \, +  \,\, a_0(x), 
\end{eqnarray}
the ``symmetric Calabi-Yau condition'' 
reads\footnote[2]{The ``symmetric Calabi-Yau condition'' 
for order-four operators can be found but is 
drastically larger than (\ref{symCab}).}:
\begin{eqnarray}
\label{symCab}
\hspace{-0.9in}&& \quad    \, \,  \qquad 
4\, a_2(x)^3 \, \, \,  -18\,  a_1(x)\, a_2(x) \, \,
 + 9  \cdot \,  {{d^2 a_2(x)} \over {dx^2}}  \, \,
+ 18 \cdot \,  a_2(x)\,  \cdot \,  {{d a_2(x)} \over {dx}}
\nonumber  \\
\hspace{-0.9in}&& \qquad \qquad  \qquad  \qquad \qquad 
\, \, \, +54\, a_0(x) \, \,  \, -27 \, {{d a_1(x)} \over {dx}}
\, \, \, \,  \, = \, \, \,\, \,  \,  0.
\end{eqnarray}
Operators satisfying this ``symmetric Calabi-Yau condition''
actually correspond to the situation described in (\ref{Joyce}). 
If their Wronskian $W(\Omega_3)$ is a $\, N$-th root of a 
rational function they are conjugated to their 
adjoint ($f(x) \, = \,\,W(\Omega_3)^{2/3}$):
\begin{eqnarray}
\label{83}
\hspace{-0.9in}&&   \quad  \, \,   
\Omega_3 \cdot \, f(x) \, \,     \, = \, \, \, 
  f(x) \cdot \, adjoint(\Omega_3)
  \quad \, \,  \hbox{with:}  \qquad  a_2(x)  \, = \, \, 
-{{3} \over {2}}  \, {{1} \over {f(x) }} \, {{d f(x) } \over {dx }}.
\end{eqnarray}

\vskip .1cm

In order to disentangle the main focus of this very paper,
namely the {\em algebraic-differential structures}, from 
other structures of more analytical, or arithmetic, 
character (series integrality~\cite{Short,Big}, MUM property, etc.), 
we concentrate, in this section, on (mostly order-four) linear 
differential operators satisfying the  Calabi-Yau 
condition (\ref{CalabiCond}), or {\em homomorphic to  operators 
satisfying} (\ref{CalabiCond}).

\vskip .1cm

\subsection{Weak and strong Calabi-Yau conditions}
\label{weak}

If one considers an operator that is homomorphic to an operator
with a {\em rational Wronskian}, 
satisfying the Calabi-Yau condition (\ref{CalabiCond}), with 
intertwiners that are of order greater or equal to 
one\footnote[1]{We must exclude intertwiners of 
order zero (namely functions): in that case, 
it is a straightforward calculation to see that the operators are conjugated
by a function, both operators satisfying the 
Calabi-Yau condition (\ref{CalabiCond}).}, the exterior square of that
operator actually {\em has a rational solution}. Unfortunately, in contrast
with (\ref{CalabiCond}), the condition for an order-four operator to 
be such that its exterior square has a rational solution, {\em cannot be 
written directly and explicitly on its coefficients}
 $\, a_n(x)$ (see (\ref{Omega4})). We will call ``weak Calabi-Yau condition''
this condition that the exterior square of an operator has a rational solution,
the Calabi-Yau condition (\ref{CalabiCond}) being the ``strong'' 
Calabi-Yau condition. Note that the weak Calabi-Yau condition is 
{\em preserved by the adjoint transformation} (see \ref{weakpreserved}).

 \vskip .1cm

As far as classifications of Calabi-Yau operators are 
concerned~\cite{Batyrev,TablesCalabi,Almkvist1,Almkvist2}, an operator  
 non-trivially\footnote[3]{With intertwiners of order greater or equal to 
one.} homomorphic to a ``Calabi-Yau operator'' is
certainly as interesting for physics as these ``Calabi-Yau operators'',
and an operator non-trivially homomorphic to an operator verifying 
the ``strong'' Calabi-Yau condition (\ref{CalabiCond}), or 
satisfying the ``weak Calabi-Yau condition'' is 
certainly as interesting as an operator verifying 
the ``strong'' Calabi-Yau condition.

\vskip .1cm

Let us explore the relation between the ``weak Calabi-Yau condition''
and the  ``strong Calabi-Yau condition''. 

\subsection{A decomposition of operators  equivalent
 to operators satisfying the Calabi-Yau condition}
\label{dressing}

Let us consider an order-four operator 
$\, \Omega_4$,  of Wronskian
 $\, w(x) \, = \, \, u(x)^2$, which satisfies 
the Calabi-Yau condition (\ref{CalabiCond}). Let us also consider 
a monic order-four operator  $\, \tilde{\Omega}_4$  
which is (non-trivially) homomorphic (equivalent in the sense
of the equivalence of operators~\cite{vdP}) to the
order-four operator $\, \Omega_4$ satisfying 
the Calabi-Yau condition (\ref{CalabiCond}).
This amounts to saying that there exist two 
intertwiners, $\, U_3$ and $\, L_3$, of order less or equal 
to three\footnote[2]{Higher order intertwiners can always be 
reduced to intertwiners with an order less, or equal, 
to three.},
such that:
\begin{eqnarray}
\label{U3L3}
\hspace{-0.1in}&&\quad \quad \quad \, 
\tilde{\Omega}_4 \cdot U_3 
 \, \,\,\,\,  = \, \, \,\, \,\,\,  L_3 \cdot \, \Omega_4
\end{eqnarray}
It is shown in \ref{appdressing} that the 
order-four operator $\, \tilde{\Omega}_4$
{\em can, in fact, be written in terms of a remarkable 
decomposition}
\begin{eqnarray}
\label{forthcoming}
\hspace{-0.1in}&& \quad  \qquad \tilde{\Omega}_4 \, \, \, \,  \, 
 = \, \, \,  \, \,  \, \,   \,  \, 
 Z_2^{s} \cdot {{1} \over {A_0}} \cdot \, A_2 \, \, \,  \,  \,
 + \, \, \,  \, A_0,   
\end{eqnarray}
 where $\, Z_2^{s}$ and $\, A_2$ are two {\em self-adjoint operators}, 
 $\, A_0$ being a function. 
\ref{appdressing} shows how to get 
$\, Z_2^{s}$, $\, A_2$ and $\, A_0$ in such a decomposition: they 
can simply be obtained as the intertwiners of  $\, \tilde{\Omega}_4$
with its adjoints (use (\ref{intertw}), (\ref{inotherwords}), 
(\ref{otherintertw}), (\ref{otherintertw2}) in \ref{appdressing}).
Experimentally we have checked that an operator (non trivially) 
homomorphic to an operator of the form (\ref{forthcoming})
(see (\ref{U3L3})) can always be 
decomposed in a form (\ref{forthcoming}):
 {\em the decomposition (\ref{forthcoming})
is preserved by operator equivalence}.

\vskip .1cm

{\bf Byproduct:}  As a byproduct one finds 
out that the left and right intertwiners of an order-four 
operator satisfying the  weak Calabi-Yau condition are 
{\em necessarily of order two}. Note, however, that this is not true
for the intertwiners of an order-four operator satisfying the 
symmetric weak Calabi-Yau condition which are of odd orders
 (see (\ref{intertwinY1Y3}) in the section \ref{anisotrop} 
on the anisotropic simple cubic lattice Green function).

\vskip .1cm
\vskip .1cm

\subsection{Rational solutions for the exterior square of
 operators satisfying the  weak Calabi-Yau condition }
\label{ratsolweak}

We have the following general result. Any order-four linear differential 
operator of the form\footnote[1]{Note that one can always restrict to 
$\, \lambda \, = \, 1$ rescaling $\, c_0(x)$ into
  $\, \lambda^{1/2} \cdot \, c_0(x)$.}
\begin{eqnarray}
\label{forthcoming2}
 \hspace{-0.95in}&& \quad \quad \qquad \qquad 
 M_4 \, \,  \,\,\, \, \, = \, \, \,\,\, \,  \,
  L_2 \cdot \, c_0(x)  \cdot \,
 M_2  \, \, \, \, \,  \,+ \, \, \,\,\,  \,
 {{ \lambda } \over {c_0(x) }},
\end{eqnarray}
where $\, L_2$ and $\, M_2$ are two (general) self-adjoint operators
\begin{eqnarray}
\label{selfop}
 \hspace{-0.95in}&& \quad \quad  \qquad \quad \qquad 
 L_2  \, \, \,  = \, \, \,   \,  \, \,
 \alpha_2(x) \cdot \, D_x^2 \,  \,  \,
+ \, \, {{d \, \alpha_2(x)} \over {dx}}  \cdot \, D_x
\,\,  \, + \, \, \alpha_0(x), \quad \\
\label{selfop2}
 \hspace{-0.95in}&& \quad \quad \qquad  \quad \qquad 
  M_2  \, \,  \, = \, \, \, \,   \,  \, \beta_2(x) \cdot \, D_x^2 \, \,  \,
 + \, \,  {{d\, \beta_2(x)} \over {dx}} \cdot \, D_x
\,\, \,  + \, \, \beta_0(x),
\end{eqnarray}
is such that {\em its exterior square has} 
$\,\, 1/\beta_2(x)\,$ {\em as a solution} (up to an overall 
multiplicative constant). This  result can be
seen to be the consequence of a non trivial identity (\ref{Extidenity}) 
given in \ref{ratsolweakappend}. 

\vskip .1cm

{\bf Byproduct:} Thus the {\em exterior square} of 
$\, \tilde{\Omega}_4$ has a {\em rational solution}, 
which is the inverse of the head polynomial
of the second order self-adjoint operator $\, A_2$
in the decomposition (\ref{forthcoming}).

\vskip .1cm 

{\bf To sum-up:} The operators, non-trivially homomorphic to 
operators satisfying the (strong) Calabi-Yau condition 
(\ref{CalabiCond}), necessarily satisfy the ``weak Calabi-Yau condition'': 
their exterior square have a {\em rational solution}. Furthermore 
this rational solution corresponds to the Wronskian
of a self-adjoint order-two operator $\,L_2$ of a remarkable
decomposition (\ref{forthcoming2}). Decomposition (\ref{forthcoming2})
is the most general form of an operator satisfying the 
``weak Calabi-Yau condition''. 

\vskip .1cm 

{\bf Conversely:} This naturally raises the reciprocal question. Is
any order-four operator satisfying the ``weak Calabi-Yau condition'' 
(its exterior square has a rational solution)
non trivially homomorphic to an  operator satisfying
 the (strong) Calabi-Yau condition 
(\ref{CalabiCond}) ?  In view of the remarkable 
decomposition (\ref{forthcoming2}), we can also ask 
the following questions. Is any  order-four operator satisfying
the ``weak Calabi-Yau condition'' necessarily of the form 
(\ref{forthcoming2}), i.e. homomorphic to its adjoint 
with {\em order-two} intertwiners ? Is any  order-four operator 
of the form (\ref{forthcoming2}) homomorphic to an operator
satisfying the (strong) Calabi-Yau condition (\ref{CalabiCond}) ? 
These questions will be revisited in a forthcoming 
publication\footnote[2]{If one switches to a representation 
in terms of {\em differential systems}, such a system 
with Galois group $\, Sp(4,\mathbb{C})$ can always be reduced, via a 
``gauge-like'' transformation~\cite{AparicioLast,AparicioLast3},
 to a system with a {\em hamiltonian} matrix $\, M$. Such a 
system is such that 
the exterior square
system associated with a $\, 6 \times 6$ matrix, 
has a constant solution, namely 
$\,\, [0, \, 1, \,0 , \, 0 , \, 1 , \, 0 ]$
 (see~\cite{AparicioLast,AparicioLast3}).
Switching back to the operator representation, one actually finds
that this operator is homomorphic to its adjoint
%with order-one and order-three intertwiners 
with order-two intertwiners 
(themselves homomorphic to their
adjoints). Consequently 
they can always be decomposed into a form (\ref{genr1}).
}.
The reason why these questions are difficult to answer in general, 
beyond specific examples, comes from the fact that such a reduction 
by operator equivalence of operators satisfying the weak Calabi-Yau 
condition to operators satisfying the strong 
Calabi-Yau condition, is {\em not unique} (an infinite number 
of equivalent operators can satisfy the Calabi-Yau 
condition (\ref{CalabiCond}), see \ref{equivCalabi}). 

\vskip .1cm 

\vskip .1cm 

\subsection{Calabi-Yau conditions and rational solutions 
of the exterior square for order-one intertwiners}
\label{ratsyst}

Let us consider an order-four operator $\, \Omega_4$,
satisfying the Calabi-Yau condition (\ref{CalabiCond}),
and let us introduce $u(x)$  the square root of its Wronskian:
 $\, w(x) \, = \, \, u(x)^2$. Let us consider the LCLM of  $\, \Omega_4$
and of the order-one operator $\, D_x$. This LCLM reads: 
\begin{eqnarray}
\label{102}
 \hspace{-0.9in}&& \, \, \,  \, \, \, \, 
L_1 \cdot \, \Omega_4 \,\,\,  = \, \, \,\, M_4 \cdot \, D_x 
 \quad \, \, \, \, \,  \, \hbox{where:} \quad \quad  \, \, 
L_1 \,\, = \, \, \,\, 
D_x \,\,  - \, {{ 1 } \over { a_0(x)  }} \cdot \, {{ d a_0(x) } \over {dx }}.
\end{eqnarray}
The order-four operator $\, M_4$ verifies 
the ``weak Calabi-Yau condition''. Its 
exterior square has $\, u(x)$ as a solution. More precisely, the 
exterior square of $\, M_4$ is, in fact, the {\em direct sum}
 of an order-one operator and of an order-five operator
\begin{eqnarray}
\label{extM4}
\hspace{-0.9in}&&  \, \,  \, \, \, Ext^2(M_4) \, \, = \, \, \, 
M_5 \, \oplus \, M_1  \quad  \, \, \, \, \, 
\hbox{where:}\,  \quad \quad 
M_1 \,\, = \, \, \,\, 
D_x \, \, - \, {{ 1 } \over { u(x)  }} \cdot \, {{ d u(x) } \over {dx }},
\end{eqnarray}
where the order-five operator is homomorphic to the exterior 
square of $\, \Omega_4$, with two order-two intertwiners 
$\, U_2$ and $\, V_2$, namely $\,  M_5 \cdot  \, U_2 \, = \, \, 
V_2  \cdot  \, Ext^2(\Omega_4)$, 
the order-two intertwiner $\, U_2$ reading
\begin{eqnarray}
\label{U2extM4}
\hspace{-0.8in}&&  \quad \, U_2 \, \,\, =  \,   \,   \,  \,
D_x^2\,  \,\,
 - \, {{ 1 } \over { u(x)  }} \cdot \, {{ d u(x) } \over {dx }} \cdot \, D_x 
\, \,\, \,   + \, r(x) \qquad  \, \, \quad \quad \hbox{where:} 
\nonumber \\
\hspace{-0.8in}&&  \qquad \quad \quad \quad  \, 
r(x)  \,\, \, = \, \, \,\, \, a_2(x) 
\,\, \, \, + \, {{ 1 } \over { u(x)  }} \cdot \, {{ d^2 u(x) } \over {dx^2 }}
\,\,  \, - 2  \cdot  \, \Bigl({{ 1 } \over { u(x)  }} 
\cdot \, {{ d u(x) } \over {dx }}\Bigr)^2. 
\end{eqnarray}

These results can be generalized to more general order-one intertwiners. 
One can easily deduce the result (here $O^F$ denotes the conjugate
by $\, F(x)$ of an 
operator $\, O$: $O^F \, = \, \, F(x) \cdot \, O  \cdot \, F(x)^{-1}$)
\begin{eqnarray}
\label{moregen}
\hspace{-0.95in}&& \qquad \qquad \quad \quad 
L_1^{F} \cdot \, \Omega_4^{F} \,\,\,\,  = \, \,\, \,\,
 M_4^{F} \cdot \,
 \Bigl(D_x \, - \, {{1} \over {F(x)}} \cdot \, {{d F(x)} \over {dx}}\Bigr),
\end{eqnarray}
where, again, $\,\Omega_4^{F}$ satisfies the Calabi-Yau 
condition (\ref{CalabiCond}), and where $\, M_4^{F}$ satisfies 
the weak Calabi-Yau condition. Along this  more general order-one
line, see (\ref{SSW}) in (\ref{weakpreserved}) and (\ref{subcase}) below. 

Switching to a linear differential system representation of an order-four 
operator satisfying the Calabi-Yau condition (\ref{CalabiCond}), one 
has the following solution for the differential system :
\begin{eqnarray}
\label{solratCalab}
\hspace{-0.95in}&& \qquad \qquad \quad \quad  \, \,
 \Bigl[0, \,\,\, 0, \,\,\, -u(x),\, \,\, \, u(x),
 \, \,\,\, {{d u(x)} \over {dx}}, \,
\, \, \, r(x) \cdot \, u(x)\Bigr]. 
\end{eqnarray}
\vskip .1cm
In that heuristic case one remarks that $\, A_0 \, = \, \, a_0(x)\,$ 
in (\ref{Omega4}), and that $\, U_2$ in (\ref{U2extM4}) is simply 
related to $\, A_2$ in the decomposition  (\ref{forthcoming}): 
$\, U_2 \, = \, \, u(x) \cdot \, A_2$. 

\vskip .1cm

If one assumes that the Wronskian of the order-four linear differential operator
is a {\em square of a rational function}, one, thus, finds a {\em rational solution
for the exterior square of the  differential system} (resp. hyperexponential 
solution~\cite{Barkatou} for a Wronskian $ \, N$-th root of a 
rational function).

\vskip .1cm

If one prefers to stick with differential {\em operators}
 instead of differential {\em systems},
one can see the emergence of a {\em rational solution} for the exterior square
of the order-four operator, exchanging the order-four operator 
satisfying the Calabi-Yau condition (\ref{CalabiCond}) for an equivalent operator 
(in the sense of the equivalence of operators~\cite{vdP}, see next subsection). 
This equivalent operator {\em does not satisfy the Calabi-Yau condition}
 (\ref{CalabiCond})
since, as we noticed, the  Calabi-Yau condition (\ref{CalabiCond})
is {\em preserved by pullback} but {\em not by operator equivalence}.

\vskip .1cm

{\bf Remark 1:} In this case of an operator, like $\, M_4$ 
in (\ref{102}), satisfying the weak Calabi-Yau condition 
i.e. homomorphic to an operator $\, \Omega_4$ satisfying the 
Calabi-Yau condition (\ref{CalabiCond}) with an {\em order-one} 
intertwiner, one can, for a given $\, M_4$, find $\, \Omega_4$
 from (\ref{extM4}), (\ref{U2extM4}). In this case ({\em order-one} 
intertwiner) the order-four operator $\, \Omega_4$ is 
(up to overall factors) unique. 

\vskip .1cm

{\bf Remark 2:} Recalling the general decomposition result (\ref{forthcoming}), 
one actually finds that an order-one intertwiner situation 
$\, L_1 \cdot \, \Omega_4 \,\, = \, \, \,\, M_4 \cdot \, (D_x\, +q_0(x))$,
 corresponds to
the {\em self-adjoint operators} $\, Z_2^{s}$ 
and $\, A_2$ in (\ref{forthcoming}) of the form
($\, Z_2^{s}$ factors in order-one 
operators\footnote[9]{Operators $\, Z_2^{s}$ 
and $\, A_2$ of the form (\ref{subcase}) are automatically 
self-adjoint.}):
\begin{eqnarray}
\label{subcase}
\hspace{-0.9in}&& \quad \, Z_2^{s} \, \, \, \, = \, \, \,  \,
 u(x) \cdot \, A_0 \cdot \, 
\Bigl( D_x \, +{{ 1 } \over { u(x)  }} 
 \cdot \, {{ d u(x) } \over {dx }} \, \,  + q_0(x)  \Bigr) 
  \cdot \, \Bigl( D_x \, +{{ 1 } \over { A_0  }} 
\cdot \, {{ d A_0 } \over {dx }} \, \, - q_0(x)   \Bigr), 
\nonumber \\ 
\hspace{-0.9in}&& \quad \,  u(x) \cdot \, A_2 
\,  \,\, \, = \, \, \, \, \,   \, D_x^2\,  \,\,  \,  
 - \, {{ 1 } \over { u(x)  }} \cdot \, {{ d u(x) } \over {dx }} \cdot \, D_x 
\, \, \,   + \, r(x),
\end{eqnarray}
where $\,r(x)$ reduces to  (\ref{U2extM4}) when $\, q_0(x) \, = \, \, 0$.

\vskip .1cm

{\bf Remark 3:} These results are specific of order-one intertwiners. One can 
consider order-two intertwiners introducing the LCLM of $\, \Omega_4$ and 
of an order-two operator $\, M_2$, yielding the intertwining relation
 $\, L_2 \cdot \Omega_4 \, =  \, \, M_4 \cdot \, M_2$. Again one still has 
a decomposition (\ref{forthcoming}) with $\, Z_2^{s}$ and $\, A_2$, 
two order-two self-adjoint operators, but where $\, Z_2^{s}$ no longer factorizes. 

\vskip .1cm

\subsection{The lattice Green function of the anisotropic simple cubic lattice}
\label{anisotrop}

At this step it is important to recall the results 
of Delves and Joyce~\cite{Delves,Delves2} 
for the lattice Green function of the 
{\em anisotropic simple cubic lattice}, generalizing the results
displayed in section (\ref{warm20}). The lattice Green function 
of that anisotropic lattice is solution of an order-four operator 
(see (14) in~\cite{Delves2}), depending on an anisotropy parameter
$\, \alpha$.  This order-four operator reads in terms of 
the homogeneous derivative $\, \theta \, = \, x \cdot D_x$:
\begin{eqnarray}
\label{rewrit}
 \hspace{-0.95in}&&  \quad \quad   
 G_4^{asc} \, = \, \, \,
 24 \cdot \, {\theta}^{3} \cdot \, (\theta-1) \, \, \,  \, 
-4 \cdot \, x \cdot \,\theta \cdot \, P_1(\theta) \, \,\, \,   
+2 \cdot \,{x}^{2} \cdot \, (2\,\theta+1) \cdot \, P_2(\theta) 
\nonumber \\
 \hspace{-0.95in}&& \quad \quad   \quad   \quad  \quad 
-A \cdot \, {x}^{3} \cdot\, (2\,\theta+3) 
 \, (2\,\theta+1) \cdot \, P_3(\theta)  \\
 \hspace{-0.95in}&& \quad \quad   \quad   \quad  \quad 
+ 5 \cdot \, (A+4)  \cdot {A}^{3} \,\cdot \, x^{4}
  \cdot \,  ( 2\,\theta+5)  \, (2\,\theta+3) 
 \, (2\,\theta+1)  \, (\theta+1), 
\nonumber 
\end{eqnarray}
where $\,\, A\,= \alpha^2 \, -4\, \, $ and 
\begin{eqnarray}
\label{rewritP}
 \hspace{-0.95in}&& \, 
 P_1(\theta) \, \, \, = \,\,  \, \, 6\,\cdot  \, (2\,\theta+1) 
 \, (10\,{\theta}^{2}+10\,\theta+3)\, \, 
 +A \cdot \, (28\,{\theta}^{3}+7\,{\theta}^{2}+16\,\theta+3), 
\nonumber \\
 \hspace{-0.95in}&& \,
 P_2(\theta) \, \,\,  = \, \,\,  \,
 12\,\cdot  \, (4\,\theta+5)  \, (2\,\theta+3) 
 \, (4\,\theta+3)        \,  +2
\,A \cdot \, \left( 172 \,{\theta}^{3} \, 
+252\,{\theta}^{2}+234\,\theta+81 \right) \, 
\nonumber \\
 \hspace{-0.95in}&& \quad \quad \qquad \qquad \qquad 
+3\, {A}^{2} \cdot \,
 (16\,{\theta}^{3} +21\,{\theta}^{2}+18\,\theta+6),
  \nonumber \\
 \hspace{-0.95in}&& \, 
    P_3(\theta) \,\,  \, = \, \, \,\, 
 40\,\cdot  \, (4\,\theta+3)  \, (4\,\theta+1) +12
\,A \cdot \, (22\,{\theta}^{2}+29\,\theta+12)\, 
 +{A}^{2} \cdot \, (36\,{\theta}^{2}+57\,\theta+31).
 \nonumber 
\end{eqnarray}
Operator (\ref{rewrit}) is 
 {\em globally nilpotent}~\cite{bo-bo-ha-ma-we-ze-09,Deitweiler}
for {\em any rational value} of the parameter $\, A$, 
the Jordan reduction of its
 $\, p$-curvature~\cite{bo-bo-ha-ma-we-ze-09,Deitweiler} reading 
\begin{eqnarray}
\hspace{-0.8in}&&\qquad \qquad \quad \quad 
J_4 \, \, = \, \, \, \,\,
\left[ \begin {array}{cccc}
0&0&0&0 \\ \noalign{\medskip}
0&0&1&0 \\ \noalign{\medskip}
0&0&0&1 \\ \noalign{\medskip}
0&0&0&0
\end {array} \right].
\end{eqnarray}
Its  characteristic polynomial $\, P(\lambda)$ is
 $\, \lambda^4$,  and its minimal polynomial is
$\, \lambda^3$ (mod. any prime $\,p$).

This order-four operator $\, G_4^{asc}$ is {\em not MUM}. 
It has two solutions analytic at $\, x\, = \, \, 0$
\begin{eqnarray}
\label{solGasc}
\hspace{-0.95in}&& 
1 \,\, + {{1} \over {2}} \, \left( {\alpha}^{2}+2 \right) \cdot  \, x \, 
+ {{3} \over {8}}\, \left( {\alpha}^{4}+8\,{\alpha}^{2}+6 \right) \cdot  \, x^2
\, +{\frac {5}{16}} \, \left( {\alpha}^{6}
+18\,{\alpha}^{4}+54\,{\alpha}^{2}+20 \right) \cdot \, {x}^{3}
 \,\, + \,\,  \cdots,  
\nonumber \\
\hspace{-0.95in}&&    
x \,\, \,+ \,  {{3} \over  {8}} \, \left( 3\,{\alpha}^{2}+11 \right)
\cdot \,   {x}^{2}\, \, \,
+{\frac {5}{48}} \, \left( 11\,{\alpha}^{4}+119\,{\alpha}^{2}+146 \right)
\cdot \,   {x}^{3}
  \\
\hspace{-0.95in}&& \quad \qquad \qquad  \quad 
+{\frac {35}{768}}\, 
\left( 25\,{\alpha}^{6}+537\,{\alpha}^{4}+2049\,{\alpha}^{2}+1217 \right)
\cdot \,  {x}^{4}
\,\, \,  + \,\,  \cdots \nonumber
\end{eqnarray}
together with a solution with a $\, \log$ and a solution  with a $\, \log^2$.
The first analytic solution is {\em  globally bounded}~\cite{Short,Big} 
for generic rational values of $\,\alpha$, or, even, generic rational values 
of $\, A$: for $\, A \, = \, \, p/q$ the rescaling
$\,\,  x \, \rightarrow \, 4 \, q \cdot \,  x\, $ changes this series
into a series with {\em integer coefficients}. 
 The second analytic solution  (\ref{solGasc}) is {\em not globally bounded}
for generic rational values of $\, \alpha$, but 
becomes  globally bounded for $\, \alpha \, = \, \pm 1$: with a rescaling
$\, x \, \rightarrow \, 4 \, x$, the series becomes
a series with {\em integer coefficients}.

The exterior square of  $\, G_4^{asc}$
(depending on the parameter $\, \alpha$) is of order six
with no rational (or hyperexponential~\cite{SingUlm}) solutions. 
The {\em symmetric square} of  $\, G_4^{asc}$ is of order {\em nine}, instead 
of the order ten one could expect. 
In other words $\, G_4^{asc}$ 
{\em verifies the symmetric Calabi-Yau condition} for order-four operators 
(see (\ref{symCab}) above for order-three symmetric condition). 

If, as previously done, we introduce an order-four 
operator $\, \tilde{G}_4^{asc}$ equivalent to $\, G_4^{asc}$
\begin{eqnarray}
\label{equivGasc}
\hspace{-0.95in}&& \quad  \qquad  \quad  \qquad 
S_1^{asc} \cdot \, G_4^{asc} \, \,\, = \, \,\, \, \tilde{G}_4^{asc} \cdot \, D_x ,
\end{eqnarray}
the   {\em symmetric square} of that equivalent order-four operator
has a {\em rational solution} $\, r(x)$:
\begin{eqnarray}
\label{ratsolL4A}
 \hspace{-0.95in}&& \quad  \quad 
r(x) \, \, = \, \, \, \, {\frac { ( \alpha^{2}-4)\cdot \, x \, \, +3}{{x}^{2} 
\cdot \, (1\,  -\alpha^{2}\cdot \,  x) 
 \cdot  \, ( 1- \, (\alpha  \, -2)^{2} \cdot \, x) 
 \cdot \, (1 \, - \,(\alpha+2)^{2} \cdot \, x) }}.
\end{eqnarray}
The order-four operator (\ref{rewrit}) can be decomposed 
in terms of two self-adjoint operators
of order one and three, $\,  Y_1^{(s)}$ and  $\,  Y_3^{(s)}$, namely
 \begin{eqnarray}
\label{decompoG4}
 \hspace{-0.95in}&&    
 G_4^{asc} \, = \, \,  
Y_1^{(s)} \cdot \, \rho(x) \cdot \,  Y_3^{(s)}
\,  + \,  {{8 \cdot \, (\alpha^2\, -1)^2} \over {\rho(x)}}, 
\quad \,
\rho(x) \,  = \, \,  
{\frac { ( (\alpha^{2}\, -4) \cdot \,x \, +3)^{4}}{ 
(5\, (\alpha^{2} \, -4) \cdot \,  x \, -3)^{3}}},
\end{eqnarray}
\begin{eqnarray}
\hspace{-0.95in}&&  \quad \hbox{where:} \quad  \quad \quad 
 \ \rho(x) \cdot \, Y_1^{(s)}  \,\, \, = \, \,  \, \, \,
 \, 2\,\cdot  \, ((\alpha^2 \, -4) \cdot \,  x \, +3) 
 \ \, (5\, \, (\alpha^2 \, -4) \cdot \, x \, -3) \cdot \,  D_x  \, 
\nonumber \\ 
\hspace{-0.95in}&&  \quad \quad \quad   \qquad  \qquad  \qquad   \quad   \quad  
 \ + \, (\alpha^2 \, -4)  \, (5\, (\alpha^2 \, -4) \cdot \, x \, +69),  
\end{eqnarray}
 $\,  Y_3^{(s)}$ being slightly more involved. 
One more time, and similarly to what has been 
done for $\, G_6^{5Dfcc}$ and $\, G_8^{6Dfcc}$
(see (\ref{finalresultrewrit}) and (\ref{finalresult8rewrit})), 
one can rewrite (\ref{decompoG4}) as
\begin{eqnarray}
\label{decompoG4rewrit}
\hspace{-0.95in}&& \quad \quad \quad  
  \rho(x) \cdot \, Y_1^{(s)} \cdot \, \rho(x) \cdot \,  Y_3^{(s)}
\, \,\,= \, \, \,  \, 
- 8 \cdot \, (\alpha^2\, -1)^2 \, \,\, \,  +  \rho(x) \cdot \, G_4^{asc}, 
\end{eqnarray}
which means that the two intertwiners $\, \rho(x) \cdot \, Y_1^{(s)}$
and $\, \rho(x) \cdot \, Y_3^{(s)}$ are {\em inverse of each other 
modulo the operator} $\,\,  \rho(x) \cdot \,G_4^{asc}$.

The order-four operator (\ref{rewrit}) is homomorphic
 to its adjoint with the
intertwining relations:
\begin{eqnarray}
\label{intertwinY1Y3}
\hspace{-0.95in}&&  \quad\quad  \quad  \qquad 
 Y_3^{(s)} \cdot \, \rho(x) \cdot \, G_4^{asc} \, \,  \, = \, \, \,  \, 
adjoint(G_4^{asc}) \cdot \, \rho(x) \cdot \,Y_3^{(s)}, 
\nonumber \\
 \hspace{-0.95in}&&  \quad \quad \quad \qquad 
 G_4^{asc} \cdot \, \rho(x)  \cdot \, Y_1^{(s)} \, \,  \,  = \, \, \, \,  
 Y_1^{(s)} \cdot \, \rho(x) \,\cdot \,adjoint(G_4^{asc}).
\end{eqnarray}
If one denotes $\, W$ the Wronskian of $\, G_4^{asc}$
one has the relation 
$\,\, r(x)^{20} \, \, = \, \, \, W^8 \cdot \, \rho(x)^5  $
$  \,(5\, \,(\alpha^2 \, -4) \cdot \, x \, -3)^7$.

Recalling the example of the order-six lattice Green 
operator $\, G_6^{5Dfcc}$, one sees that the fact that 
it is the {\em symmetric square} (and not the exterior square) 
of that order-four operator which has a rational solution, 
is related to the {\em odd order} 
of the two intertwiners. This anisotropic example shows that all the 
differential algebra structures we display in this paper  
{\em can be generalized, mutatis mutandis,  to
problems with more than one variable} (see also~\cite{Wu}). 

\vskip .1cm

\section{Exceptional differential Galois groups}
\label{except}

Recently a set of Calabi-Yau type operators whose differential
Galois group is $\, G_2(C)$, 
the {\em exceptional}\footnote[2]{The compact form of $\, G_2$, 
{\em subgroup of} $\,SO(7)$, can be described as the automorphism 
group of the octonion algebra.}
subgroup~\cite{Agricola} of $\, SO(7)$, were explicitly
 given~\cite{BognerGood,DettReit}.
These examples  read 
(see page 18 section 4.3 of~\cite{BognerGood}, $\, \theta$ 
denotes the homogeneous derivative 
$\, \theta \, = \, \, x \cdot \, D_x$): 
\begin{eqnarray}
\label{except1}
 \hspace{-0.95in}&& \quad \quad   \quad 
E_1  \, \,= \, \, \, \theta^7 \, -128 \cdot \, x \cdot \, 
(48 \, \theta^4\, +96 \,\theta^3 \, + 124 \, \theta^2 \, +76 \, \theta \, +21) 
\, (2 \, \theta \, +1)^3 
\nonumber \\ 
 \hspace{-0.95in}&& \quad  \quad   \quad \qquad 
+4194304 \cdot \, x^2 \cdot \, (\theta \, +1) \cdot \,
 (12 \, \theta^2 \, +24 \, \theta \, +23)
 \cdot \, (2  \, \theta \, +1)^2 \cdot \, (2  \, \theta \, +3)^2
\nonumber \\ 
 \hspace{-0.95in}&& \quad \quad   \quad \qquad 
-34359738368 \cdot \, x^3 \cdot \, (2  \, \theta \, +5)^2 \cdot \,
 (2  \, \theta \, +1)^2 \cdot \, (2  \, \theta \, +3)^3, 
\end{eqnarray}
and:
\begin{eqnarray}
\label{except2}
 \hspace{-0.95in}&& \quad \quad \quad    
E_2 \, \,= \, \, \, \theta^7 \, -128 \cdot \, x \cdot \, 
(8 \, \theta^4\, +16 \,\theta^3 \, + 20 \, \theta^2 \, +12 \, \theta \, +3) 
\, (2 \, \theta \, +1)^3 \nonumber \\ 
 \hspace{-0.95in}&& \quad \quad \quad \quad   \quad \qquad 
+1048576 \, x^2 \cdot \, (2  \, \theta \, +1)^2 \cdot \, 
(2  \, \theta \, +3)^2 \cdot \, (\theta \, +1)^3, 
 \nonumber 
\end{eqnarray}
\begin{eqnarray}
\label{except3}
 \hspace{-0.95in}&&  
E_3 \, \,= \, \, \,\theta^7 \, \,
\nonumber \\ 
 \hspace{-0.95in}&&  \qquad  
-3^3 \cdot \, x \cdot \, 
(81 \, \theta^4\, +162 \,\theta^3 \, + 198 \, \theta^2 \, +117 \, \theta \, +28) 
 \,  (2 \, \theta \, +1)  \, (3 \, \theta \, +1)  \, (3 \, \theta \, +2) 
\nonumber \\ 
 \hspace{-0.95in}&&  \qquad  \quad 
+3^{12} \, x^2 \cdot \, (3  \, \theta \, +5) \cdot \, (3  \, \theta \, +1) \cdot \, 
(\theta \, +1) \cdot \, (3 \, \theta \, +2)^2 \cdot \, (3 \, \theta \, +4)^2, 
\end{eqnarray}
\begin{eqnarray}
\label{except4}
 \hspace{-0.95in}&& 
E_4 \, \,= \, \, \,\, \,  \theta^7 \,\, \, \, \, 
\nonumber \\
\hspace{-0.95in}&&  
 \qquad - 2^{7} \, \cdot \, x \cdot \, 
(128 \, \theta^4\, +256 \,\theta^3 \, + 304 \, \theta^2 \, +176 \, \theta \, +39) 
 \,  (4 \, \theta \, +1)  \, (4 \, \theta \, +3)  \, (2 \, \theta \, +1)
\nonumber \\ 
 \hspace{-0.95in}&&  \qquad 
+ 2^{26} \, x^2 \cdot \, (4  \, \theta \, +7) 
 \, (4 \, \theta \, +5) \, (4  \, \theta \, +3) 
\, (4 \, \theta \, +1)  \, 
(2\,\theta \, +1)  \, (2 \, \theta \, +3) \,  \, (\theta \, +1),
 \nonumber 
\end{eqnarray}
\begin{eqnarray}
\label{except5}
 \hspace{-0.95in}&&  
E_5 \, \,= \, \, \, \theta^7 \,\, 
\nonumber \\ 
 \hspace{-0.95in}&&  
 - 2^{7} \,\, 3^{3} \cdot \, x \cdot \, 
(648 \, \theta^4\, +1296 \,\theta^3 \, + 1476 \, \theta^2 \, +828 \, \theta \, +155) 
 \,  (6 \, \theta \, +5) \, (6 \, \theta \, +1) \, (2 \, \theta \, +1)
\nonumber \\ 
 \hspace{-0.95in}&&  \quad \quad \, 
+2^{20} \,\, 3^{12} \, \, x^2 \cdot \, (6  \, \theta \, +11) \,
 (6  \, \theta \, +7)  \, (6  \, \theta \, +5)  \, (6  \, \theta \, +1)\, 
 (3  \, \theta \, +5)  \,  (3  \, \theta \, +1)  \, (\theta \, +1).
 \nonumber 
\end{eqnarray}

\vskip .1cm

Note that  for the five $\, \,  E_n$, their conjugate
$\, x^{-1/2} \cdot \,  E_n  \cdot \,  x^{1/2}$ are 
{\em self-adjoint operators}, and of course\footnote[1]{If an operator 
$\, \Omega$ is self-adjoint, the 
operator $\, f(x) \cdot \, \Omega\cdot \, f(x)$
is also self-adjoint for any function  $\, f(x)$.} 
$\, x^{-1} \cdot \,  E_n$ are 
{\em also self-adjoint operators}. Also note that 
the homogeneous derivative 
$\,\, \theta \, = \, \, x \cdot \, D_x \,$
is just shifted by $\, 1/2$ by this conjugation:
\begin{eqnarray}
\label{shift}
\hspace{-0.95in}&&  \qquad \qquad \quad \qquad  \quad 
x^{-1/2} \cdot \, \theta  \cdot \,  x^{1/2} 
\, \,\, = \, \,\, \,\, \, \theta \, + \, {{1} \over {2}}.
\end{eqnarray}
Therefore one gets easily the expressions of these new self-adjoint
operators by changing $\, \theta$ into  $\, \theta \, + \, 1/2$ 
in the previous definitions.

The Wronskians $\, w_n$ of these order-seven operators $\,  E_n$
read respectively:
\begin{eqnarray}
\label{wronskEn}
\hspace{-0.95in}&&  \qquad  \,  w_1 \, \, = \,\,  \, 
\Bigl({{1} \over {p^{w}_1}} \Bigr)^{21/2} 
\quad \quad \, \, \hbox{and:}\quad  \quad \, \, 
 w_n \,\,  = \,\,  \,\Bigl({{1} \over {p^{w}_n}} \Bigr)^7 
\qquad n \, = \, 2, \,3, \, 4, \, 5,
\end{eqnarray}
where the $\, p^{w}_n$'s are the following polynomials:
\begin{eqnarray}
\label{polwronskEn}
\hspace{-0.9in}&&  
 \, p^{w}_1 \, \, = \,\,  \, (16384\,x \, -1) \cdot \, x^2,  \,\quad 
 p^{w}_2 \, \, = \,\,  \, (4096\,x \, -1) \cdot \, x^3,  \, \quad 
 p^{w}_3 \, \, = \,\,  \, (19683\,x-1) \cdot \, x^3,
\nonumber \\
\hspace{-0.9in}&&  \qquad \quad 
p^{w}_4 \, \, = \,\,  \, (262144\,x \, -1) \cdot \, x^3,
 \quad \quad \quad 
p^{w}_5 \, \, = \,\,  \,  (80621568\,x \, -1)  \cdot \, x^3. 
\end{eqnarray}

The solution-series, analytic at $\, x\, = \, \, 0$, 
of these order-seven operators $\,  E_n$
are actually series with {\em integer coefficients}. These 
series are displayed in \ref{analysisG2}.
These order-seven operators are MUM and are {\em globally nilpotent} 
(see (\ref{pcurvJ7}) in \ref{analysisG2}).
The corresponding nomes 
(called ``special coordinates''
in~\cite{BognerGood}) defined as
 $\, q^{(n)} \, \, = \, \, \exp(y_1^{(n)}/y_0^{(n)})$
$ \, \, = \, \, x \cdot \, \exp(\tilde{ y}_1^{(n)}/y_0^{(n)})$,
as well as the various {\em Yukawa couplings}~\cite{Short,Big} 
of these order-seven operators,
correspond to series with {\em integer coefficients }
(see \ref{analysisG2}).

\vskip .1cm 

\vskip .1cm 

Note that, after performing the following rescalings 
$\, x \, \rightarrow \, \, x/4096, \, x/19683,$
$ \, x/262144, $ $\,x/80621568$
on the  $\,  E_n$'s for 
$\, n \, = \, \, 2, \, 3, \, 4, \, 5$,  
these four rescaled $\,  E_n$'s have now the same Wronskian: 
 $\,(1-x)^{-7} \cdot \, x^{-21}$.
The homogeneous derivative being invariant by these rescalings
the previous rescalings just amount to modifying the coefficients 
in front of the $\, x^n$'s in the
previous definitions, for instance:
\begin{eqnarray}
\label{except2bis}
 \hspace{-0.95in}&& \quad   E_2
\quad   \longrightarrow \, \quad   \quad 
{\hat  E}_2 \, \, \, = \, \, \,  \,\, 
2^5 \cdot \, \theta^7 \,  \,  \, \, 
-\, x \cdot \, (8 \, \theta^4\, +16 \,\theta^3 \,
 + 20 \, \theta^2 \, +12 \, \theta \, +3) 
\, (2 \, \theta \, +1)^3
 \nonumber \\ 
 \hspace{-0.95in}&& \quad \quad 
 \quad \quad \quad   \quad \qquad  \qquad \quad 
+2 \, x^2 \cdot \, (2  \, \theta \, +1)^2 \cdot \, 
(2  \, \theta \, +3)^2 \cdot \, (\theta \, +1)^3. 
 \nonumber 
\end{eqnarray}
After these rescalings these rescaled operators $\,{\hat  E}_i$
for $\, i\, = 2, \, \cdots \, 5$,  have, now, their singularities
in $\, 0$, $\, 1$ and $\, \infty$. They read:
\begin{eqnarray}
\label{rescal}
\hspace{-0.95in}&& \quad \quad \quad 
 {\hat E}_i \, \, \,\, = \, \,\,\, \,   \, 
(x-1)^2 \cdot \, x^7 \cdot \, D_x^7 \,  \, \, 
+ 7 \, \cdot \, (4\, x\, -3) \, (x-1) \cdot \, x^6  \cdot \,D_x^6
 \, \, \, \,  + \, \cdots 
\end{eqnarray}
while for $\, {\hat E}_1$ one has:
\begin{eqnarray}
\label{rescal1}
\hspace{-0.95in}&& \quad \quad \quad 
 {\hat E}_1 \, \, \,\, = \, \, \, \,\,  \, 
(x-1)^3 \cdot \, x^7  \cdot \,D_x^7 \, \, \, 
+ {{21} \over {2}}  \, (3\, x\, -2) \, (x-1)^2  \cdot \, x^6  \cdot \,D_x^6 
\, \, \, \,  + \,\cdots 
\end{eqnarray}

Combining these rescalings with the shift of $\, \theta$ 
by $\, 1/2$ one easily deduces self-adjoint operators, for instance: 
\begin{eqnarray}
\label{except2ter}
\hspace{-0.95in}&& \quad \quad \,    E_2
\, \, \quad  \longrightarrow \quad \, \,   (2 \, \theta \, +1)^7
 \, \, \,  - 16 \, x \cdot \, (16 \, \theta^4\,
 +64 \,\theta^3 \, + 112 \, \theta^2 \, +96 \, \theta \, +33) 
\, (\theta \, +1)^3
 \nonumber \\ 
 \hspace{-0.95in}&& \quad \quad \quad 
 \quad \quad \quad   \quad \qquad  \qquad \quad 
+\,16 \,  x^2 \cdot \, (\theta \, +1)^2 \cdot \, (\theta \, +2)^2  \cdot \, 
(2  \, \theta \, +3)^3. 
 \nonumber 
\end{eqnarray}

\vskip .1cm 

From now on, let us consider the ``rescaled''
 $\, {\hat  E}_n$. The  exterior squares 
of these order-seven operators $\, {\hat  E}_n$ are of order 14 instead of 
the order 21 one could expect generically.
The {\em exterior cube} of these order-seven operators are 
of order 27 (instead of order 35). The symmetric squares 
of these order-seven operators are of order 27, instead of the order 28 
 one could expect generically (see (\ref{partcombsym2R1})).

Note that any operator such that its symmetric square
is of order less than the generic expected order (here 28) 
is such that its solutions verifies a quadratic relation. For instance,
the seven formal solutions of the order-seven 
operator $\, {\hat  E}_1$  verifies
the simple quadratic identity:
\begin{eqnarray}
\label{partcombsym2R1}
2\, S_1\, S_7\,\, \,  -2 \,S_6\, S_2\, \, \, +2 \,S_5\, S_3\,\,\,  -S_4^2
\,\,\,  \,=  \,\,\,\, 0. 
\end{eqnarray}

Let us consider the operators $\, E_n^{(m)}$ non trivially homomorphic to 
the $\,   {\hat  E}_n$'s:
\begin{eqnarray}
\label{equivEn}
\hspace{-0.95in}&&  \qquad  \qquad \qquad 
   E_n^{(m)}  \cdot \, D_x^m  
    \, \, \, \, = \,  \, \, \, \,   L_m \cdot \,   {\hat  E}_n,  
\end{eqnarray}
where $\, L_m$ is an order-$m$ operator.
For $\, m\, = \, \, 1$ the {\em exterior squares} of the $ \,  E_n^{(m)}$'s 
are of order 21 (as expected generically), but the symmetric  
squares of the $ \, E_n^{(m)}$'s  are still of order 27.
The exterior cube of the  $ \,  E_n^{(m)}$'s are of order 34
instead of the order 35 expected generically.  

For $\, m\, = \, \, 2$  the symmetric  
squares of the $ \, E_n^{(m)}$'s  are still of order 27,
however for $\, m\, = \, \, 3$  the symmetric  
squares of the $ \,  E_n^{(m)}$'s is of the order 28 
expected generically. The {\em exterior cube} of the  $ \, E_n^{(m)}$'s 
are of order 35 expected generically.  

\vskip .1cm 

The  exterior squares of the $ \,  E_n^{(1)}$'s
are actually a {\em direct sum} of an order-fourteen operator
and an order-seven operator
\begin{eqnarray}
\label{m=1}
\hspace{-0.95in}&&  \qquad  \qquad  \qquad 
Ext^2(E_n^{(1)}) \, \, \,  = \, \, \, \,  \, 
L_{14}^{(n)} \oplus  \,  L_{7}^{(n)},
\end{eqnarray}
where the order-seven operators are actually 
simply conjugated to the $ \, {\hat  E}_n$'s:
\begin{eqnarray}
\label{m=1conj}
\hspace{-0.95in}&& \quad     
   L_{7}^{(1)} \,  = \, \,   
 {{1} \over { (1-x)^{3/2}\, x^3}} \cdot
 \,   {\hat  E}_1 \cdot \, (1-x)^{3/2}\, x^3, 
\quad     \, \,  \, \, \, 
L_{7}^{(n)} \,  = \, \, 
 {{1} \over { (1-x)\, x^3}} \cdot
 \, {\hat  E}_n \cdot \, (1-x)\, x^3. 
\nonumber 
\end{eqnarray}

The {\em symmetric squares} of the $ \, E_n^{(3)}$'s
are actually a {\em direct sum} of an order-27 operator
and an order-one operator
\begin{eqnarray}
\label{m=3}
\hspace{-0.95in}&&  \qquad  \qquad  \qquad 
Sym^2(E_n^{(3)}) \, \, \, \,   = \, \,\, \,   \, 
L_{27}^{(n)} \oplus  \,  L_{1}^{(n)},
\end{eqnarray}
where the order-one operators $\ L_{1}^{(n)}$ have the 
following {\em rational} solutions $\, r_n$:
\begin{eqnarray}
\label{m=3rat}
\hspace{-0.95in}&& \quad    \quad \,\,
 r_1 \,  = \, \,   {\frac {1}{ (x-1)^{3} \cdot \, {x}^{6}}},
 \quad  \, \quad  \, \, \, 
 r_n \,  = \, \,   {\frac {1}{(x -1)^{2} \cdot \, {x}^{6}}} 
\qquad n \, = \,\, 2, \, \cdots, \,  5.
\end{eqnarray}

The {\em exterior cubes} of the $ \,  E_n^{(2)}$'s
are actually a {\em direct sum} of an order-27 operator 
$\, M_{27}^{(n)}$, an order-seven operator $\, M_{7}^{(n)}$
and an order-one $\ M_{1}^{(n)}$ operator
\begin{eqnarray}
\label{m=2}
\hspace{-0.95in}&&  \qquad  \qquad  \qquad 
Ext^3(E_n^{(2)}) \, \,  \,\,\, = \, \, \, \,  \, \, 
 M_{27}^{(n)} \oplus \,  M_{7}^{(n)} \oplus \,  M_{1}^{(n)},
\end{eqnarray}
where the order-one operators $\ M_{1}^{(n)}$ 
have an {\em algebraic} solution for $ \, E_1^{(2)}$:
\begin{eqnarray}
\label{m=3alg}
 a_1 \, \,\, = \, \, \, \, {\frac {1}{ (x-1) ^{9/2} \cdot \, x^{9}}}, 
\end{eqnarray}
and the following {\em rational} solutions 
for the  exterior cube of the other $ \, E_n^{(2)}$'s:
\begin{eqnarray}
\label{m=3alg}
\hspace{-0.9in}&&  \quad \quad  \quad \quad  \rho_n \, \, = \, \,  \, 
{\frac {1}{(x \, -1)^{3} \cdot \,{x}^{9} }}, 
\quad \qquad n \, = \,\, 2,\, \cdots, \,  5.
\end{eqnarray}

\vskip .1cm 

{\bf Remark 1:} The emergence of the exceptional group $\, G_2$ 
corresponds to the 
appearance of rational (or square root of rational\footnote[9]{
In that case, the group is not connected but its Lie algebra is still 
$\mathfrak{g}_2$, i.e the connected component of the group which
 contains the identity is $ \, G_2$.
}) solutions for
the {\em symmetric square and exterior cube} of these operators. 
This is reminiscent (see page 320 of 
Chapter 9 of~\cite{Katz}) of the (non-Fuchsian) order-seven operator 
$\,  D_x^7 \,\, -x \cdot D_x \,\, -\, 1/2$, 
which has the differential Galois group $\, G_2$, 
namely the {\em exceptional} subgroup~\cite{Agricola} 
of $\, SO(7)$. If we had only  
rational (or square root of rational) solutions of the symmetric square
of the operators we would have $\, SO(7)$ differential Galois groups:
the appearance of the rational (or square root of rational) solutions 
for the {\em exterior cube} of these operators 
explains the emergence of this exceptional subgroup of  $\, SO(7)$.

\vskip .1cm 

{\bf Remark 2:} 
Throughout the paper, we see the systematic emergence
 of decompositions as {\em direct sums} 
(see for instance (\ref{m=1}), (\ref{m=3}), (\ref{m=2})) instead
 of just a factorization, 
each time we find a rational solution for some symmetric or exterior power.
This should not be seen as a surprise.
Indeed, the linear differential operator 
$ \, E_n^{(m)}$ 
 is {\em irreducible}. This implies that its differential 
Galois group is {\em reductive}, i.e. all its 
{\em representations are semi-simple} 
(i.e. decompose as a {\em direct sum of irreducible representations}): 
see section 2.2, specially discussion 
before lemma 2.3 in~\cite{Singer}. In practice, this means that if 
we perform any construction like $Sym^m$, $Ext^r$, etc,
and if the corresponding operator factors, then it decomposes as an LCLM of 
{\em irreducible} operators (because the differential Galois group 
acts on the solution space of this operator, so the above applies).

\vskip .1cm 

{\bf Remark 3:} All these results on  symmetric squares, exterior squares
and exterior cubes of (equivalent) order-seven operators have {\em not} been 
obtained using the Maple's DEtools command ``ratsols'' and ``expsols'', the corresponding 
algorithms being {\em not powerful enough} to cope with such examples of 
too large order. Furthermore, if one switches to differential systems 
representations, using the package~\cite{Secret} one finds\footnote[1]{Use the 
commands with(TensorConstructions);
with(IntegrableConnections);
then companion-system(*), 
exterior-power-system(*,N), symmetric-power-system(*,N),
 RationalSolutions([*],[x]), 
HyperexponentialSolutions([*],[x]).}, again the corresponding 
algorithms\footnote[2]{We try to promote, in this paper the idea that
switching to differential systems is a more intrinsic and powerful method
that working on the operators (seen at first 
sight by physicists, as simpler). With these examples we see that 
even switching to differential systems is not enough: one needs 
to switch to $\theta$-systems.} are {\em not powerful enough} 
to cope with such examples of too large order. One needs to go 
a step further, {\em switching to $\theta$-systems},
a method that yields systematically simple poles.

Seeking for rational solutions of differential systems (for regular systems like these),
one (roughly) needs to find a transformation that cast them in simple poles. 
Then one finds the exponents, and then one reduces to polynomial solutions.
Switching to  $\theta$-systems\footnote[8]{An alternative way, if one does not want to
switch to $\theta$-systems, would be to perform  a "Moser-reduction"
on the companion matrix {\em before} calculating 
the symmetric or exterior powers (these  powers preserving the order of the poles).}, 
one  automatically has simple poles: this bypasses the first reduction step. However, 
when one considers the companion matrix, one needs to perform a reduction 
of each singularity: this can, in general, yield a lot of reductions.

Using the TensorConstruction package, the command to be used is 
Theta-companion-system
or full-theta-companion-system. One needs to perform a ``reduction at $\, \infty$''
in order to find polynomial solutions. Doing all these tricks, one finally 
finds these results almost immediately for the symmetric squares, and  for 
the exterior cubes. 

\vskip .1cm 

\subsection{Two-parameter and three-parameter 
deformations of $\,  {\hat E}_1$ }
\label{deformtexttwo}

Another operator $\, \Omega(p,\,q)$  (generalizing $\, {\hat E}_1$),
depending on {\em two} parameters $\, p$ and $\, q$, is given 
in \ref{deform2} (see also the 
operator $\, L$ in section 5.2 of~\cite{DettReit}).

Let us denote $ \, \tilde{\Omega}(p,\,q)^{(m)}$
the  order-seven linear differential 
operator homomorphic to $\,\Omega(p,\,q)$ 
with a $\, D_x^m$ intertwiner (perform the LCLM of 
$\,\Omega(p,\,q)$  and $\, D_x^m$
and rightdivide by $\, D_x^m$).  

One has the following {\em direct sum} decompositions, 
{\em for arbitrary values} of $\, p$ and $\, q$,
for the {\em symmetric square} and {\em exterior cube} of these 
equivalent operators: 
\begin{eqnarray}
\label{m=3pq}
\hspace{-0.95in}&&  \qquad  \qquad  \qquad 
Sym^2(\tilde{\Omega}(p,\,q)^{(3)})
 \, \,  \, \, = \, \,\, \,   \, 
L_{27}\oplus  \,  L_{1},
\end{eqnarray}
and
\begin{eqnarray}
\label{m=2pq}
\hspace{-0.95in}&&  \qquad  \qquad  \qquad 
Ext^3(\tilde{\Omega}(p,\,q)^{(2)})
 \, \,  \,\,  = \, \, \, \,  \,
 M_{27}^{(n)} \oplus \,  M_{7} \oplus \,  M_{1},
\end{eqnarray}
where the order-one operator $\, L_{1}$ has the rational solution 
$\, 1/x^6/(x\, -1)^3$, and where  $\, M_{1}$ 
has the square root of rational solution
$\, 1/x^9/(x\, -1)^{9/2}$.
Furthermore one has 
\begin{eqnarray}
\label{m=1pq}
\hspace{-0.95in}&&  \,   \quad  \quad \qquad  
Ext^2(\tilde{\Omega}(p,\,q)^{(1)})
  \, \, \, = \, \, \, \,  \, 
L_{14} \oplus  \,  L_7,
\quad \quad \quad \quad \quad \hbox{where:}  \\
\hspace{-0.95in}&& \quad  \quad  \quad  \quad \qquad 
  L_7 \,\,  = \, \, \, 
{{1} \over {(x\, -1)^{3/2} \cdot \, x^3 }} \cdot \, 
\Omega(p, \, q) 
\cdot \, (x\, -1)^{3/2} \cdot \, x^3.
\nonumber
\end{eqnarray}

We have the same results as (\ref{m=3pq}), (\ref{m=2pq}) and  (\ref{m=1pq})
 for (the equivalent of) a simple one-parameter 
deformation of $\, {\hat E}_1$, namely:
\begin{eqnarray} 
\label{P1text}
\hspace{-0.95in}&& \quad \qquad  \qquad   \quad \qquad 
{\hat E}_1(r) \,\, \,   = \, \, \,\,  \, 
 {\hat E}_1 \, \,  \,  + \, \, \, r \cdot \, x  \cdot \, (2\, \theta \, +1).
\end{eqnarray} 

Note, however, that combining the previous deformation (\ref{P1text}) 
on $\, \Omega(p,\,q)$ does  yield a {\em three parameters} deformation 
satisfying direct sum decompositions like (\ref{m=3pq}), 
(\ref{m=2pq}) and (\ref{m=1pq})   
\begin{eqnarray} 
\label{generthree}
\hspace{-0.95in}&& \quad   \qquad   \quad 
\Omega(p,\,q, \, r) \, \, \,  = \, \, \,  \,
\Omega(p,\,q) \,   \, 
+ \,  r \cdot \,x  \cdot \, (2\, \theta \, +1), 
 \qquad   \qquad  \hbox{where}
\end{eqnarray} 
\begin{eqnarray}
\label{m=3pqthree}
\hspace{-0.95in}&&  \qquad  \qquad  \quad \quad 
Sym^2(\tilde{\Omega}(p,\,q, \, r)^{(3)})
 \, \,  \, \, \, = \, \, \,\, \,   \, 
L_{27}\oplus  \,  L_{1},
\end{eqnarray}
\begin{eqnarray}
\label{m=2pqthree}
\hspace{-0.95in}&&  \qquad  \qquad  \quad \quad 
Ext^3(\tilde{\Omega}(p,\,q, \, r)^{(2)})
 \, \,  \,\,  = \, \, \, \,  \,
 M_{27}^{(n)} \oplus \,  M_{7} \oplus \,  M_{1},
\end{eqnarray}
where $\, L_{1}$ and  $\, M_{1}$ have respectively the solutions
$\, 1/x^6/(x\, -1)^3$ and 
$\, 1/x^9/(x\, -1)^{9/2}$.

Furthermore one has 
\begin{eqnarray}
\label{m=1pqr}
\hspace{-0.95in}&&  \,   \quad  \quad  \quad \qquad 
Ext^2(\tilde{\Omega}(p,\,q, \, r)^{(1)})
  \, \, \, = \, \, \, \,  \, 
L_{14} \oplus  \,  L_7,
\quad \quad \quad \quad \quad \hbox{where:}  \\
\hspace{-0.95in}&& \quad  \quad  \quad  \quad \quad \qquad 
  L_7 \,\,  = \, \, \, 
{{1} \over {(x\, -1)^{3/2} \cdot \, x^3 }} \cdot \, 
\Omega(p,\,q, \, r)
\cdot \, (x\, -1)^{3/2} \cdot \, x^3.
\nonumber
\end{eqnarray}

\vskip .1cm 

{\bf Remark:} If one wants to get rid of algebraic solution like
 $\, 1/x^9/(x\, -1)^{9/2}$, just perform a conjugation of $\, {\hat E}_1$:
$\, {\hat E}_1 \, \, \rightarrow \, \,  (x-1)^{-1/2} \cdot \, {\hat E}_1 \cdot \, (x-1)^{1/2}$,
the rational solution of $\, L_1$ will become $\, 1/x^6/(x\, -1)^4$ 
and the algebraic solution of $\, M_1$ will become $\, 1/x^9/(x\, -1)^6$ 

This three-parameter conjugated operator\footnote[1]{For
 $\,\Omega(p,\,q, \, r)$ the group is  the exceptional 
group $\, G_2(C)$ {\em up to a center}, as a consequence
of the emergence of a square root $\, 1/x^9/(x\, -1)^{9/2}$.}
 $\, (x-1)^{-1/2} \cdot \, \Omega(p,\,q, \, r) \cdot \, (x-1)^{1/2}$
 probably also has the exceptional group $\, G_2(C)$
as its differential Galois group.

\vskip .1cm 

\subsection{Three-parameter family of order-seven operators 
with exceptional Galois groups}
\label{threeparam}

The order-seven operators ${\hat  E}_i$ for $\, i \, = \, \, \, 2 \, \cdots 5$ 
can also be seen as special cases of another order-seven operator 
 $ \, \Omega_{a,b, c}$ depending on three parameters (see \ref{familly}, see also
operator $\, P_1$ in section 5.1 of~\cite{DettReit}).

Let us denote again $ \, \tilde{\Omega}^{(m)}_{a,b, c}$
the  order-seven linear differential 
operator homomorphic to $\, \Omega_{a,b, c}$ 
with a $\, D_x^m$ intertwiner (perform the LCLM of 
$\, \Omega_{a,b, c}$  and $\, D_x^m$
and rightdivide by $\, D_x^m$).  
The operator $\, \Omega_{a,b, c}$ is generically irreducible. Again this implies 
{\em direct sum decompositions} for any construction $Sym^m$, $Ext^r$.

The {\em symmetric square} of $\, \tilde{\Omega}^{(3)}_{a,b, c}$ 
is actually a {\em direct sum} of an order-27 operator
and an order-one operator, $\, L_1$, with the rational solution $\, 1/(x -1)^2/x^6$
\begin{eqnarray}
\label{m=3symP1}
\hspace{-0.95in}&&  \qquad  \qquad  \quad  \qquad 
Sym^2(\tilde{\Omega}^{(3)}_{a,b, c}) \,\,\, \, = \,  \,  \,  \, \,
L_{27} \oplus  \, L_1
\end{eqnarray}
and the  {\em exterior cube} of $\, \tilde{\Omega}^{(2)}_{a,b, c}$
is actually a {\em direct sum} of an order-27 operator $\, M_{27}^{(n)}$,
 an order-seven operator $\, M_{7}^{(n)}$,
and an order-one $\ M_{1}^{(n)}$ operator
which has the rational solution $\, 1/(x -1)^3/x^9$:
\begin{eqnarray}
\label{m=2extcubeP1}
\hspace{-0.95in}&&  \quad  \quad  \qquad \qquad 
Ext^3(\tilde{\Omega}^{(2)}_{a,b, c}) \, \,  \, = \, \, \, \,  \,
 M_{27}^{(n)} \oplus \,  M_{7}^{(n)} \oplus \, M_{1}^{(n)}.
\end{eqnarray}

Furthermore one has\footnote[1]{Note that these results
(\ref{m=3symP1}), (\ref{m=2extcubeP1}), (\ref{m=1abc}),
 are obtained for {\em arbitrary values of the three parameters}
$\, a$, $\, b$, $\, c$,  of $ \, \Omega_{a,b,c}$ independently 
of the fact that $\, \Omega_{a,b,c}$ is MUM or not.} 
\begin{eqnarray}
\label{m=1abc}
\hspace{-0.95in}&&  \, \quad  \quad  \quad  \quad  \qquad 
Ext^2(\tilde{\Omega}_{a,\,b, \, c}^{(1)})
  \, \, \, = \, \, \, \,  \, 
L_{14} \oplus  \,  L_7,
\quad \quad \quad \quad \quad \quad  \hbox{where:}  \\
\hspace{-0.95in}&& \quad  \quad \quad  \quad  \quad  \quad \qquad  
  L_7 \,\,  = \, \, \, 
{{1} \over {(x\, -1) \cdot \, x^3 }} \cdot \, 
\Omega_{a,\,b, \, c}
\cdot \, (x\, -1) \cdot \, x^3.
\nonumber
\end{eqnarray}

This three-parameter operator probably also has the exceptional group $\, G_2(C)$
as its differential Galois group.

\vskip .1cm 

\section{Comments and speculations: diagonal of rational functions}
\label{comments}

Let us recall that the (minimal) linear differential operators 
for the $\, \chi^{(n)}$'s,  the $\, n$-particle
contributions of the magnetic susceptibility of the square Ising model,
are not irreducible, but {\em factor into many irreducible operators} of various 
orders~\cite{High,bernie2010,Khi6,ze-bo-ha-ma-05c,bo-gu-ha-je-ma-ni-ze-08} 
(two, three, four, ...). For all the factors for which the calculations can be 
performed\footnote[3]{There are factors of order 12 or 23, that are too large
to see, by brute-force calculations, if these operators are homomorphic 
to their adjoint, or such that their exterior or symmetric square could 
have a rational solution.} we have seen that these irreducible factors 
are {\em actually homomorphic to their adjoint}.  Thus, the interesting 
question is  to see whether all the factors of these (minimal) 
operators for the $\, \chi^{(n)}$'s are homomorphic to their adjoint, 
i.e. have a ``special'' differential Galois group, possibly as a consequence 
of the fact that  the $\, \chi^{(n)}$'s are 
{\em diagonals of rational functions} (see~\cite{Short,Big} for a definition).

In this paper we underline selected linear differential operators having  
selected differential structures (special differential Galois groups)
characterized in a differential algebra way (homomorphisms to their adjoint, 
rational, or hyperexponential~\cite{Barkatou}, solutions 
of their exterior or symmetric powers). The idea is to 
disentangle these selected geometrical properties from other selected structures 
of a more arithmetic properties (globally bounded series 
solutions~\cite{Short,Big}), both kinds of selected properties occurring 
simultaneously with the concept of ``modularity''. It is important to 
understand the relationship between these two kinds of properties. Operators 
with selected differential Galois groups do not necessarily correspond to 
globally bounded solution series~\cite{Short,Big}. It is thus
 natural to see whether operators with globally bounded 
solution series~\cite{Short,Big} necessarily correspond to selected 
differential Galois groups. This question
being probably too difficult to address, let us ask
 the following question: if a linear differential 
operator has solutions that are {\em diagonals of rational 
functions}\footnote[1]{Diagonals of rational functions
are necessarily solutions of linear differential operators
 (see Lipshitz~\cite{Lipshitz,Lairez}).}, does it 
necessarily correspond to selected  differential Galois groups, 
or, more simply, are such operators 
homomorphic to their adjoint (possibly in an algebraic extension) ?
Note that we have accumulated a quite large number of operators with solutions 
that are {\em Hadamard products}~\cite{Hadamard,Necer} of {\em algebraic functions} 
 (and are thus simple examples of diagonals 
of rational functions~\cite{Short,Big}). They all have been seen 
to be homomorphic to their adjoint 
(sometimes up to algebraic extensions). Let us recall that 
{\em diagonals of rational functions} are (most of the time transcendental)
 functions that are the 
{\em simplest extensions of algebraic functions}~\cite{Short,Big} (modulo 
each prime, they are algebraic functions). It is worth noting that
 linear differential operators with 
algebraic solutions 
are always homomorphic to their adjoint (up to an algebraic 
extension). It is thus tempting to 
see whether (the factors of minimal) differential operators with solutions that 
are {\em diagonal of rational functions} are necessarily 
{\em homomorphic to their adjoint} (possibly in an algebraic extension). In 
order to get some hint on this question, we 
% have considered 
consider a set of simple but,
hopefully, generic enough\footnote[2]{We try to avoid operators with
 hypergeometric or Hadamard product 
solutions.},  
diagonals of rational functions, 
% finding 
find the minimal operators that 
annihilate them, 
% in order to 
and  check  wether the factors of these operators 
could all be homomorphic to their adjoint (up to algebraic extensions). 

\subsection{Diagonal of rational function: a heuristic simple example 
of an arbitrary number of variables}
\label{genericdiag}

Let us first consider one of the simplest example of diagonal of rational 
functions of $\, N$ variables, namely the diagonal of the rational function
\begin{eqnarray}
\label{simplest}
\hspace{-0.95in}&&  \, \quad  \quad  \quad  
S_N \,\, = \, \, \, \,
Diag\Bigl({{1} \over { 1 \,  \, -x_1 \, -x_2 \,  \,  \cdots \, -x_N }}\Bigr) 
\, \,\,\, = \, \, \, \, \, \, 
 \sum_{k=0}^{\infty} \, {{(k\, N)! } \over {(k!)^N }} \cdot x^k. 
\end{eqnarray}
The series $\, S_N$ are solutions of the 
order $\, N-1$ linear differential operators $\, L_{N-1}$
\begin{eqnarray}
\label{LN-1}
\hspace{-0.95in}&&  \, \quad  \quad L_{N-1} \, \, = \, \, \, 
\sum_{k=0}^{N-1} \, x^{k} \cdot \,  (\alpha_k \cdot \, x \, -\, \beta_k) \cdot \, D_x^k 
\, \, \, = \, \, \, \, x^{N-2} \cdot \, (N^N \cdot \, x \, -1) \cdot D_x^{N-1} \,
\\
\hspace{-0.95in}&&  \, \quad  \quad  \quad  \quad  \quad \, \, 
  +x^{N-3}  \cdot \,  {{N-1} \over {2}}  \cdot \,  
\Bigl((N-1) \cdot \, N^N \cdot \,x \, -\, (N-2)\Bigr) \cdot \, D_x^{N-2} 
\,\, \, \, + \cdots \, 
\nonumber 
\end{eqnarray}
which, are remarkably, {\em self-adjoint} operators. The 
first operators read:
\begin{eqnarray}
\hspace{-0.95in}&&  \, \,   
L_4 \, = \, \, (3125\, x-1)\cdot \, x^3 \, D_x^4 \, \, 
+2\, (12500\, x\, -3) \cdot \, x^2 \, D_x^3 \, 
 \, +(45000\, x \, -7)\cdot \, x \, D_x^2 
\nonumber \\ 
\hspace{-0.95in}&& \quad \quad \quad       
\,+(15000\,x-1)\cdot \, Dx\,\,  +120,
\nonumber \\ 
\hspace{-0.95in}&&  \, \, 
L_5 \, = \, \, (46656\,x\, -1)\cdot \, x^4 \, D_x^5 \,\, 
+10 \, (58320\, x\, -1) \cdot \, x^3 \, D_x^4 \,\, 
+25 \, (79056\, x \, -1)\cdot x^2 \, D_x^3
\nonumber \\ 
\hspace{-0.95in}&& \quad \quad \quad      
+15 \, (126360\, x\, -1) \cdot \, x \,  D_x^2
\,\,  +(331920\, x \, -1) \cdot \, Dx
\,\,   +720, 
\nonumber 
\end{eqnarray}
or, more simply in $\, \theta \, = \, \, x \, D_x$: 
\begin{eqnarray}
\hspace{-0.95in}&&  \, \,     
x \cdot \, L_4 \, \, = \, \,  \, 
  5 \,x \cdot  \,(5\,\theta +1)\cdot \, (5\,\theta +2)\cdot \, 
(5\,\theta +3) \cdot \, (5\,\theta +4)\,  -\theta^4, 
\nonumber \\ 
\hspace{-0.95in}&&  \, \, 
x \cdot \, L_5 \, \, = \, \,  \, 
  6 \,x \cdot  \,(6\,\theta +1)\cdot \, (6\,\theta +2) \cdot \, 
(6\,\theta +3) \cdot \, (6\,\theta +4) \,\cdot  \,
 (6\,\theta +5)\,  -\theta^5,   
 \nonumber \\ 
\hspace{-0.95in}&&  \, \,   
x \cdot \, L_6 \, \, = \, \,  \,  
 7 \,x \cdot  \,(7\,\theta +1)\cdot \, (7\,\theta +2) \cdot \, 
(7\,\theta +3) \cdot \, (7\,\theta +4) \,
\cdot  \, (7\,\theta +5) \,\cdot  \, (7\,\theta +6)\, 
 -\theta^6,   
\nonumber 
\end{eqnarray}
and generally 
\begin{eqnarray}
\label{LNm1}
\hspace{-0.95in}&&  \quad \,   
x \cdot \, L_{N-1} \, \, = \, \,    \,  \,   
N \cdot \,x \cdot  \,(N\,\theta +1)\cdot \, (N \,\theta +2) 
 \, \, \,  \cdots \, \, 
  \, (N\,\theta + N-1)\,  \,  \,\,  -\theta^{N-1},  
\end{eqnarray}
which makes crystal clear that these operators are 
hypergeometric operators with $\, _{N-1}F_{N-2} \,$ solutions.
For instance, for $\, L_4$, we recover the $\, _4F_3$ 
hypergeometric solution 
occurring in Candelas et al. paper~\cite{Candelas}:
\begin{eqnarray}
\hspace{-0.95in}&&  \quad \quad   \qquad   \qquad  
 _4F_3\Bigl([
{{1} \over {5}}, \, {{2} \over {5}}, \,  {{3} \over {5}}, \, {{4} \over {5}}], \, 
[1, \, 1, \, 1], \, \, 5^5 \cdot \, x\Bigr). 
\end{eqnarray}
For arbitrary values of $\, N$ we get, for $\, L_{N-1}$,
 the  $\, _{N-1}F_{N-2}$ hypergeometric solution:
\begin{eqnarray}
\label{N-1FN}
\hspace{-0.95in}&&  \quad   \quad   \qquad  
  _{N-1}F_{N-2}\Bigl([
{{1} \over {N}}, \, {{2} \over {N}}, \, \cdots, \,  \,  \, {{N-1} \over {N}}], \, 
[1, \, 1, \, \cdots,  \, \, 1], \,\,  N^N \cdot \, x\Bigr). 
\end{eqnarray}
It is worth noting, for larger values of $\, N$, that the $\, L_{N-1}$
operators are such that, not only the  series-solution, associated 
with (\ref{N-1FN}), is a globally bounded series~\cite{Short,Big}
(with the $\, N^N$ factor in the argument of the hypergeometric
function it is even a series with {\em integer} coefficients), but that the 
series for the nome and {\em all the Yukawa couplings}, are all 
 series with {\em integer} coefficients, thus corresponding to 
a {\em modularity} of the operators. These results can be seen 
to be a consequence of~\cite{JRoques} which gives the special parameters 
of generalized hypergeometric equations leading to mirror maps 
with integral Taylor coefficients at $\,x\, = \,  0$. 
For instance, the nome $\, q(L_{N-1})$ of the first $\, L_{N-1}$'s read:
\begin{eqnarray}
\hspace{-0.95in}&&  \,   
q(L_3) \, \, = \, \, \,\, x \,\, +104\,{x}^{2}\, +15188\,{x}^{3}\, 
+2585184\,{x}^{4}\, +480222434\,{x}^{5}\, 
 + \,\, \cdots 
\nonumber \\
\hspace{-0.95in}&&  \,  
q(L_4) \, \, = \, \,\, \,x \,\, +770\,{x}^{2}\,+1014275\,{x}^{3}\,
+1703916750\,{x}^{4}\, +3286569025625\,{x}^{5} \,\, 
 \, + \,\, \cdots 
\nonumber \\
\hspace{-0.95in}&&  \,   
q(L_5) \, \, = \, \,\, \,x \,\,  +6264\,{x}^{2}\, +87103188\,{x}^{3}\, 
+1736438167584\,{x}^{4}\, 
\nonumber \\
\hspace{-0.95in}&&  \qquad \qquad \qquad 
+42329034160944354\,{x}^{5} \,  \,
 + \,\, \cdots 
\nonumber \\
\hspace{-0.95in}&&  \, 
q(L_6) \, \, = \, \,\, \,x \,\, +56196\,{x}^{2}\, +9646450758\,{x}^{3}\, 
+2718983725393656\,{x}^{4}
\nonumber \\
\hspace{-0.95in}&&  \, \qquad \qquad \qquad 
\, +1002323538601613453169\,{x}^{5} \,\, 
\, + \,\, \cdots  
\nonumber 
\end{eqnarray}
\begin{eqnarray}
\hspace{-0.95in}&&  \, 
q(L_7) \, \, = \, \,\, \, x \, \, +554112\,{x}^{2} \,\, 
 +1360868807232\,{x}^{3}\, 
+6344223280197623808\,{x}^{4}
 \nonumber \\
\hspace{-0.95in}&&  \, \qquad \qquad \qquad 
\, +41288465594250793127633184\,{x}^{5}
\, \,  + \,\, \cdots   
\nonumber \\
\hspace{-0.95in}&&  \,
q(L_8) \, \, = \, \,\, \, x \,\,  +5973264\,{x}^{2}\,\, 
 +240205268638728\,{x}^{3}\, 
+21222454448347058876544\,{x}^{4}\,
\nonumber \\
\hspace{-0.95in}&&  \, \qquad \qquad \qquad 
  +2781115919369621686237935319524\,{x}^{5}\, 
\, \,  + \,\, \cdots   
\nonumber \\
\hspace{-0.95in}&&  \,
q(L_9) \, \, = \, \, \, \,x \,\,  +69998400\,{x}^{2}
 \,\,  +52035672968460000\,{x}^{3} \, 
+98009214813679052640000000\,{x}^{4} \, 
\nonumber \\
\hspace{-0.95in}&&  \, \qquad \qquad \qquad 
+289691284689365345860892113743750000\,{x}^{5} \, 
\, \,  + \,\, \cdots   
\end{eqnarray}
and all the  Yukawa series, including 
the ``higher order Yukawa couplings\footnote[1]{See appendic C, and especially C.2,
 in~\cite{Short,Big} 
for the definition of these ``higher order'' Yukawa couplings $\, K_n$.}'', $\, K_n$,
are globally bounded, and, even, series with {\em integer coefficients}. 
We actually found the
following relations between the Yukawa couplings:
\begin{eqnarray}
\hspace{-0.95in}&&  \,  \quad   
K_4(L_4) \, \, = \, \, K_3(L_4)^2,  \quad  
\nonumber 
\end{eqnarray}
\begin{eqnarray}
\hspace{-0.95in}&&  \,  \quad   
K_4(L_5) \, \, = \, \, K_3(L_5)^3,  
\quad \quad   \quad  \hbox{and:} \quad \quad   \quad \quad 
K_5(L_5) \, \, = \, \, K_3(L_5)^5, 
\nonumber 
\end{eqnarray}
\begin{eqnarray}
\hspace{-0.95in}&&  \,\quad    
K_5(L_6) \, \, = \, \, K_4(L_6)^2, 
\quad \quad  \quad   \hbox{and:} \quad \quad   \quad \quad 
K_6(L_6) \, \, = \, \, K_4(L_6)^3, 
\qquad  \quad
\nonumber 
\end{eqnarray}
\begin{eqnarray}
\hspace{-0.95in}&&  \,\quad    
K_7(L_7) \, \, = \, \, \Bigl({{K_4(L_7)} \over {K_3(L_7)}}\Bigr)^7, 
\quad 
K_6(L_7) \, \, = \, \, \Bigl({{K_4(L_7)} \over {K_3(L_7)}}\Bigr)^5,
 \quad \hbox{and:} \quad \quad
K_5(L_7)\, \, = \, \, \Bigl({{K_4(L_7)} \over {K_3(L_7)}}\Bigr)^3,
\nonumber 
\end{eqnarray}
\begin{eqnarray}
\hspace{-0.95in}&&  \,\quad      
K_8(L_8) \, \, = \, \,  \Bigl({{K_5(L_8)} \over {K_3(L_8)}}\Bigr)^4,  \quad 
K_7(L_8)  \, \, = \, \, \Bigl({{K_5(L_8)} \over {K_3(L_8)}}\Bigr)^3 
 \quad \hbox{and:} \quad \quad
K_6(L_8)  \, \, = \, \, \Bigl({{K_5(L_8)} \over {K_3(L_8)}}\Bigr)^2,
 \nonumber 
\end{eqnarray}
\begin{eqnarray}
\hspace{-0.95in}&&  \,\quad     
K_9(L_9) \, \, = \, \, \,  \Bigl({{K_5(L_9)} \over {K_4(L_9)}}\Bigr)^9,
  \quad  \quad 
K_8(L_9)  \, \, = \, \,\,  \Bigl({{K_5(L_9)} \over {K_4(L_9)}}\Bigr)^7,
  \quad  \quad 
K_7(L_9) \, \, = \, \, \, \Bigl({{K_5(L_9)} \over {K_4(L_9)}}\Bigr)^5,
\nonumber \\ 
 \hspace{-0.95in}&&  \,\quad   \qquad  \qquad   
 \quad \hbox{and:} \quad \quad \qquad  \quad   
K_6(L_9) \, \, = \, \,  K_3(L_9) \cdot \,  \Bigl({{K_5(L_9)} \over {K_4(L_9)}}\Bigr)^3, 
\nonumber 
\end{eqnarray}
\begin{eqnarray}
\hspace{-0.95in}&&  \,\, \,  \,      
K_{10}(L_{10}) \, \, = \, \, 
 \Bigl({{K_6(L_{10})} \over {K_4(L_{10})}}\Bigr)^5,   \quad 
K_{9}(L_{10})  \, \, = \, \,
 \Bigl({{K_6(L_{10})} \over {K_4(L_{10})}}\Bigr)^4, \quad
K_{8}(L_{10}) \, \, = \, \,
 \Bigl({{K_6(L_{10})} \over {K_4(L_{10})}}\Bigr)^3, 
\nonumber \\ 
 \hspace{-0.95in}&&  \,\quad   \qquad  \qquad   
 \quad \hbox{and:} \quad   \quad \quad \qquad  
K_7(L_{10})  \, \, = \, \, 
 K_3(L_{10}) \cdot \, \Bigl({{K_6(L_{10})} \over {K_4(L_{10})}}\Bigr)^2,
 \nonumber 
\end{eqnarray}
where these Yukawa couplings read respectively 
\begin{eqnarray}
\hspace{-0.95in}&&  \,  \quad   
K_3(L_4) \, \, = \,\, \, \, \,  
1 \, \,\, +575\,x\, \, +1418125\,{x}^{2}\,  \, +3798200625\,{x}^{3}\, \,
+10597067934375\,{x}^{4}
\nonumber \\
\hspace{-0.95in}&&  \, \qquad \qquad \quad   \,
+30287765070550575\,{x}^{5}
\,+ \,\,\, \cdots,
\nonumber 
\end{eqnarray}
\begin{eqnarray}
\hspace{-0.95in}&&  \,  \quad  
K_3(L_5) \, \, = \, \, \,\,  1 \,\, +10080\,x\, +357073920\,{x}^{2}\,
 +13943124679680\,{x}^{3}
 \nonumber \\
\hspace{-0.95in}&&  \,  \quad  \quad   \quad  \quad \quad  \qquad 
+570470634728386560\,{x}^{4}
\, +23986351416805190461440\,{x}^{5}\,
  + \,\,\,\cdots
\nonumber 
\end{eqnarray}
For $\, L_6$ and $\, L_7$ one has {\em two} independent Yukawa couplings.  
For $\, L_8$ and $\, L_9$  one  has {\em three} independent Yukawa couplings,
for $\, L_{10}$ and  $\, L_{11}$   one  has {\em four} independent Yukawa couplings, ...
These series are displayed in \ref{Yukawaof}.

\vskip .1cm

For $\, N$ an {\em odd} integer the {\em exterior square} of 
the $\, (N-1)$-order operator (\ref{LN-1}) is 
of order $\,(N-1)\, (N-2)/2\, -1$ instead of $\,(N-1)\, (N-2)/2$. 
For  $\, N$ an {\em even} integer the {\em symmetric square} of the $\, (N-1)$-order 
operator (\ref{LN-1}) is of order $\,N\,(N-1)/2\, -1$ instead 
of $\,N\,(N-1)/2$. Similarly to (\ref{G1Broad6}) or (\ref{G1Broad8}),
introducing an equivalent operator $\, \tilde{L}_{N-1}^n$, such that 
$\, S^n  \cdot \, L_{N-1} \, = \, \tilde{L}_{N-1}^n \cdot \, D_x^n$, 
the exterior or symmetric square of that equivalent operator 
has, for a well-suited value of $\, n$ the rational solution 
$\, 1/x^{N-2}/(N^N \, x \, -1)$.

\vskip .1cm

\subsection{Diagonal of rational function: heuristic simple examples 
of three variables}
\label{genericdiag2}

Increasing the degree of the rational functions, 
another example corresponds to the diagonal of 
\begin{eqnarray}
\label{algeb}
\hspace{-0.95in}&&  \, \quad  \quad  \quad  \quad  \qquad   \qquad 
R(x,\, y, \, z)\, \,  \,= \, \, \, \,
 {{1  \, -x \, + \, y \, z} \over { 1 \, -3\, x \,z \,    -5\, y^2 }}.
\end{eqnarray}
Unfortunately this diagonal is solution 
of an order-two operator with {\em algebraic solutions}, 
$\,(4\,-1215\,x^2)\cdot \, D_x^2\,-3645\cdot \,x\,D_x\,-1080$
(which is conjugated to its adjoint). 

\vskip .1cm 

We have a similar algebraic result with
\begin{eqnarray}
\label{algeb2}
\hspace{-0.95in}&&  \, \quad  \quad \qquad  \quad  \quad \quad  \quad 
R(x,\, y, \, z)\, \,  \,= \, \, \, \,
 {{1  \, -7\, x \, + \, 2  \, y \, z} \over { 1 \, +3\, x \,z \,    -5\, y^3 }}.
\end{eqnarray}
Its diagonal is the lacunary series
\begin{eqnarray}
\label{algeb2ser}
\hspace{-0.95in}&&  \,  \quad  \quad  
1\,\,\, -540\,{x}^{3}\,\,+510300\,{x}^{6}\,\,-541282500\,{x}^{9}\,\,
+604514137500\,{x}^{12} 
\nonumber \\
\hspace{-0.95in}&&  \,  \quad \quad \quad  \quad  \quad 
-695204544150000\,{x}^{15}\,
+814769502147562500\,{x}^{18} \, \,\, + \, \, \cdots 
\end{eqnarray}
which is an {\em algebraic function} solution of an order-three
operator (homomorphic to its adjoint). 

Another example corresponds to the diagonal of 
\begin{eqnarray}
\label{hypergeoma}
\hspace{-0.95in}&&  \, \quad \quad  \quad  \quad  \quad \quad   \qquad 
R(x,\, y, \, z)\, \,  \,= \, \, \, \,
 {{1  \, -x} \over { 1 \, -3\, x \, + z \,  -5\, y^2 }}.
\end{eqnarray}
which also corresponds to the series expansion of a hypergeometric function:
\begin{eqnarray}
\label{hypergeomser}
\hspace{-0.95in}&&  \,   \quad    \quad 
{{2 } \over {15}} \, \, \, + \, \, \, {{13} \over {15}} \, \cdot \,
 _4F_3\Bigl( [ {{1 } \over {5}}, \, 
{{2 } \over {5}}, \, {{3 } \over {5}}, \,{{4 } \over {5}}], \, \,
 [{{1 } \over {2}}, \,{{1 } \over {2}}, \, 1], \,  \, \,
 5^6 \cdot ({ {3} \over {4}} \cdot \, x)^2\Bigr) 
\nonumber \\
\hspace{-0.95in}&&  \,  \quad \quad  \quad  \qquad 
\, = \, \, \, \,\, \,  1 \, \,\, +1170\,{x}^{2} \,\, +5528250\,{x}^{4} \, \,
+33202669500\,{x}^{6}  \\
\hspace{-0.95in}&&  \,  \quad \quad  \quad  \quad \quad  \quad \qquad 
\, +221602408706250\,{x}^{8}\,
\, +1569831463275075000\,{x}^{10} \,\, \, +  \,  \,  \cdots
 \nonumber
\end{eqnarray}
The corresponding order-five operator is a {\em direct sum} 
$\, D_x \oplus \, M_4$ where the order-four operator $\, M_4$,
which annihilates the $\,_4F_3$ in (\ref{hypergeomser}),
 is homomorphic to its adjoint
and such that its exterior square has a rational solution
namely $\, 1/x/(9 \cdot \,5^6 \, x^2 \, -16)$. The order-four operator 
$\, M_4$ is of the form (\ref{forthcoming2}), namely:
\begin{eqnarray}
\hspace{-0.95in}&&  \, \, \, \,   
M_4  \,  \,= \, \, \, L_2 \cdot c_0(x) \cdot \, M_2 
\, + \, \, {{9} \over{5^4}} \cdot \, {{1} \over{c_0(x)}}, \qquad 
c_0(x) \, \,  = \, \,  \,  \,
 x^2 \cdot \, \Bigl(x^2 - {{16} \over {9 \cdot \, 5^6 }}\Bigr),  
\end{eqnarray}
where $\, L_2$ and $\, M_2$ are two order-two {\em self-adjoint} 
operators\footnote[2]{$\, M_2$ is the product of two order-one operators 
the right factor having the polynomial solution 
$\, x^2  \, (9 \cdot \,5^6 \, x^2 \, -16)$.}.

Noting that the diagonal of 
\begin{eqnarray}
\label{hypergeoma2}
\hspace{-0.95in}&&  \, \quad  \quad  \quad  \quad  \qquad     
\tilde{R}(x,\, y, \, z)\, \,  \,= \, \, \, \,
 {{1} \over { 1 \, -3\, x \, + z \,  -5\, y^2 }}.
\end{eqnarray}
is solution of the order-four operator $\, M_4$, and is nothing but 
\begin{eqnarray}
\label{hypergeomser2}
\hspace{-0.95in}&&  \,   \quad  \qquad    \quad  \quad 
 _4F_3\Bigl( [ {{1 } \over {5}}, \, 
{{2 } \over {5}}, \, {{3 } \over {5}}, \,{{4 } \over {5}}], \, \,
 [{{1 } \over {2}}, \,{{1 } \over {2}}, \, 1], \,  \,\,
{{9 \cdot \, 5^6} \over {16}} \cdot x^2\Bigr), 
\end{eqnarray}
and introducing
 $\,\hat{R}(x,\, y, \, z)  \,= \,  \, 15 \, R(x,\, y, \, z)\, $
$ -\, 13 \, \tilde{R}(x,\, y, \, z) \, -2$, 
one easily deduces, from (\ref{hypergeomser2}), that the diagonal of 
\begin{eqnarray}
\label{hypergeoma3}
\hspace{-0.95in}&&  \, \quad  \quad  \,     
\hat{R}(x,\, y, \, z)\, \,  \,= \, \, \, \,
 {{10 \,y^2\,-9\,x\,-2\,z } \over { 1 \, -3\, x \, + z \,  -5\, y^2 }} 
\quad \quad     \hbox{or}
\quad \, {{2  \, - 15 \, x } \over { 1 \, -3\, x \, + z \,  -5\, y^2 }},
\end{eqnarray}
is equal to zero. 

These cases, reducible to algebraic or hypergeometric  situations, 
are still too simple to be representative of the ``generic'' situation. 

\vskip .1cm

\subsection{Towards a ``generic'' diagonal of rational function example}
\label{genericdiag}

Trying to avoid these too simple cases\footnote[3]{When the operators
annihilating diagonal of rational functions are of order two, one often finds
modular forms, the corresponding nome being seen to be a globally 
bounded series~\cite{Short,Big}. A set of examples of diagonal 
of Szego's rational functions 
can be found in~\cite{Krauers}.} reducible to 
hypergeometric functions (or Hadamard product of algebraic functions), 
we have considered the operator annihilating the diagonal of a rational 
function of three variables, hopefully involved enough, with no symmetry 
between the three variables, to be seen as a ``generic'' diagonal 
of a rational function. 

 \vskip .1cm

\subsubsection{Towards a ``generic'' diagonal of rational function: a first example\newline }
\label{genericdiagfirst}

The rational function
we have considered reads:
\begin{eqnarray}
\label{generic}
\hspace{-0.95in}&&  \, \quad  \quad  \quad  \quad 
R(x,\, y, \, z)\, \, = \, \, \, \,
 {{1} \over { 1 \, -3\, x \, -5\, y \, -7\, z \,+x\, y \,+2\,y\,z^2\, +3\,x^2\,z^2}}.
\end{eqnarray}
The diagonal of this rational function 
reads\footnote[2]{Use the maple command mtaylor(F, [x,y,z], terms), 
to get the series in three variables, then take the
diagonal. Other method, in Mathematica install the risc package 
Riscergosum~\cite{Mathematica}, and in HolonomicFunctions`
use the command FindCreativeTelescoping.}:
\begin{eqnarray}
\label{genericdiag}
\hspace{-0.95in}&&  \, \quad 
S^{(0)}_0 \, \, = \, \,\, \,\, Diag(R(x,\, y, \, z))
\, \, = \, \, \,  \,\, 1 \,\, \,  +616 \, x \, \, 
 \, +947175 \, x^2 \, \, + \,1812651820 \, x^3 
\nonumber \\ 
\hspace{-0.95in}&&  \,\quad  \quad   \quad  
\, \, + \,3833011883965 \, x^4  \, 
+ \, 8582819380142616 \, x^5 \, + \,19946071353510410136\, x^6    
\nonumber \\ 
\hspace{-0.95in}&&  \, \quad \quad  \quad   
 \, \, + \, 47578122531207001944168 \, x^7
 \, + \,115702070514540009854741415 \, x^8 \,
\nonumber \\ 
\hspace{-0.95in}&&  \, \quad  \quad \quad   \,\,  
 + \, 285583642613093627090885877280 \, x^9 
\nonumber \\ 
\hspace{-0.95in}&&  \, \quad  \quad  \quad  \,\, 
\, + \,713269435359072253352128013072035\, x^{10}
\,\, + \, \,  \cdots 
\end{eqnarray}
The minimal order operator that annihilates the diagonal 
of this rational function (\ref{generic})
is a {\em quite large order-six}
 linear differential operator\footnote[1]{We thank 
Alin Bostan for providing this order six operator from a 
creative telescopic code (not by guessing).}. Again, this operator is too 
large to check that it is {\em homomorphic to its adjoint}.
We can, however, check that its {\em exterior square} is of order 15.
However, switching to the associated differential theta-system, 
we have been able to see that it is actually  
{\em homomorphic to its adjoint}: one actually finds  the {\em exterior square} 
of the associated differential system has a rational solution (but 
not its symmetric square). The differential Galois group thus corresponds 
to a {\em symplectic structure}. 

In fact this operator is {\em not MUM}. It has 
four solutions, analytic at $\, x \, = \, \, 0$, 
namely $\, S^{(0)}_0$ given by (\ref{genericdiag}) 
and
\begin{eqnarray}
\label{othersol}
\hspace{-0.95in}&&  \, \quad  \quad  
S^{(1)}_0 \, \, = \, \, \,\, x \,\, \,\,
-{\frac {947569825302083891091227422045}{3191686441638931584990008514}}\,{x}^{4} \, 
 \nonumber \\ 
\hspace{-0.95in}&&  \, \quad \quad  \qquad \quad \quad  \quad    \quad  
-{\frac {13038344513942350315758249091274688499}{19626034561464639086279672353532}}\,{x}^{5}
\,\, + \, \, \cdots,  
 \nonumber \\ 
\hspace{-0.95in}&&  \, \quad  \quad  
S^{(2)}_0 \, \, = \, \, \, \,  \, x^{2} \,\,\,  +{\frac {60}{7^3}}\,{x}^{4}
 \, \,\,
 -{\frac {576}{7^4}}\,{x}^{5}\, \, + \, \, \cdots , 
\nonumber \\ 
\hspace{-0.95in}&&  \,  \quad  \quad  
S^{(3)}_0 \, \, = \, \, \, \, {x}^{3} \, \,
 +{\frac {30608172563777847511388970395}{14474768442806945963673508}}\,{x}^{4}
\,\nonumber \\ 
\hspace{-0.95in}&&  \,  \quad \quad  \quad \quad \quad \quad  \quad   \quad  
 +{\frac {6637738302888023001730565011179544651}{
1401859611533188506162833739538}}\,{x}^{5} \,\,\, + \,\, \cdots 
\end{eqnarray}
the last series $\,S^{(3)}_0 $ 
 being {\em not globally bounded}. The two other solutions 
have a log (but no $\, log^2$, $\, log^3$, ...):
\begin{eqnarray}
\label{twoothersol}
\hspace{-0.95in}&&  \, \quad  \quad   \quad  \,
  S^{(0)}_1 \, = \, \, \, S^{(0)}_0 \cdot \, \ln(x) \, + \, \,  T^{(0)}_0, 
\qquad \, \, S^{(2)}_1 \,\, = \, \, \,
  S^{(2)}_0 \cdot \, \ln(x) \, + \, \,  T^{(2)}_0, 
\end{eqnarray}
the two series $\, T^{(0)}_0$ and  $\, T^{(2)}_0$  being analytic 
at $\, x \, = \, \, 0$, for instance:
\begin{eqnarray}
\label{twoothersolmore}
\hspace{-0.95in}&&      
T^{(0)}_0 \, = \, \, \, \,
{\frac {1769904090259426475015551868948047756831494229112489}{
6347493572699380825284454014187955842800}}\,{x}^{4} 
 \\ 
\hspace{-0.95in}&&  \, \,   \,   
+{\frac {21577983707661117706708514436988691858431632715744973527227853}{
21340441599994994868198204433731283524323434200}}\,{x}^{5} 
\, + \, \, \cdots \nonumber 
\end{eqnarray}
Of course there is an ambiguity in all these solutions, 
except $\, S^{(3)}_0$ which is well defined: 
one can add $\, S^{(3)}_0$ to $\, S^{(2)}_0$, etc ...  There is 
an ambiguity on  $\, S^{(0)}_0$,  and  $\,S^{(2)}_0$, but the match of $\, S^{(0)}_0$
diagonal of the rational function and the form $\, S^{(2)}_1\,\, = \, \, \,
  S^{(2)}_0 \cdot \, \ln(x) \, + \, \,  T^{(2)}_0$ fixes the normalization of 
 $\, S^{(0)}_0$,  and  $\,S^{(2)}_0$. Thus $\, T^{(0)}_0$ is defined 
by (\ref{twoothersolmore}),
up to the solutions  $\, S^{(0)}_0$,  $\, S^{(1)}_0$ and  $\,S^{(2)}_0$. 
Trying to go further the fact that the operator is not MUM,
 one can try to define two nomes by 
\begin{eqnarray}
\hspace{-0.95in}&&  \, \quad  \qquad  \quad   \quad  \quad 
 q_1 \,\,\,  = \, \, \, \,
 \exp\Bigl( {{S^{(0)}_1 } \over {S^{(0)}_0 }}   \Bigr), \qquad \quad 
q_2 \,\,\,  = \, \, \, \,
 \exp\Bigl( {{S^{(2)}_1 } \over {S^{(2)}_0 }}   \Bigr), 
\end{eqnarray}
and, using this ambiguity, seek for  $\, T^{(0)}_0$ and $\, T^{(2)}_0$ 
such that the two nomes are globally bounded series. Unfortunately, 
it is almost impossible to see if one can build nomes, such that 
their series  are globally bounded series~\cite{Short,Big}.

\vskip .1cm 

With this example that is not MUM, we exclude any simple modularity
property for the operator, where the series for the nome, Yukawa
couplings, etc ... would be globally bounded. Diagonal of rational 
functions do not necessarily yield modularity. 

\vskip .1cm 

\subsubsection{Towards a ``generic'' diagonal of rational function: a second example\newline }
\label{genericdiagsec}

\vskip .1cm 

Let us consider another  simpler example with the diagonal of 
 another rational function of three variables:
\begin{eqnarray}
\label{generic}
\hspace{-0.95in}&&  \, \quad  \quad  \quad  \quad  \quad 
R(x,\, y, \, z)\, \, \, = \, \, \, \,
 {{1} \over { 1 \, +\, z \,- x\, y \,+\,x\,z\, +\, y^2}}.
\end{eqnarray}
The diagonal of this rational function reads:
\begin{eqnarray}
\label{genericdiag2}
\hspace{-0.95in}&&  \, \quad 
S^{(0)}_0 \, \, = \, \, \,\, Diag(R(x,\, y, \, z))\, \, = \, \, \,\,\,  
1 \,\,\, -2\,x\,  +3\,{x}^{2}\,  +40\,{x}^{3}\,  -545\,{x}^{4}\,  
+3948\,{x}^{5}\, 
\nonumber \\ 
\hspace{-0.95in}&&  \,\quad  \quad  \quad   \qquad  
\, \,-14910\,{x}^{6}\,  -55176\,{x}^{7}\,  \,  +1544895\,{x}^{8}\, 
-14999270\,{x}^{9}\, 
+82528303\,{x}^{10}\,
\nonumber \\ 
\hspace{-0.95in}&&  \, \quad  \quad \quad \quad  \quad \qquad  
\,\,  -29585712\,{x}^{11}\,  -5093494406\,{x}^{12}  
 \,\,\, + \, \,  \cdots 
\end{eqnarray}
It is solution of an irreducible order-four operator $\, L_4$ 
that is {\em not} MUM. This operator is non-trivially 
{\em homomorphic to its adjoint}, with order-two intertwiners 
and is actually of the form (\ref{forthcoming2}):
\begin{eqnarray}
\label{genericdiagform2}
\hspace{-0.95in}&&  \,   
 L_4 \, \, = \, \, \,\,   L_2 \cdot \, c_0(x)  \cdot \,
 M_2  \, \, \, \, \,  \,+ \, \, \,\,\,  \,
 {{ \lambda } \over {c_0(x) }},  \quad  
\lambda \, = \, \, \,  \Bigl({{26 } \over {639 }}\Bigr)^2, \quad   
c_0(x) \, = \, \, \, 
{{\lambda} \over {9}} \cdot {{ p_3(x)^2 \cdot p_4(x) } \over {p_6(x)}},  
 \nonumber \\
\hspace{-0.95in}&&  \,  
p_6(x) \, = \, \, \,676\,{x}^{6}\,+10514\,{x}^{5}\,
-2047\,{x}^{4}\,+82424\,{x}^{3}\,-15796\,{x}^{2}\,+10304\,x\, +448,
 \nonumber \\
\hspace{-0.95in}&&  \,  
p_4(x) \, = \, \,\, \,729\,{x}^{4}-1568\,{x}^{3}+984\,{x}^{2}+192\,x+16,
\quad p_3(x) \, = \, \, \, \, (71\,x+14)  \, (x-2) \cdot \, x, 
 \nonumber
\end{eqnarray}
where $\, L_2$ and $\, M_2$ are two {\em self-adjoint} operators, 
their Wronskian reading respectively 
\begin{eqnarray}
\hspace{-0.95in}&&  \,   \qquad \quad  \quad  \quad  \quad 
{{x^2 \cdot \, p_3(x) \cdot \, p_4(x)^2} \over {p_6(x) }},
 \quad \quad \qquad 
 {{ p_3(x) } \over {x^2 \cdot \, p_4(x) }}. 
\end{eqnarray}
The {\em exterior square} of $\, L_4$ is an order-six operator 
with a {\em rational function solution} $\, R(x)$, corresponding to the 
direct sum decomposition:
\begin{eqnarray}
\hspace{-0.95in}&&  \,   \quad  \quad 
Ext^2(L_4) \, \, = \, \, \, 
L_5  \, \oplus \, \Bigl(D_x \, -{{d \ln(R(x)} \over {dx}}   \Bigr), 
\qquad 
R(x)  \, \, = \, \, \,{{ p_3(x)} \over {x^2 \cdot \, p_4(x) }}, 
\end{eqnarray}
where  $\, L_5$ is an irreducible order-five operator. 
The operator  $\, L_4$ has the following four solutions:
the diagonal  $\,S^{(0)}_0 $ (see (\ref{genericdiag2})), an 
analytic solution $\, S^{(1)}_0$  
\begin{eqnarray}
\label{secondanal}
\hspace{-0.95in}&&  \,  
S^{(1)}_0 \, \, = \, \, \,\,\,
 x \,\,\, -{\frac {35}{2^2}}\,{x}^{2}\,\, \,
 +{\frac {3185}{2^{6}}}\,{x}^{3} \, \,
-{\frac {35035}{2^{8}}}\,{x}^{4} \,\, -{\frac {16207191}{2^{14}}}\,{x}^{5} \, \,
+{\frac {1217957741}{2^{16}}}\,{x}^{6} \,
\nonumber \\
\hspace{-0.95in}&&   \, \, \,    \, \,  \,\, \,
\, -{\frac {165312417127}{2^{20}}}\,{x}^{7}
\, +{\frac {3091190741925}{2^{22}}}\,{x}^{8} \, 
  +{\frac {1071079996954825}{2^{30}}}\,{x}^{9}
 \,\, \, + \, \, \cdots 
\end{eqnarray}
and two formal series solution with a log, namely 
$\, \, S^{(0)}_1 \, + \, \ln(x) \cdot \, S^{(0)}_0$ 
and also $\, \, S^{(1)}_1 \, + \, \ln(x) \cdot \, S^{(1)}_0$
\begin{eqnarray}
\hspace{-0.95in}&&  \,   \quad  \quad 
S^{(0)}_1 \, \, = \, \,\, \,\, 2 \,\,\,\,
 -{\frac {35}{2}}\,x \,\, +{\frac {5703}{56}}\,{x}^{2} \, \,
-{\frac {321597}{896}}\,{x}^{3} \, \,-{\frac {2659681}{3584}}\,{x}^{4}
\,\, +{\frac {29836311703}{1146880}}\,{x}^{5}
 \nonumber \\
\hspace{-0.95in}&&   \, \quad \quad  \quad  \quad 
-{\frac {1156839045933}{4587520}}\,{x}^{6} \,\,
 +{\frac {722563886554257}{513802240}}\,{x}^{7} \, \,
-{\frac {550307089986855}{411041792}}\,{x}^{8}
  \nonumber \\
\hspace{-0.95in}&&   \, \quad  \quad \quad  \quad \,
 -{\frac {22561115451957783769}{315680096256}}\,{x}^{9}
\,\,\, + \, \, \cdots  
\nonumber
\end{eqnarray}
and:
\begin{eqnarray}
\hspace{-0.95in}&&  \,   \quad  
S^{(1)}_1 \, \, = \, \,\, \,\,   {{11} \over {2}} \,{x}^{2} \, \,  \,\, 
-{\frac {30163}{576}}\,{x}^{3} \, \,  +{\frac {792323}{2304}}\,{x}^{4}
\, \,   -{\frac {219473079}{163840}}\,{x}^{5}
\, \,  -{\frac {16188947647}{5898240}}\,{x}^{6} \,
 \nonumber \\
\hspace{-0.95in}&& \quad  \quad   \,
\, +{\frac {70828996802681}{660602880}}\,{x}^{7} \, 
-{\frac {27874560487367}{25165824}}\,{x}^{8} \,
 +{\frac {5457381128456532577}{811748818944}}\,{x}^{9}
\, \,  + \, \, \cdots 
 \nonumber
\end{eqnarray}
If one introduces the nome, its series expansion does not seem to 
be globally bounded~\cite{Short,Big}: 
\begin{eqnarray}
\hspace{-0.95in}&&  \,   \, 
q(L_4) \,\, = \, \, \,
 x \cdot  \, exp\Bigl( {{ S^{(1)}_1} \over {S^{(1)}_0 }}  \Bigr)
\, = \, \, \,\,  
x \,\,\, + {{11} \over {2}} \,{x}^{2}\,\,  +{\frac {6269}{576}}\,{x}^{3}\, 
+{\frac {43165}{1152}}\,{x}^{4}\,  +{\frac {1040535941}{13271040}}\,{x}^{5}\,  
 \nonumber \\
\hspace{-0.95in}&& \quad  \quad     \quad    \qquad      
-{\frac {11364935021}{26542080}}\,{x}^{6}\,\, 
 +{\frac {851517278314609}{160526499840}}\,{x}^{7}\, \,
-{\frac {1854100924158503}{64210599936}}\,{x}^{8}
 \nonumber \\
\hspace{-0.95in}&& \quad  \quad  \quad  \quad  \quad   \qquad       \, 
+{\frac {790034414470824586787}{14794122225254400}}\,{x}^{9}\,
\, \, + \, \, \cdots 
 \nonumber
\end{eqnarray}

Do note that the second analytic series is 
actually globally bounded. With a rescaling 
$\, x \, \rightarrow \, 2^4 \, x$, the series 
(\ref{secondanal}) becomes a series with integer coefficients:
\begin{eqnarray}
\hspace{-0.95in}&&  \quad  \,  16\,x \, \, 
-2240\,{x}^{2} \,\,  +203840\,{x}^{3} \,\,  -8968960\,{x}^{4} \,\,  -1037260224\,{x}^{5}
\,\,  +311797181696\,{x}^{6} 
\nonumber \\
\hspace{-0.95in}&&  \quad  \quad  \quad  \qquad  \, -42319978784512\,{x}^{7}  
\, + \, 3165379319731200\,{x}^{8} \, \, + \, \, \cdots    
\end{eqnarray}

\vskip .1cm 

\subsubsection{Towards a ``generic'' diagonal of rational function: a third example\newline }
\label{genericdiagthird}

Let us consider another example with the diagonal of 
 another rational function of three variables:
\begin{eqnarray}
\label{generic3}
\hspace{-0.95in}&&  \, \quad \qquad  \quad  \quad  \quad  \quad 
R(x,\, y, \, z)\, \, \, = \, \, \, \, \,
 {{1 \, -x \, -y \, +x \, y \, z} \over { 
1 \,\, -x \, -y \, - x\, y \, -\, y^2 \, z^3}}.
\end{eqnarray}
The diagonal of this rational function reads: 
\begin{eqnarray}
\label{genericdiag3}
\hspace{-0.95in}&&  \, \quad 
S^{(0)}_0 \, \, = \, \, \, \,\, Diag(R(x,\, y, \, z))
\, \, \, = \, \, \,\,  \,
1\, \, \, +x\,\,  +10\,{x}^{3}\,\,  +32\,{x}^{4}\,\,  +966\,{x}^{6}\, \, 
+3192\,{x}^{7}\, 
\nonumber \\ 
\hspace{-0.95in}&&  \,\quad  \quad   \qquad   \qquad  
+120340\,{x}^{9}\, +401720\,{x}^{10}\, \,+16712150\,{x}^{12}\,
 +56066920\,{x}^{13}\, 
\nonumber \\ 
\hspace{-0.95in}&&  \,\quad  \quad \quad   \qquad   \qquad  
\, \,  +2466298800\,{x}^{15}
+8298484992\,{x}^{16} \, +378403867380\,{x}^{18}
\\ 
\hspace{-0.95in}&&  \, \quad  \quad \qquad   \qquad  \quad \quad 
\,\,\, \,  +1275714885984\,{x}^{19}\, +59651272137600\,{x}^{21}
 \,\,\, + \, \,  \cdots 
\nonumber 
\end{eqnarray}
It is solution of an {\em order-five} operator $\, L_5$ 
which factors as $\, L_5 \, = \, \, L_4 \cdot \, D_x$,
where  $\, L_4$ is an {\em irreducible} order-four operator
that is {\em not MUM}. The {\em exterior square} of $\, L_4$ is an 
order-six operator with a {\em rational function solution} $\, R(x)$, 
corresponding to the {\em direct sum} decomposition:
\begin{eqnarray}
\hspace{-0.95in}&&  \,   \quad  \quad 
Ext^2(L_4) \, \, = \, \, \,
 L_5  \, \oplus \, \Bigl(D_x \, -{{d \ln(R(x)} \over {dx}}   \Bigr), 
\qquad \quad 
R(x)  \, \, = \, \, \,{{ p_{10}(x)} \over {x^2 \cdot \, p_6(x)^2 }}, 
\nonumber \\
\hspace{-0.95in}&&  \, \quad    \quad  \quad 
p_{10}(x)\, \, = \, \, \,11008\,{x}^{10} \, +165760\,{x}^{9} \, 
-637392\,{x}^{8} \, +383388\,{x}^{7} \, +196287\,{x}^{6}
\nonumber \\
\hspace{-0.95in}&&  \, \quad   \quad  \quad  \qquad  \quad  \quad  \,
 -281004\,{x}^{5} \, 
-66582\,{x}^{4} \,  -45360\,{x}^{3} \, +15660\,{x}^{2} \, -810\,x \, +162, 
\nonumber \\
\hspace{-0.95in}&&  \,   \quad  \quad  \quad 
p_{6}(x)\, \, = \, \, \,1024\,{x}^{6} \, -9909\,{x}^{3} \, +54. 
\end{eqnarray}
This order-four operator $\, L_4$ is non-trivially homomorphic to its adjoint,
with order-two intertwiners 
and is actually of the form (\ref{forthcoming2}):
\begin{eqnarray}
\label{genericdiagform3}
\hspace{-0.95in}&&  \,   
 L_4 \, \, = \, \, \,\,   L_2 \cdot \, c_0(x)  \cdot \,
 M_2  \, \, \, \, \,  \,+ \, \, \,\,\,  \,
{\frac {1305}{29584}} \cdot \,  {{ 1 } \over {c_0(x) }},  \quad   \quad  
c_0(x) \, = \, \, \, {\frac {145}{473344}} \cdot {{ p_{10}(x)^2  } \over {p_{16}(x)}},  
 \nonumber \\
\hspace{-0.95in}&&    
p_{16}(x) \,\, = \, \, \, 37120\,{x}^{16}\,+1255680\,{x}^{15}\,-48870560\,{x}^{14}\,
+594756560\,{x}^{13}\,-31084335\,{x}^{12}
 \nonumber \\
\hspace{-0.95in}&& \quad  \quad \quad  \qquad  
   +2785358960\,{x}^{11}\,+4430975954\,{x}^{10}\,
-8858296096\,{x}^{9}\,-1107376429\,{x}^{8} 
 \nonumber \\
\hspace{-0.95in}&& \quad \quad \quad  \qquad   -369545240\,{x}^{7}\, 
+4215494304\,{x}^{6}\, -1487095128\,{x}^{5}\,-466418052\,{x}^{4}\,
\nonumber \\
\hspace{-0.95in}&& \quad  \quad \quad  \qquad  +228523680\,{x}^{3}\,  
 -21096612\,{x}^{2}\, +2737800\,x\, -717336, 
 \nonumber
\end{eqnarray}
where $\, L_2$ and $\, M_2$ are two self-adjoint operators, 
their Wronskian reading respectively 
\begin{eqnarray}
\hspace{-0.95in}&&  \,   \qquad \qquad  \quad 
{{x^2 \cdot \, p_{10}(x) \cdot \, p_6(x)^2} \over {p_{16}(x) }}, 
\qquad \qquad 
 {{ p_{10}(x)  } \over {x^2 \cdot \, p_6(x)^2 }}. 
\end{eqnarray}

The operator $\, L_4$ is not MUM: it has {\em three} solution
 analytic at $\, x\, = \, 0$, namely 
the derivative of diagonal (\ref{genericdiag3}) 
\begin{eqnarray}
\label{deuxdeplus}
\hspace{-0.95in}&&  \,  \, \, \,   
{{d S^{(0)}_0} \over {dx}}  \, \, = \, \, \,\, 
 1 \, \, +30\,{x}^{2} \, +128\,{x}^{3} \, +5796\,{x}^{5}
 \, +22344\,{x}^{6} \, +1083060\,{x}^{8} \,\, +  \, \, \cdots,   
\end{eqnarray}
and the two series solutions
\begin{eqnarray}
\label{deuxdeplus}
\hspace{-0.95in}&&  \,     \quad  \quad 
x \, \, \,\,  -{\frac {595}{1107}}\,{x}^{2}\,\, \,
  +{\frac {3515617}{1225449}}\,{x}^{3}\, \, 
+{\frac {227188435}{1225449}}\,{x}^{4}\,\, 
  -{\frac {2520602}{15129}}\,{x}^{5}\, \, 
+{\frac {8346429274}{11029041}}\,{x}^{6}  
\nonumber \\
 \hspace{-0.95in}&&  \, \quad   \quad   \quad   \quad \quad \quad 
\, +{\frac {2633989297550}{77203287}}\,{x}^{7} \, \, 
 -{\frac {42751323143}{1225449}}\,{x}^{8} \,\,\,   + \, \, \cdots,
 \nonumber \\
 \hspace{-0.95in}&&  \,    \quad  \quad 
{x}^{2} \,\,  \, \, -{\frac {595}{1107}}\,{x}^{3} \,\, \,
+{\frac {4523}{1107}}\,{x}^{4} \,\,  +{\frac {37758}{205}}\,{x}^{5} \,\, 
-{\frac {1412590}{9963}}\,{x}^{6}
 \,\,  +{\frac {63250564}{69741}}\,{x}^{7} 
\nonumber \\
 \hspace{-0.95in}&&  \,   
\quad \quad \quad   \quad   \quad \quad  \quad 
+{\frac {187516948}{5535}}\,{x}^{8} \,\,\,  + \, \, \cdots,
\end{eqnarray}
One also has a formal series solution with a log, namely
\begin{eqnarray}
\hspace{-0.95in}&&  \,   \qquad \qquad  \quad \quad 
 S_1(x) \,  \, \, + \, \,  \ln(x) \cdot \, {{d S^{(0)}_0} \over {dx}},  
\end{eqnarray}
where $\, S_1(x)$ is a series analytic at $\, x \, = \, \, 0$: 
\begin{eqnarray}
\label{SS1}
\hspace{-0.95in}&&  \,   
S_1(x) \, \, \,  = \, \, \,  \,  {{1} \over {5 \, x}} 
\, \, \, +{\frac {3083}{2214}} \, +{\frac {5222887}{4084830}}\,x \, 
+{\frac {956031447781}{22609534050}}\,{x}^{2} \, 
+{\frac {661652916345161}{2502875419335}}\,{x}^{3} 
\nonumber \\ 
\hspace{-0.95in}&&  \, \,  \quad 
\, +{\frac {2061248440531939}{12514377096675}}\,{x}^{4}
\, +{\frac {25090312744949777}{3089969653500}}\,{x}^{5} \, 
+{\frac {43029529492015002359}{901035150960600}}\,{x}^{6} \, 
\nonumber \\ 
\hspace{-0.95in}&&  \,\,   \quad \quad \quad 
+{\frac {21734062670504361827}{788405757090525}}\,{x}^{7} \, 
+{\frac {5314508101399026611659}{3504025587069000}}\,{x}^{8}
\, \,\, + \, \,\,  \cdots 
\end{eqnarray}

The last series in (\ref{deuxdeplus}), as well as $\, S_1(x)$ 
(see (\ref{SS1})),  are {\em not globally bounded} 
series~\cite{Short,Big}.

\vskip .1cm 

{\bf Remark 1:} The series  (\ref{deuxdeplus}) is, of course, 
also a diagonal of a rational function:
\begin{eqnarray}
\hspace{-0.95in}&&  \,   \quad  \qquad \qquad 
x \cdot \, {{d S^{(0)}_0} \over {dx}}  \, \, = \, \, \,\, 
Diag\Bigl( x \cdot \, {{ \partial R(x,\, y, \, z)} \over {\partial x}} \Bigr) 
 \\ 
\hspace{-0.95in}&&  \, \qquad   \qquad \quad  \quad \qquad  \, \, = \, \, \,\, 
Diag\Bigl( y \cdot \, {{ \partial R(x,\, y, \, z)} \over {\partial y}} \Bigr) 
 \, \, = \, \, \,\, 
Diag\Bigl( z \cdot \, {{ \partial R(x,\, y, \, z)} \over {\partial z}} \Bigr). 
\nonumber  
\end{eqnarray}

\vskip .1cm 

{\bf Remark 2:} With this order-five example, we see that the 
minimal order operator $\, L_5$, that annihilates 
the diagonal of a rational function, is not necessarily
irreducible. Let us recall the results of~\cite{Short,Big}
where we have shown that the $\, \tilde{\chi}^{(n)}$'s 
of the susceptibility of the square Ising model 
are {\em actually diagonals of rational functions}. The corresponding 
(globally nilpotent) linear differential operators   
 annihilating the $\, \tilde{\chi}^{(n)}$'s are not irreducible, on the contrary they
factor into many  linear differential operators, of various 
orders~\cite{bo-bo-ha-ma-we-ze-09,High,bernie2010,Khi6,ze-bo-ha-ma-05c,ze-bo-ha-ma-04} 
(one, two, three,  ..., 12, 23, ...). 
The interesting property we must focus on, is not that the (minimal order) 
linear differential 
operators annihilating the $\, \tilde{\chi}^{(n)}$'s are homomorphic 
to their adjoint, but that {\em all their factors} could be homomorphic 
to their adjoint. It is the differential Galois group of {\em all 
these factors} that we expect to be ``special''.

\vskip .1cm 

{\bf To sum up:} One may consider the following conjecture: all 
the irreducible factors of the minimal order linear differential 
operator annihilating a diagonal of a rational function 
should be homomorphic to their adjoint (possibly 
on an algebraic extension).
 
\vskip .1cm 

{\bf Remark 3:} Let us recall that the series 
of the hypergeometric function considered in~\cite{Short,Big}
\begin{eqnarray}
\label{totalrecal}
\hspace{-0.95in}&&  
_3F_2\Bigl([{{1} \over {9}}, \, {{4} \over {9}}, \, {{5} \over {9}}], \,
 [{{1} \over {3}},\, 1], \, 729 \, x\Bigr)
 \,\,  = \,\,  \,  \, 1 \, \,  +60\, x \,  \, 
+20475\, x^2 \, +9373650\, x^3 \, \, 
 \, + \, \, \cdots 
\end{eqnarray}
still remains a series with {\em integer} coefficients such that 
one cannot prove that it is the diagonal of a rational function, 
or discard that option. The minimal order operator annihilating 
this series is an order-three operator $\, L_3$ which 
is {\em not}\footnote[2]{It is not even 
homomorphic up to algebraic extensions. The order-two intertwining
operator $\, M_2$ such that 
$\, M_2 \cdot \, L_3 \, = \, \, adjoint(L_3) \cdot \, adjoint(M_2)$
has {\em transcendental} coefficients.} homomorphic to its adjoint. 

If our conjecture above was correct, this would be a way to 
show that the series (\ref{totalrecal}) {\em cannot be the diagonal of a 
 rational function}. 

\vskip .1cm 

\section{Conclusion}
\label{conclu}

Selected differential Galois groups correspond to symmetric square 
or exterior square, and possibly higher powers
(as seen with order-seven operators
with exceptional differential Galois groups of section (\ref{except})) 
 of operators, or equivalent operators, having rational solutions
(or $\, N$-th root of rational solutions,
 i.e. hyperexponential~\cite{SingUlm,Barkatou} 
 solutions).  
We have focused, in this paper, on a concept of ``Special Geometry''
corresponding to operators {\em homomorphic to their adjoint}. 
In a forthcoming publication, more focused on {\em differential systems},
and on demonstrations, we will show the equivalence 
of the homomorphism of an operator 
with its adjoint (possibly with algebraic extension), and the 
fact that its symmetric, or exterior,
square of the corresponding differential systems have 
a rational (resp. $\, N$-th root of rational) solution.

\vskip .1cm 

In~\cite{Bogner,BognerGood} Bogner has been able, from the very existence of 
underlying Calabi-Yau varieties, to show that the Calabi-Yau differential 
operators are actually conjugated to their adjoint (Poincar\'e pairing). If one 
does not assume strong hypotheses like this one, it is not simple to 
disentangle the differential algebra structures (corresponding to selected 
differential Galois groups) we have addressed in this paper,
and more ``arithmetic'' concepts associated to the notion of {\em modularity},
and the {\em integrality}, or {\em globally bounded} properties~\cite{Short,Big} 
of the various series occurring with these differential operators 
(solution of the operator, the nome, the Yukawa couplings, ...). Recalling 
section \ref{comments}, it is clear that  the concept of ``Special Geometry'',  
which we address in this paper, does not necessarily 
yield\footnote[2]{Katz's book~\cite{Katz} provides
 examples of {\em self-adjoint} operators with special 
differential Galois groups
 that are {\em not even Fuchsian} (see also one of our first (hypergeometric) 
examples (\ref{nonfuchs}).} 
arithmetic properties like the globally bounded~\cite{Short,Big} 
character of various series associated with the
operators. Conversely, we know that globally 
bounded series~\cite{Short,Big}
do not necessarily correspond to holonomic functions (see the 
example of the {\em non-holonomic} susceptibility of the
Ising model and its series with {\em integer} 
coefficients~\cite{bo-gu-ha-je-ma-ni-ze-08}).
Along such ``modularity'' line, the idea that 
operators annihilating {\em diagonals of rational functions} 
should always correspond to a modularity property that the 
corresponding nome and all the Yukawa's~\cite{Short,Big}
are globally bounded series, has been ruled out (see, for 
instance, the example of subsection \ref{genericdiagfirst}). From 
a mathematics viewpoint, there is still a lot 
of work to be performed to clarify the relations between 
these various neighboring concepts around the notion of ``modularity''.
Along this line, it is probably useful to keep in mind all the
simple examples\footnote[1]{For instance the order-six and order-eight 
operators $\, G_6^{5Dfcc}$ and $\, G_8^{6Dfcc}$ of sections \ref{Order-six} 
and \ref{ordereight} are not MUM.} of section \ref{comments}. From 
a physics viewpoint, one would like to identify, 
more specifically, what kind of 
``Special Geometry'' we encounter (Calabi-Yau, selected 
hypergeometric functions up to pull-backs~\cite{CalabiYauIsing}, ...). 

\vskip .1cm 

In this paper, the emergence of selected differential Galois groups 
has been seen, in a down-to-earth physicist's viewpoint, 
as {\em differential algebra} properties: one calculates 
various exterior, or symmetric, powers,
and looks (up to operator equivalence) for their  
rational solution (or hyperexponential~\cite{SingUlm,Barkatou}  solution), 
and {\em one calculates the homomorphisms of an operator with its adjoint}. 
We have shown that quite involved lattice Green operators of 
order six and eight are non trivially homomorphic to their adjoint, 
and that this yields the non trivial decompositions (\ref{finalresult}) 
and (\ref{finalresult8}),
where their intertwiners emerge in a crystal clear way 
(see also (\ref{decompoG4})
in section \ref{anisotrop}). 
Such decompositions enable to understand why the  
lattice Green operator (\ref{Broad6}) has a differential Galois group
included in the orthogonal group $\, O(6, \, \mathbb{C})$
instead of the symplectic $\, Sp(6, \, \mathbb{C})$
 differential Galois group, that one might expect 
for an order-six operator: the intertwiners are of {\em odd orders}.

Decompositions such as 
(\ref{genr1}), (\ref{genr2}) can be generalized 
for  linear differential operators 
of {\em any even} order. In fact, one can actually use 
the decompositions (\ref{genr1}), (\ref{genr2}) 
as an {\em ansatz} to provide linear differential operators of 
{\em any even} order, that will {\em automatically} have
selected differential Galois groups. 

With these lattice Green operators, we see that the simple 
generalization of the {\em Calabi-Yau condition} (\ref{CalabiCond}) 
for operators of order $\, N > 4$ (namely the condition that their 
exterior square is of order less than
 the generic $ \,N \cdot \, (N-1)/2$ order), 
is a {\em too restrictive concept for physics}. These lattice Green 
operators do not satisfy such  higher-order generalization of the 
 Calabi-Yau condition (\ref{CalabiCond}), but must be seen as 
higher-order generalization of a ``weak Calabi-Yau condition'' 
(see section \ref{weak}) which amounts to saying that 
their exterior or symmetric squares have rational solutions,
and that they are non-trivially homomorphic to their adjoint.

For order-four operators, Calabi-Yau operators are defined, among 
several other conditions (see Almkvist et al.~\cite{TablesCalabi}), 
essentially by the {\em Calabi-Yau condition} (\ref{CalabiCond}).
It is, however, quite clear that any equivalent operator (in the sense
of the equivalence of operator, i.e. homomorphic to
the Calabi-Yau operator), is also a selected operator interesting 
for physics. We have shown, in this paper, that any order-four operator,
non-trivially homomorphic to an irreducible operator satisfying 
the {\em Calabi-Yau condition} (\ref{CalabiCond}),
has the following properties: it is 
{\em homomorphic to its adjoint} with {\em order-two} intertwiners,
it has a simple decomposition (\ref{forthcoming}),
and it is such that its exterior square necessarily has
a rational solution. Conversely, showing that ``irreducible order-four 
operators such that their exterior square have a rational solution, 
or, even, have a  decomposition (\ref{forthcoming})'' 
are necessarily equivalent to irreducible operators satisfying 
the {\em Calabi-Yau condition} (\ref{CalabiCond}) is a difficult
question.

To illustrate the differential algebra structures corresponding to 
higher-order symmetric or exterior powers, 
we have also analysed some families of order-seven self-adjoint 
operators with exceptional differential Galois groups, where one sees,
very clearly, the emergence of rational solutions for 
symmetric square and {\em exterior cube} of equivalent operators.
Finally, since among the Derived From Geometry $\, n$-fold 
integrals (``Periods'') occurring in physics, we have seen that they are
quite often {\em diagonals of rational functions}~\cite{Short,Big}, we have 
also addressed many examples of (minimal order) operators annihilating
diagonals of rational functions, and remarked that they 
have {\em irreducible factors homomorphic to their adjoint}.

\vskip .1cm 

The $\, n$-fold integrals we encounter in theoretical physics
are solutions of {\em Picard-Fuchs} differential equations, 
or in a more modern mathematical language~\cite{Voisin,Steenbrink}, 
variation of Hodge 
structures\footnote[1]{Corresponding to the integrands in the
$\, n$-fold integrals, namely one-parameter
families of algebraic varieties.} and Gauss-Manin 
systems~\cite{DettReit,Bogner,Gri70,Gri84}. According to mathematicians
one should necessarily have for such variation of Hodge 
structures, a ``{\em polarization}''\footnote[2]{Which is a 
non-degenerated bilinear map (dual to an intersection mapping, 
see also the Poincar\'e duality~\cite{Movasati,Cattani}).} 
necessarily yielding to a ``{\em duality}'' 
which would send
differential operators into their adjoint\footnote[3]{
The Picard-Fuchs linear differential operators associated with a family of 
smooth projective manifolds are homomorphic to their adjoint. This can be
seen using the Poincar\'e duality~\cite{Christolprivate}. }. In our 
physical examples one seems to systematically inherit
 this ``duality'' on {\em each factor} of the minimal order operator, 
each irreducible factor being homomorphic to its adjoint. Along this line,
 section \ref{comments} strongly suggests 
to consider the conjecture that 
{\em (minimal) operators annihilating diagonal of rational functions 
solutions, always factor into irreducible operators 
homomorphic to their adjoint}, may be on algebraic extensions 
(these factors thus corresponding to ``special'' differential Galois groups). 

This paper tries to promote the idea that, before  deciphering 
the obfuscation of mathematicians 
on this subject, physicists should, in a down-to-earth way, use all the
differential algebra tools\footnote[9]{DEtools in Maple.} they 
have at their disposal, checking systematically 
if the linear differential operators
 they work on, have factors which are homomorphic to their 
adjoint, or are such that, up to operator equivalence,
their exterior (resp. symmetric) square have a rational 
solution. The emergence of this ``duality'' on all the irreducible
factors of a large class of differential operators of physics
needs to be understood.

\vskip .4cm 

\vskip .3cm 

{\bf Acknowledgments:} We thank A. Bostan, G. Christol
 and P. Lairez for fruitful
 discussions on diagonals of rational functions.
 We thank  Y. Andr\'e, D. Bertrand, 
F. Beukers, L. Di Vizio, C. Voisin,
 for fruitful discussions on
 differential Galois groups and (self-adjoint) dualities
in geometry. We also thank P. Boalch for fruitful
isomonodromic discussions.
As far as physicists authors 
are concerned, this work has been performed without
 any support of the ANR, the ERC or the MAE. 

\vskip .1cm 

\vskip .1cm 

\appendix

\vskip .1cm

\section{A decomposition of operators  equivalent
 to operators satisfying the Calabi-Yau condition }
\label{appdressing}

Let us again consider an order-four operator 
$\, \Omega_4$ which satisfies 
the Calabi-Yau condition (\ref{CalabiCond}), of Wronskian
 $\, w(x) \, = \, \, u(x)^2$.

Let us consider a monic order-four operator  $\, \tilde{\Omega}_4$  
which is (non-trivially) equivalent 
to the order-four operator $\, \Omega_4$ satisfying 
the Calabi-Yau condition (\ref{CalabiCond}).
This amounts to saying that there exist two  (at most) order-three
intertwiners $\, U_3$ and $\, L_3$
\begin{eqnarray}
\hspace{-0.9in}&& \quad \quad \quad 
U_3 \, \, \, \,= \, \, \, \, \,\,\, \,
b_3(x)\cdot \, D_x^3 \,\,\, \,\,  +b_2(x)\cdot \, D_x^2\, \,\, \,
 +b_1(x) \cdot \, D_x\,\, \, \, +b_0(x),  
\end{eqnarray}
such that\footnote[2]{Given $\, \Omega_4$
and $\, U_3$, in order to build the equivalent
operator $\,  \tilde{\Omega}_4$, just perform,
in Maple, the LCLM of $\, \Omega_4$ and $\, U_3$, and, 
then, right-divide by $\, U_3$.}:
\begin{eqnarray}
\label{U3L3}
\hspace{-0.1in}&&\quad \quad \quad
\tilde{\Omega}_4 \cdot U_3 
 \, \,\,\,  = \, \, \,\, \,\,\,  L_3 \cdot \, \Omega_4
\end{eqnarray}
We choose $\, L_3$ such that $\, \tilde{\Omega}_4$ is monic 
($\, \tilde{\Omega}_4 \, = \, \, D_x^4 \, + \, \, \cdots$).
Of course, we also have the (adjoint) relation: 
\begin{eqnarray}
\label{U3L3adj}
\hspace{-0.6in}&& \quad \quad
 adjoint(U_3)  \cdot \, adjoint(\tilde{\Omega}_4) 
 \, \,\,\,  = \, \, \,\, \,\,\,
  adjoint(\Omega_4)\cdot \,  adjoint(L_3).
\end{eqnarray}
Furthermore, it is shown in \ref{self} that any operator  satisfying 
the Calabi-Yau condition (\ref{CalabiCond}), is homomorphic to its
adjoint, up to a conjugation by the square root of its Wronskian:
\begin{eqnarray}
\label{self1}
\hspace{-0.6in}&& \quad \quad \quad  \quad 
u(x) \cdot \,  adjoint(\Omega_4) 
 \, \,\,\,  = \, \, \,\, \,\,\,  \Omega_4  \cdot \, u(x).
\end{eqnarray}
Combining (\ref{U3L3}), (\ref{U3L3adj}) and (\ref{self1})
one straightforwardly deduces:
\begin{eqnarray}
\label{deduce}
\hspace{-0.6in}&& \quad \quad \quad  \quad 
\tilde{\Omega}_4 \cdot \, Y_6  \,\,\,  = \, \, \,\, \,\,\, 
 adjoint(Y_6) \cdot \, adjoint(\tilde{\Omega}_4),  
\end{eqnarray}
where  the order-six operator $\, Y_6$ reads:
\begin{eqnarray}
\hspace{-0.1in}&&\quad  \quad Y_6 \, \, \, \, \,  = \,\,  \, \,  \, \, \,
 U_3 \cdot \, u(x) \cdot \, adjoint(L_3). 
\end{eqnarray}

Let us introduce the two operators $\, N_2$ and $\, Z_2$ 
corresponding to the euclidean division of $\, Y_6$
 by $\, adjoint(\tilde{\Omega}_4)$:
\begin{eqnarray}
\label{Y6decomp}
\hspace{-0.1in}&& \quad  \quad Y_6 \, \, \, \,  = \, \,\,  \,  \, \,
 N_2 \cdot \, adjoint(\tilde{\Omega}_4) \, \,\,\,   + \, \, Z_2.  
\end{eqnarray}
 $\, N_2$ is of course an order-two operator, but, 
noticeably,  $\, Z_2$ is {\em also}  an {\em order-two} operator
instead of an order-three operator one could expect generically.

Furthermore, and noticeably,  $\, N_2$ is an order-two 
{\em self-adjoint} operator such that:
\begin{eqnarray}
\hspace{-0.9in}&& \quad    \quad \quad  \quad 
{{1}  \over {b_3(x)}} \cdot \, N_2 \cdot \, {{1}  \over {b_3(x)}} 
 \,  \, \,  \,  \,  = \, \,\,  \,  \,  \,  \, 
u(x) \cdot \Bigl(D_x^2 \, \,
 - {{d \ln(1/u(x))} \over {dx}} \cdot \, D_x \Bigr)  \, \, 
\nonumber \\
\hspace{-0.9in}&& \quad  \quad    \qquad \qquad  \qquad  \quad 
- \,u(x) \cdot \, 
\Bigl({{d b_2(x) } \over {dx}} \, + \, b_2(x)^2 \, -2\, b_1(x)\Bigr)
\nonumber \\
\hspace{-0.9in}&& \quad  \quad  \quad \quad  
 \qquad  \qquad  \qquad  \qquad  \quad
- \,u(x) \, - \, {{d^2 u(x) } \over {dx^2}}\, 
+ \,  {{2} \over {x}} \cdot \,\Bigl({{d u(x) } \over {dx}}\Bigr)^2.
\end{eqnarray}

A consequence of the self-adjoint character of  $\, N_2$
is that one also has the ``adjoint'' 
relation\footnote[1]{One uses the fact that
the adjoint of the sum of an order-six and an order-two 
operator is the sum of these adjoints.} of (\ref{Y6decomp}):
\begin{eqnarray}
\label{adjY6decomp}
\hspace{-0.1in}&& adjoint(Y_6) \, \, \, \,  = \, \,\,  \,  \, \,
   \tilde{\Omega}_4  \cdot \, N_2  \, \,\,\,   + \, \, adjoint(Z_2).  
\end{eqnarray}
Combining (\ref{deduce}), (\ref{Y6decomp}) and (\ref{adjY6decomp})
one deduces the following homomorphisms of $\, \tilde{\Omega}_4$
with its adjoint, with an {\em order-two} intertwiner: 
\begin{eqnarray}
\label{intertw}
\tilde{\Omega}_4 \cdot \, Z_2 \, \,   = \, \,  \,    \,  
adjoint(Z_2) \cdot \, adjoint(\tilde{\Omega}_4),
\end{eqnarray}

Let us now perform the euclidean division of
 $\, adjoint(\tilde{\Omega}_4)$ by $\, Z_2$:  
\begin{eqnarray}
\label{decomptildeOmeg}
\hspace{-0.1in}&&\quad adjoint(\tilde{\Omega}_4)
 \, \, \, \,  \, = \, \, \, \,\,   \, 
A_2 \cdot \, Z_2 \,\,  \,\,  + \, \, A_0 
\end{eqnarray}
where $\, A_2$ is an order-two operator and, surprisingly,  $\, A_0$
is not an order-one operator, {\em but a function} (order zero). 
Of course (and using the fact that the adjoint of two even order 
operators is the sum of the adjoints) one also has the 
``adjoint relation''  of (\ref{decomptildeOmeg}), namely
\begin{eqnarray}
\label{adjdecomptildeOmeg}
\hspace{-0.1in}&&\quad \tilde{\Omega}_4
 \, \, \, \,  \, = \, \, \, \,\,   \, 
adjoint(Z_2) \cdot \, adjoint(A_2)   \,\,  \,\,  + \, \, A_0. 
\end{eqnarray}

In fact, and noticeably $\, A_2$
 is a {\em self-adjoint operator}.  Combining (\ref{intertw}),
(\ref{decomptildeOmeg}) and (\ref{adjdecomptildeOmeg}),
one immediately deduces that $\, Z_2$ is 
{\em conjugated to its adjoint}, or equivalently, that  the 
following order-two operator $\, Z_2^{s}$ {\em  is self-adjoint}:
\begin{eqnarray}
\label{inotherwords}
\hspace{-0.5in}&&\quad \quad \qquad \,  
 Z_2^{s}  \, \, \, \,  = \, \, \,  \, \,  \, 
A_0 \cdot \, Z_2 \, \, \,  =  \, \, \, \, adjoint(Z_2) \cdot \, A_0.
\end{eqnarray}

One finds out that the order-four operator $\, \tilde{\Omega}_4$
can, in fact, be written in terms of a remarkable decomposition with two 
 {\em order-two  self-adjoint operators}:
\begin{eqnarray}
\label{forthcomingappend}
\hspace{-0.1in}&& \qquad \tilde{\Omega}_4 \, \, \, \,  \, 
 = \, \, \,  \, \,  \, \,   \,  \, 
 Z_2^{s} \cdot {{1} \over {A_0}} \cdot \, A_2 \, \, \,  \,  \,
 + \, \, \,  \, A_0.  
\end{eqnarray}

One then deduces the  homomorphisms of $\,\tilde{\Omega}_4$
with its adjoint:
\begin{eqnarray}
\label{otherintertw}
\hspace{-0.5in}&&\quad \quad  \qquad 
A_2 \cdot \, {{1} \over {A_0}} \cdot \,
 \tilde{\Omega}_4 \, \, \, = \, \, \,  \,
adjoint(\tilde{\Omega}_4) \cdot \,  {{1} \over {A_0}} \cdot \, A_2. 
\end{eqnarray}
to be compared with
\begin{eqnarray}
\label{otherintertw2}
\hspace{-0.5in}&&\quad \quad  \qquad 
 \tilde{\Omega}_4 \cdot \, {{1} \over {A_0}} \cdot \, Z_2^{s}
 \, \, \,\,  = \, \, \,  \,
Z_2^{s}    \cdot \,  
{{1} \over {A_0}}  \cdot \,   adjoint(\tilde{\Omega}_4). 
\end{eqnarray}

\vskip .1cm

Note that one also has a relation similar to (\ref{deduce})
but on $\, \Omega_4$.  If one introduces 
\begin{eqnarray}
\hspace{-0.5in}&&\quad \quad \quad  \quad
Z_8 \, \, \, = \, \, \, \,  \, 
adjoint(L_3) \cdot \, {{1} \over {A_0}}
 \cdot \, A_2  \cdot \, U_3 \cdot \, u(x),
\end{eqnarray}
one has 
\begin{eqnarray}
\hspace{-0.5in}&&\quad \quad  \qquad 
adjoint(Z_8)  \cdot \, \Omega_4 \cdot \, u(x) 
  \, \, \, = \, \, \, \,  \, \Omega_4 \cdot \, u(x)  \cdot \, Z_8 
\end{eqnarray}
where  $\,\Omega_4 \cdot \, u(x)$ 
also verifies the Calabi-Yau condition
 and is {\em actually self-adjoint}. Denoting
$\,\Omega_4^{(s)} \, = \, \Omega_4  \cdot \, u(x)$
this gives
\begin{eqnarray}
\hspace{-0.5in}&&\quad \quad  \qquad 
adjoint(Z_8)  \cdot \,  \Omega_4^{(s)} 
  \, \, \, = \, \, \, \,  \, \Omega_4^{(s)}   \cdot \, Z_8. 
\end{eqnarray} 
One discovers that 
\begin{eqnarray}
\hspace{-0.5in}&&\quad \quad  \qquad 
Z_8  \, \, \, = \, \, \, \,  \, 
 X_4 \cdot \, \Omega_4^{(s)} \, \, -1.
\end{eqnarray}
where $\, X_4$ is  {\em actually self-adjoint}.
One straightforwardly deduces
\begin{eqnarray}
\hspace{-0.5in}&&\quad \,  \qquad  
 adjoint(X_4) \cdot \, adjoint(Z_8)  
  \, \, \, = \, \, \, \,  \,  \, 
\nonumber \\
\hspace{-0.5in}&&\quad \quad  \qquad  \quad  \quad \qquad \qquad
X_4 \cdot \, adjoint(Z_8)  
\, \, \, = \, \, \, \,  \,  \, Z_8 \cdot \, X_4. 
\end{eqnarray} 

\vskip .2cm

If one does not normalize the order-four operators 
$\, \tilde{\Omega}_4$ and $\, \Omega_4$ to be monic,
these results are easily modified, mutatis mutandis:
\begin{eqnarray}
\hspace{-0.9in}&&\quad 
\Omega_4 \, \, \rightarrow \, \, \,  \alpha(x) \cdot \, \Omega_4,
\qquad \tilde{\Omega}_4 \, \, \rightarrow \, \, \, 
 \beta(x) \cdot \,\tilde{\Omega}_4,
\qquad L_3 \, \, \rightarrow \, \, \,
  \beta(x) \cdot \,  L_3 \cdot \,\alpha(x)^{-1}
\nonumber \\
\hspace{-0.9in}&& 
 u(x) \, \, \rightarrow \, \, \,  \alpha(x) \cdot \, u(x), 
\qquad Y_6 \, \, \rightarrow \, \, \, \,  Y_6  \cdot \, \beta(x), \,
\qquad Z_2 \, \, \rightarrow \, \, \,  Z_2  \cdot \,  \beta(x),
 \\
\hspace{-0.9in}&& 
 SZ_2 \, \, \rightarrow \, \, \, 
 \beta(x) \cdot \,   SZ_2  \cdot \,  \beta(x),
\quad A_0 \, \, \rightarrow \, \, \, \beta(x) \cdot \, A_0, \quad 
(A_2, \, N_2, \,U_3) \, \, \rightarrow \, \, \, (A_2, \, N_2, \,U_3).
\nonumber
\end{eqnarray}

\vskip .1cm

These calculations could have been performed, in another way, 
considering the other homomorphism relation
between $\, \tilde{\Omega}_4$ and $\, \Omega_4$, instead of (\ref{U3L3}):
\begin{eqnarray}
\label{otherhomo}
\hspace{-0.9in}&&\quad \qquad \qquad \qquad 
 V_3 \cdot \, \tilde{\Omega}_4 
 \, \, \, \,  =  \, \, \,\, \,\Omega_4 \cdot \, M_3, 
\end{eqnarray}
and its ``adjoint'' relation:
\begin{eqnarray}
\label{adjotherhomo}
\hspace{-0.9in}&&\quad \qquad \qquad
  adjoint(\tilde{\Omega}_4)   \cdot \, adjoint(V_3)
 \, \,\,\,  =  \, \, \, \,\,   adjoint(M_3)  \cdot \, adjoint(\Omega_4).
\end{eqnarray}

Combining (\ref{otherhomo}), (\ref{adjotherhomo}) and (\ref{self1}), 
one easily deduces:
\begin{eqnarray}
\label{otherZ6}
\hspace{-0.9in}&&\quad \qquad \qquad 
adjoint(\tilde{\Omega}_4)   \cdot \, Z_6
 \, \, \,\,  =  \, \, \, \, \, 
adjoint(Z_6)  \cdot \, \tilde{\Omega}_4, 
\end{eqnarray}
where:
\begin{eqnarray}
\label{defZ6}
\hspace{-0.9in}&&\quad \qquad \qquad \qquad 
Z_6 \, \,\,  =  \, \, \, \, \, 
adjoint(V_3)  \cdot \,  {{1} \over {u(x)}}  \cdot \, M_3.
\end{eqnarray}
The rightdivision of this order-six operator by $\, \tilde{\Omega}_4$ yields:
\begin{eqnarray}
\label{rightdefZ6}
\hspace{-0.9in}&&\quad \qquad \qquad \qquad 
Z_6 \, \, \,\,  =  \, \, \, \, \, \, 
  P_2 \cdot \tilde{\Omega}_4 \, \, \, \, 
+ \, \, A_2 \cdot \, {{1} \over {A_0}},  
\end{eqnarray}
where $\, P_2$ is an order-two self-adjoint operator
(much more involved than $\, N_2$ ...).

\vskip .1cm

\section{Calabi-Yau conditions}
\label{calab}

\vskip .1cm

\subsection{Calabi-Yau conditions and self-adjoint operators}
\label{self}

Since most of the linear differential operators of order four
with polynomial coefficients, we have encountered in lattice 
statistical 
mechanics~\cite{ze-bo-ha-ma-04,High,Khi6,bo-gu-ha-je-ma-ni-ze-08,mccoy3,CalabiYauIsing},
enumerative combinatorics, are {\em globally nilpotent}~\cite{bo-bo-ha-ma-we-ze-09},
and thus their Wronskian are $\, N$-th root of {\em rational functions}, let us 
write, without any loss of generality, the coefficient $\, a_3(x)$ of 
operator $\, \Omega_4$ (see (\ref{Omega4})),
 in the log-derivative form:
\begin{eqnarray}
\label{logder}
\hspace{-0.8in}&&  \quad  \quad  \quad  \,  
a_3(x)\, \,\, \,   = \,\,\, \, \,   -\,  \,  \, {{ d \, \ln(w(x))} \over {dx}} 
\qquad \,    \hbox{with:} \qquad \,   w(x) \,   = \,\,\,u(x)^2.
\end{eqnarray}
Global nilpotence, even being Fuchsian, corresponds to the Wronskian 
$\, w(x) \, = \, \,  u(x)^2$ being 
 $\, N$-th root of a {\em rational function}.

It is straightforward to verify that if $\, \Omega_4$ satisfies 
the Calabi-Yau condition (\ref{CalabiCond}), then:  
\begin{eqnarray}
\label{conj}
 \hspace{-0.1in}&&  \quad  \quad   u(x) \cdot \, adjoint(\Omega_4)
 \,  \,  \,   \,\, = \, \,\, \, \,\,  \, \,    \,  \Omega_4 \cdot \, u(x).
\end{eqnarray}
In other words, the Calabi-Yau condition (\ref{CalabiCond}) 
necessarily means that the
order-four operator (\ref{Omega4}) is, not only homomorphic to its adjoint,
but conjugated to its adjoint.
The following conjugate of $\, \Omega_4$ is {\em self-adjoint}:
\begin{eqnarray}
\label{conj}
 \hspace{-0.1in}&&  \quad  \quad   
  {\tilde \Omega}_4 \,  \,   \,\, = \, \,\, \, \,\,  \, \,  
   u(x)^{-1/2}  \cdot \,\Omega_4 \cdot \, u(x)^{1/2}.
\end{eqnarray}

This can be easily checked on all the ODEs of the large list of
Calabi-Yau ODEs displayed in~\cite{TablesCalabi}.

Conversely, let us impose that an order-four operator  (\ref{Omega4})
is conjugated to its adjoint. Again it is straightforward to see 
that relation (\ref{conj}) yields:
\begin{eqnarray}
\label{C1}
\hspace{-0.9in}&&\quad \quad \quad a_1(x) \, \,  \, \,  
 = \,  \,  \, \, \, \,  \, {{d a_2(x)} \over {dx}}  \,  \,  \, \,
 \,   - {{a_2(x)} \over {u(x)}} \cdot \,  {{d u(x)} \over {dx}} \, \, 
 \, \,  + {{1} \over {u(x)}} \cdot \,  {{d^3 u(x)} \over {dx^3}}\, \,
 \\
\hspace{-0.9in}&& \quad \quad \qquad \quad \quad \quad \, \, \, \,\, \,
 + 6 \cdot \,  \Bigl(\Bigl({{1} \over {u(x)}} \cdot \,
  {{d u(x)} \over {dx}}\Bigr)^3  \,
- {{1} \over {u(x)^2}} \cdot \,  {{d u(x)} \over {dx}} 
 \cdot \,  {{d^2 u(x)} \over {dx^2}}\Bigr),  \nonumber
\end{eqnarray}
which is {\em nothing but the Calabi-Yau condition}
 (\ref{CalabiCond})  taking into account (\ref{logder}).

In other words, the {\em Calabi-Yau condition}
 (\ref{CalabiCond}) {\em is equivalent to say that an order-four operator 
is conjugated to its adjoint}. This result 
can also be found in Bogner (see ~\cite{Bogner}). 

\subsubsection{Strong Calabi-Yau conditions versus self-adjoint conditions
on operators: higher order operators \newline}
\label{versus}
An operator of order five is self-adjoint if it is of the form:
\begin{eqnarray}
\label{selfadj5}
\hspace{-0.9in}&&\quad \quad \quad L_5  \, \, \,  = \, \, \,   \,  \, \, \, 
 a_5(x) \cdot \, D_x^5 \,  \, 
 + \, {{5} \over {2}} \cdot \, {{d a_5(x)} \over {dx}}\cdot \, D_x^4 
\,\,  \,  + \, a_3(x) \cdot \, D_x^3
 \,  \, \,\,\nonumber \\ 
\hspace{-0.9in}&&\quad\quad \quad \quad \quad \quad  \quad 
 + \,  \, \Bigl( {{3} \over {2}}  \, {{d a_3(x)} \over {dx}} \,
 -  {{5} \over {2}}  \, {{d^3 a_5(x)} \over {dx^3}}
   \Bigr) \cdot \, D_x^2 \, \,\,\, + \, a_1(x) \cdot \, D_x
\\ 
\hspace{-0.9in}&&\quad \quad \quad \quad 
\quad \quad  \quad \quad  \quad \quad  \quad 
 \, + \, \Bigl({{1} \over {2}} \cdot \, {{d a_1(x)} \over {dx}} \,
 + {{1} \over {2}} \cdot \, {{d^5 a_5(x)} \over {dx^5}}
\, -{{1} \over {4}} \cdot \, {{d^3 a_3(x)} \over {dx^3}} \Bigr).
\nonumber 
\end{eqnarray}
Its {\em symmetric square is of order fourteen}
 instead of the order fifteen one could expect generically.
In other words this (exactly) self-adjoint operator, or an
order-five operator conjugated of (\ref{selfadj5})
by an arbitrary function, satisfies 
the symmetric Calabi-Yau condition (that its symmetric square 
is of order fourteen).

An operator of order six is self-adjoint if it is of the form:
\begin{eqnarray}
\label{selfadj6}
\hspace{-0.9in}&&\quad \quad \quad 
L_6  \, \, \,  = \, \, \,   \,  \, \,
 a_6(x) \cdot \, D_x^6 \,  \, 
\,\, + \, 3 \cdot \, {{d a_6(x)} \over {dx}}\cdot \, D_x^5 \,  \, 
\,\, + \, a_4(x) \cdot \, D_x^4
\nonumber \\ 
\hspace{-0.9in}&&\quad \quad \quad \quad  \quad \quad \quad 
+\Bigl(2 \cdot \, {{d a_4(x)} \over {dx}}\, 
 -5  \cdot \, {{d^3 a_6(x)} \over {dx^3}}\Bigr) \cdot \, D_x^3
\,\,\,  \,  + \, a_2(x) \cdot \, D_x^2 \\ 
\hspace{-0.9in}&&\quad \quad \quad \quad \quad \,\, \,
   \quad \quad \quad \quad \quad 
+\Bigl({{d a_2(x)} \over {dx}}\,  -\, {{d^3 a_4(x)} \over {dx^3}}\, 
+3  \cdot \, {{d^5 a_6(x)} \over {dx^5}}\Bigr) \cdot \, D_x 
\,\,\,\,  + \, a_0(x).\nonumber 
\end{eqnarray}
Its {\em exterior square is of order fourteen}
 instead of the order fifteen one could expect generically.
In other words this (exactly) self-adjoint operator, or an
order-six operator conjugated of (\ref{selfadj6})
by an arbitrary function, satisfies 
the Calabi-Yau condition (that its exterior square 
is of order fourteen), generalization to order six
of the order-four Calabi-Yau condition (\ref{CalabiCond}). 

It is straightforward to verify that an operator 
 conjugated of a self-adjoint operator of order $\, N$ 
verifies, for {\em any} even order $\, N$, the generalization to 
order $\, N$ of the order-four Calabi-Yau condition (\ref{CalabiCond})
and for {\em any} odd order $\, N$, the generalization to order $\, N$ 
of the order-three {\em symmetric Calabi-Yau condition}
 (\ref{symCab}). 

Of course the reciprocal, which is true for order-three and four operators 
(see (\ref{83}) and (\ref{C1})),  is not true for higher orders. For 
instance, let us introduce the order-five operator $\, M_5$ 
non-trivially homomorphic to the self-adjoint operator (\ref{selfadj5}):
\begin{eqnarray}
\hspace{-0.9in}&&\quad \quad \quad  \quad  
  M_5 \cdot \,  D_x  \, \, = \, \, \, {{1} \over {a_5(x)}} \cdot \,\Bigl( 
D_x \, - {{1} \over {W(x)}} \cdot \, {{d W(x)} \over {dx}}
\Bigr) \cdot \, L_5, 
\\
\hspace{-0.9in}&&\quad \quad \quad \,  \quad 
\hbox{where:} \quad  \quad \quad  \quad   \,  \, 
  W(x)  \, \, = \, \, \,
2  \, {{d a_1(x)} \over {dx}}  
 \, -  \, {{d^3 a_3(x)} \over {dx^3}} \, + \, 2  \, {{d^5 a_5(x)} \over {dx^5}},
\nonumber 
\end{eqnarray}
This operator also verifies the order-five {\em symmetric Calabi-Yau condition}
 (\ref{symCab}): its symmetric square is also of order fourteen.  
This result generalizes with $\, M_5$ 
\begin{eqnarray}
\hspace{-0.9in}&&\quad \quad \quad  \quad \qquad 
  M_5 \cdot \,  (D_x \, +\rho(x))
 \, \, \, = \, \, \, \, \, {{1} \over {a_5(x)}} \cdot \,\Bigl( 
D_x \, - z(x)
\Bigr) \cdot \, L_5, 
\end{eqnarray}
where $\, z(x)$ is a quite involved rational expression of $\, a_1(x)$, 
$\, a_3(x)$, $\, a_5(x)$, $\, \rho(x)$ and their derivatives. 

Similarly the order-six operator $\, M_6$ equivalent of the self-adjoint operator
(\ref{selfadj6}):
\begin{eqnarray}
\hspace{-0.9in}&&\quad \quad \quad  \quad \qquad 
  M_6 \cdot \,  D_x  \, \, \, = \, \, \, \,\,
 {{1} \over {a_6(x)}} \cdot \,\Bigl( 
D_x \, - {{1} \over {a_0(x)}} \cdot \, {{d a_0(x)} \over {dx}}
\Bigr) \cdot \, L_6, 
\nonumber 
\end{eqnarray}
verifies the order-six {\em Calabi-Yau condition}
 (\ref{CalabiCond}): its exterior square is also of order fourteen.  

This result generalizes with $\, M_6$ given by
\begin{eqnarray}
\hspace{-0.9in}&&\quad \quad \quad  \quad \qquad 
  M_6 \cdot \,  (D_x \, +\rho(x))
 \, \, \, = \, \, \, \, {{1} \over {a_6(x)}} \cdot \,\Bigl( 
D_x \, - z(x)
\Bigr) \cdot \, L_6, 
\end{eqnarray}
where $\, z(x)$ is a quite involved rational expression 
of $\, a_0(x)$, $\, a_2(x)$, $\, a_4(x)$, $\, a_6(x)$, $\, \rho(x)$ 
and their derivatives. 

The order-seven operator $\, M_7$ equivalent of the order-seven 
self-adjoint operator $\, L_7$:
\begin{eqnarray}
\hspace{-0.9in}&&\quad  \quad   
  M_7 \cdot \,  (D_x^2 \, +\rho_1(x) \cdot \, D_x \, +\rho_2(x))
 \, \, \, = \, \, \, \, {{1} \over {a_7(x)}} \cdot \, 
(D_x^2 \, +c_1(x) \cdot \, D_x \, +c_2(x)) \cdot \, L_7, 
\nonumber 
\end{eqnarray}
verifies the order-seven {\em Calabi-Yau condition} that 
its symmetric square is of order $\, 27$ 
(instead of the generic order $\, 28$).

Similarly the order-eight operator $\, M_8$ equivalent of the order-seven 
self-adjoint operator $\, L_8$:
\begin{eqnarray}
\hspace{-0.9in}&&\quad  \quad   
  M_8 \cdot \,  (D_x^2 \, +\rho_1(x) \cdot \, D_x \, +\rho_2(x))
 \, \, \, = \, \, \, \, {{1} \over {a_7(x)}} \cdot \, 
(D_x^2 \, +c_1(x) \cdot \, D_x \, +c_2(x)) \cdot \, L_8, 
\nonumber 
\end{eqnarray}
verifies the order-seven {\em symmetric Calabi-Yau condition} that 
its symmetric square is of order $\, 27$ 
(instead of the generic order $\, 28$).
These last results can easily be generalized. For instance for 
the order-nine and order-ten self-adjoint operators $\, L_9$, $\, L_{10}$
the corresponding equivalent operators $\, M_9$, $\, M_{10}$ 
obtained from the LCLM of  $\, L_9$ or $\, L_{10}$ with an {\em order-three}
operator  verify respectively the order-nine 
{\em symmetric Calabi-Yau condition}  
(namely the symmetric square of $\, M_9$ is of order 44 instead of 45) 
and the order-ten Calabi-Yau condition (namely that the symmetric and  
exterior squares of $\, M_9$ and $\, M_{10}$ are of order 44 instead of 45), 
and so on ... 

\vskip .1cm 

\vskip .1cm 

\subsection{Equivalence of operators satisfying the Calabi-Yau conditions}
\label{equivCalabi}

Let us recall some examples (see equation (O.37) in~\cite{Big})
 of  operators satisfying the Calabi-Yau
 conditions (\ref{CalabiCond}). The order-four ($\mu$-dependent) 
linear differential operator 
\begin{eqnarray}
\label{theexample7}
\hspace{-0.9in}&&\quad \quad   
{\cal C}(\mu) \,\,\, = \, \,  \,\,\, 
 16 \cdot \theta^2 \cdot \, (\theta\, -1)^2 \,\, 
\\
\hspace{-0.9in}&&\quad \quad \quad \qquad 
 -\, x \cdot  \,  ( 2 \theta \, \, +1\, -\mu) \cdot \,  ( 2 \theta \, \, +1\, +\mu) 
\cdot \, ( 2 \theta \, \, -1\, -\mu) \cdot \, ( 2 \theta \, \, -1\, +\mu),
 \nonumber 
\end{eqnarray}
is such that {\em its exterior square is actually of order five}.
These operators are {\em irreducible} for even integer values  of $\, \mu$,
but factor in a product of order-two and two order-one operators
for odd  integer values  of $\, \mu$. Operator (\ref{theexample7}) 
has simple hypergeometric solutions 
for {\em any value} of $\, \mu$:
\begin{eqnarray}
\label{other4F3}
\hspace{-0.9in}&& 
 \quad \quad \quad \quad \quad 
 x \cdot \, _4F_3\Bigl([{{\mu \, +3} \over {2}}, \, {{-\mu \, +3} \over {2}}, \,
  {{\mu \, +1} \over {2}},\, {{-\mu \, +1} \over {2}}],
  \, [1, \,  2, \, 2]; \,\,  x). 
\end{eqnarray}

These operators (\ref{theexample7}) are, for
 different {\em even integer values}  of $\, \mu$,
 non-trivially homomorphic (and similarly, these operators (\ref{theexample7}) are, 
 non-trivially homomorphic for different odd integer values  of $\, \mu$). 
One of the simplest examples of homomorphism reads:
\begin{eqnarray}
\label{simplest}
\hspace{-0.9in}&&\quad \quad 
{\cal C}(0)  \cdot U_3\,\,\, \, = \, \,\,   \,\,\,  V_3 \cdot {\cal C}(2),
 \quad \qquad  \qquad \hbox{with:}
\nonumber \\
\hspace{-0.9in}&&\quad \quad 
 U_3\,\,\, = \, \,  \,\, 
(32 \, \theta^3 \, -80 \,  \theta^2 \, +72 \, \theta  \, -27) \, \,  \,  \, 
-\, 4 \cdot \, x \cdot \, (2 \, \theta \, +1)
 \, (2 \, \theta \, -1) \, (2 \, \theta \, -3),  
\nonumber \\
\hspace{-0.9in}&&\quad \quad  
V_3\,\,\, = \, \,  \, \,
\, (32 \, \theta^3 \, -80 \,  \theta^2 \, +72 \, \theta  \, -27) \, \,  \,  \, 
-\, 4 \cdot \, x \cdot \, (2 \, \theta \, +1) \, (2 \, \theta \, -1)^2.  
\nonumber
\end{eqnarray}
More generally, for $\, \mu \, = \, 2 \, N$ one has
\begin{eqnarray}
\label{simplest}
\hspace{-0.9in}&&\quad \quad \quad \quad \quad \quad \quad 
{\cal C}(0)  \cdot U_3\,\,\, \, = \, \,\,   \,\,\,  V_3 \cdot {\cal C}(2 \, N)
\end{eqnarray}
in the order-three operators $\, U_3$ and $\, V_3$,
the degree in $\, x$ being $\, N$. We have an {\em infinite number 
of equivalent operators} satisfying the (strong) Calabi-Yau condition. This 
is a consequence of the following homomorphism 
between $\, {\cal C}(\mu)$ and $\, {\cal C}(\mu+2)$
valid for {\em any value of} $\, \mu$ ($\mu$ being not even a rational number):
\begin{eqnarray}
\label{recursion}
\hspace{-0.6in}&&\quad \quad \quad \quad \, \,\,
 {\cal C}(\mu) \cdot \, X_3 
\,\, \, \,\,   = \, \,\,   \,\,   \,\,  
Y_3 \cdot \, {\cal C}(\mu+2),
\end{eqnarray}
where $\, X_3$ reads
\begin{eqnarray}
\label{X3}
\hspace{-0.9in}&& \quad 
4\,\,  x \cdot \, 
\left( 2\, \left( 3\,{\mu}^{2}+6\,\mu-1 \right)\, \cdot \, \theta
 \, +4\,{\mu}^{3}+15\,{\mu}^{2}+14\,\mu-1 \right) \, 
 (2\,\theta-\mu-1) \,  (2\,\theta-\mu-3) \, \, 
 \nonumber \\ 
\hspace{-0.9in}&&\quad \quad \quad \quad  
-32\, \left( 3\,{\mu}^{2}+6\,\mu-1 \right)\cdot \,  {\theta}^{3} \, 
+16\, \left( 2\,{\mu}^{3}+15\,{\mu}^{2}+20\,\mu-5 \right)\cdot \,  {\theta}^{2} 
\,  \nonumber \\  
\hspace{-0.9in}&&\quad \quad \quad \quad \quad  \quad  \quad 
-8\, \left( {\mu}^{2}+2\,\mu-1 \right)  \, (\mu+3)^{2} \cdot \, \theta \,  \, 
+  \, (\mu+3)^{3} \, (\mu^2-1),  
\end{eqnarray}
and $\, Y_3$ reads:
\begin{eqnarray}
\label{Y3}
\hspace{-0.9in}&& \quad 
4\, \, x \cdot \, 
\left( 2\, \left( 3\,{\mu}^{2}+6\,\mu-1 \right) \,\cdot \, \theta \, 
+4\,{\mu}^{3}+9\,{\mu}^{2}+2\,\mu+1 \right) \,  \, (2\,\theta-\mu+1) \, 
 (2\,\theta-\mu-1) \, \nonumber \\ 
\hspace{-0.9in}&&\quad \quad \quad \quad  
 -32\, \left( 3\,{\mu}^{2}+6\,\mu-1 \right)\cdot \,  {\theta}^{3} \, 
+16\, 
\left( 2\,{\mu}^{3} +15\,{\mu}^{2}+20\,\mu-5 \right)\cdot \,  {\theta}^{2} \,  
\nonumber \\ 
\hspace{-0.9in}&&\quad \quad \quad \quad \quad  \quad  \quad 
-8\, \left( {\mu}^{2}+2\,\mu-1 \right)
  \, (\mu+3)^{2} \,\cdot \,  \theta \,  \, 
+ \, (\mu+3)^{3}  \, (\mu^2-1), 
\end{eqnarray}
the two order-three operators being themselves 
homomorphic\footnote[1]{We have the following ``tower'' of homomorphisms:
$\, X_3 \cdot \, X_2  \, \, = \, \, \, Y_2 \cdot   \, Y_3,  \,
X_2 \cdot \, X_1  \, \, = \, \, \, Y_1 \cdot   \, Y_2, 
\, \,
X_1 \cdot \, (x-1)  \, \, = \, \, \,  (x-1)  \cdot   \, Y_1$.}.

{\bf To sum-up:} This shows that a solution of the 
reduction (by equivalence of operators) 
of operators satisfying the weak Calabi-Yau condition to operators 
satisfying the (strong) Calabi-Yau condition is {\em not unique} since 
one can find, for some examples, an {\em infinite number} of homomorphic 
irreducible operators satisfying the (strong) Calabi-Yau condition.

\vskip .1cm 

\subsubsection{Decomposition \newline}
\label{decompo}

Let us consider, for instance, the order-four operator obtained 
by the rightdivision by $\, \theta$ 
of the LCLM of $\, {\cal C}(\mu)$ and  $\, \theta$. This operator 
$\,{\cal M}(\mu)$ satisfies the weak Calabi-Yau condition: 
its exterior square
has a rational solution $\, 1/(x-1)$ {\em independent of} $\, \mu$, 
and homomorphic to
the infinite number of  operators 
$\, {\cal C}(\mu \, +2 \, N)$ ($N$ is any integer), satisfying the 
(strong) Calabi-Yau condition that 
their exterior square is of order five.

One has a decomposition similar to (\ref{forthcoming2}), namely:
\begin{eqnarray}
\label{decompMmu}
\hspace{-0.9in}&& \quad {\cal M}(\mu) \, \, = \, \, \, \, 
L_2 \cdot \, a(x) \cdot \, M_2 
 \, \, + \, \, {{ (\mu^2-1)^2} \over {16 \, a(x) }}, 
\qquad \quad \quad 
a(x) \, = \, \, x^3\cdot \, (x-1), 
\nonumber \\
\hspace{-0.9in}&& \quad \quad  \quad  L_2  \, \, = \, \, \, \,
 {\frac {1}{ (x-1)^4 \, \, x^{5} }}
\cdot \, \Bigl(x \cdot \, (\theta \, -4) \, \, -(\theta \, -2)\Bigr) 
\cdot \, \Bigl( x \cdot \, (\theta \, -3) \, \,  -(\theta \, -2)  \Bigr),
\nonumber \\
\hspace{-0.9in}&& \quad \quad \quad  M_2  \, \, = \, \, \, \,
 {\frac {1}{ x^2}} \cdot \, 
\Bigl( x \cdot \, (\theta^2 \, -{{\mu^2 \, +1 } \over {2}} ) 
 \, \, - \theta \cdot  \, (\theta -1) \Bigr),
\end{eqnarray}
where $\, L_2$ and $\, M_2$ are self-adjoint operators.

Let us consider, instead of $\, {\cal C}(\mu)$, the self-adjoint
operator $\, C_s(\mu) \, = \, \, 1/x^2 \cdot \, {\cal C}(\mu)$,
which satisfies the Calabi-Yau condition
and instead of  $\, {\cal M}(\mu)$ the operator 
$\, M(\mu)\, = \, \, a(x) \cdot \, {\cal M}(\mu)$:
\begin{eqnarray}
\label{decompMmu}
\hspace{-0.9in}&& \quad  \quad \quad \quad \quad
M(\mu) \, \, = \, \, \, \, 
 -16 \cdot \, a(x) \cdot \, L_2 \cdot \, a(x) \cdot \, M_2 
 \, \, - \, \, (\mu^2-1)^2.
\end{eqnarray}
One has the following homomorphism between $\, C_s(\mu)$
and the adjoint of  $\, M(\mu)$:
\begin{eqnarray}
\hspace{-0.9in}&& \quad \quad \quad \quad \quad
(\theta\, +1) \cdot \, adjoint(M(\mu)) 
 \, \, \,  = \, \, \, \,  \,  \, \,
C_s(\mu)  \cdot \, x  \cdot \, (\theta\, +1).
\end{eqnarray}
One notes that the exterior square of $\, adjoint(M(\mu))$, 
as well as the exterior square of 
$\, (\theta\, +1) \cdot \, adjoint(M(\mu))$, 
have the rational solution $\, 1/x^3$.
This is in agreement with the fact that
the exterior square of  $\, adjoint({\cal M}(\mu))$
is nothing but the Wronskian of the 
self-adjoint operator $\, L_2$, namely, 
$\, x^3 \cdot \, (x-1)^2 \, = \, \, a(x)^2/x^3$. 
One thus sees that, if the  exterior square of $\, C_s(\mu)$
has no rational solution, 
$\, C_s(\mu)  \cdot \, x  \cdot \, (\theta\, +1)$
has a rational solution, namely $\, 1/x^3$.

\vskip .1cm 

\subsection{Calabi-Yau conditions preserved by the formal adjoint}
\label{weakpreserved}

Let us consider a (monic) order-four operator with a rational 
(resp. $\, N$-th root of rational) Wronskian $W(x)$:
\begin{eqnarray}
\label{ratwronsk}
 \hspace{-0.95in}&& \quad \quad \, \,  
\Omega_4 \,\,\, \, \, = \, \, \,\,\,\,
 D_x^4 \, \, \, \,  
-\, {{d \ln(W(x))} \over {dx}} \cdot \, D_x^3
 \, \, \,  + \, a_2(x) \cdot \, D_x^2 
 \, \, \, + \, a_1(x) \cdot \, D_x\, \, \,  + \, a_0(x). 
\nonumber 
\end{eqnarray}
One has the following {\em conjugation relation between the exterior 
square of this operator and the  exterior square of its adjoint}:
\begin{eqnarray}
\label{conjugExt}
\hspace{-0.95in}&& \quad \quad \qquad 
W(x) \cdot  \, Ext^2(adjoint(\Omega_4))
 \,\, \, \, \, = \, \, \,\,\, \, 
 Ext^2(\Omega_4) \cdot  \, W(x).  
\end{eqnarray}
From this conjugation relation it is straightforward to deduce that
if $\, \Omega_4$ satisfies the {\em Calabi-Yau condition} (\ref{CalabiCond}),
i.e. its exterior square is of order five instead of six, the
exterior square of its adjoint will also be of  order five: the Calabi-Yau 
condition (\ref{CalabiCond}) is thus
{\em preserved by the adjoint transformation}.
From this conjugation relation (\ref{conjugExt}) it is also 
straightforward to deduce that
if the operator satisfies the {\em weak Calabi-Yau condition}, i.e.
its exterior square has a rational (resp. $\, N$-th root of rational)
solution, this will also be the case for its adjoint: 
the exterior square of the adjoint of this operator
will also have a rational solution, which is nothing but the previous 
rational (resp. $\, N$-th root of rational) solution divided by 
$\, W(x)$ the Wronskian of the operator.
In other words, the  {\em weak Calabi-Yau condition is preserved by 
the adjoint}. 

\vskip .1cm 

\subsubsection{Decomposition of order-four operators \newline}
\label{decompo}

Let us consider an operator $\, M_4$ of the form
(\ref{forthcoming2})
where $\, L_2$ and $\, M_2$ are two (general) self-adjoint operators
(\ref{selfop}) and (\ref{selfop2}). Note that
  $\, a(x)  \cdot \, M_2 \cdot \, a(x)$ is also  
a self-adjoint operator, so up to an overall factor $\, \rho(x)$,
one can consider, without any restriction, the form
\begin{eqnarray}
\label{forthcoming2ter}
 \hspace{-0.95in}&& \quad \quad \qquad \qquad 
 M_4 \, \,  \,\,\, \, \, = \, \, \,\,\, \,  \,
 M_2 \cdot \, L_2  \, \, \, \, \,  \,+ \, \, \,\,\,  \,
 \lambda,
\end{eqnarray}
where $\, L_2$ and $\, M_2$ are two (general) 
self-adjoint operators (see (\ref{selfop}) and (\ref{selfop2})).

The exterior square of an order-four operator of the form 
(\ref{forthcoming2ter}) (up to an overall factor $\, \rho(x)$) 
has the rational solution $\, 1/\alpha_2(x)$. Switching to the adjoint
amounts to permuting the two self-adjoint operators $\, L_2$ and 
$\, M_2$. The order-four operator $\, M_4$ is not monic. Denoting 
  $\, M_4^{(u)}$ the order-four operator $\, M_4$ in a monic
form we have $\, M_4 \, = \, \, \rho(x) \cdot \, M_4^{(u)}$
with $\, \rho(x)  \, = \, \,  \alpha_2(x) \cdot \, \beta_2(x)$.
Thus one has the following relations\footnote[1]{Since by definition
the exterior square of an operator is normalized to be a {\em monic}
 operator.}
 between  their adjoints:
\begin{eqnarray}
 \hspace{-0.9in}&& \quad \quad
  adjoint(M_4) \,\,  = \, \, \, \, adjoint(M_4^{(u)}) \cdot \, \rho(x), 
\nonumber \\
  \hspace{-0.9in}&& \quad \quad
  Ext^2(adjoint(M_4)) \, \,  = \, \, \, \, 
 Ext^2(adjoint(M_4^{(u)}) \cdot \, \rho(x))
 \nonumber \\
  \hspace{-0.9in}&& \quad \quad \quad \quad \quad  \qquad \qquad \quad 
 \, = \, \, \,   
 {{1} \over { \rho(x)^2}} \cdot \,  Ext^2(adjoint(M_4^{(u)})) \cdot \, \rho(x)^2.
\end{eqnarray}
One has the previous relation (\ref{conjugExt}) between the exterior square of 
the {\em monic} operator, its Wronskian, and the  adjoint of the 
{\em monic} operator:
\begin{eqnarray}
\label{conjugExtbis}
\hspace{-0.9in}&& \quad \quad 
 W(x)  \cdot  \, Ext^2(adjoint(M_4^{(u)})) \,\, \, \, = \, \, \,\,\,
 Ext^2(M_4^{(u)}) \cdot  \, W(x),  \qquad  \qquad \hbox{where:}
\nonumber \\
\hspace{-0.9in}&& \quad \quad  \qquad  \qquad 
W(x) \,  \, \, \, = \, \, \,\,\, \, 
 {{\beta_2(x) } \over { \rho(x)^2 \cdot \, \alpha_2(x)}}
\, \, \, \, = \, \, \,\,\, \, { {1 } \over { \beta_2(x) \cdot \, \alpha_2(x)^3}}, 
\end{eqnarray}
which can be rewritten on the exterior square of 
$\, M_4$ and its adjoint: 
\begin{eqnarray}
\label{last}
  \hspace{-0.9in}&& \quad \quad \qquad 
 { {\beta_2(x) } \over { \alpha_2(x)}} \cdot  \, Ext^2(adjoint(M_4)) 
\,\, \, \, = \, \, \,\,\,\,\,
 Ext^2(M_4) \cdot  \, { {\beta_2(x) } \over { \alpha_2(x)}}. 
\end{eqnarray}

Since the exterior square of $\, M_4$ of the 
form (\ref{forthcoming2ter}) has the  
rational solution $\, 1/\alpha_2(x)$, 
the last relation (\ref{last}) is compatible 
with the fact that the exterior square of 
the adjoint of $\, M_4$ (of the 
form (\ref{forthcoming2ter})) has the rational 
solution $\, 1/\beta_2(x)$.

\vskip .1cm 
\vskip .1cm 

{\bf Remark:} Let us consider an operator $\, C_4$ 
with head coefficient $\, A_4(x)$ and Wronskian $\, W(x)$, 
satisfying the (strong) 
Calabi-Yau condition that its exterior square
is of order five.
If one has an intertwining relation 
\begin{eqnarray}
\label{SSW}
\hspace{-0.9in}&& \quad  \quad  
M_4 \cdot \, \tilde{R}(x) \cdot \,
\Bigl(D_x \, -{{d \ln(R(x))} \over {dx}}\Bigr)
  \,\,\, \, \, = \, \, \,\,\,
\tilde{S}(x) \cdot \,
\Bigl(D_x \, -{{d \ln(S(x))} \over {dx} } \Bigr) 
\cdot \, C_4, 
\nonumber 
\end{eqnarray}
one has the simple relations:
\begin{eqnarray}
\hspace{-0.9in}&&     \quad \qquad  \qquad  
\tilde{S}(x) \,\,\, \, \, = \, \, \,\,\, \,\,
{{ \beta_2(x)\cdot \, a(x)^2 \cdot \,\alpha_2(x) \cdot \, \tilde{R}(x)
 } \over {A_4(x) }},
\nonumber \\
\hspace{-0.9in}&&   \quad \qquad  \qquad  
W(x) \,\,\, \, \, = \, \, \,\,\,\,\,
 {{A_4(x) \cdot \, R(x) } \over {
 \tilde{R}(x)^4 \cdot \, \beta_2(x)\cdot \,
\alpha_2(x)^3\cdot \,a(x)^2 \cdot \, S(x) }}. 
\end{eqnarray}

\vskip .1cm 

Therefore there are some simple relations between 
$\, \alpha_2(x)$ and $\, \beta_2(x)$ that will correspond to the rational 
solutions of the exterior square of $\, M_4$,
or of the adjoint of $\, M_4$.

\vskip .1cm 

\subsection{Rational solutions for the exterior square of
 operators satisfying the  weak Calabi-Yau condition }
\label{ratsolweakappend}

The fact that an order-four operator satisfying the  weak Calabi-Yau condition,
namely having a decomposition of the form (\ref{forthcoming2}), is such that 
its exterior square has a rational solution which is the inverse of the 
head coefficient of the right most order-two self-adjoint operator is, in fact,  
the consequence of an identity on the difference of two exterior squares. 

A non-trivial identity exists between the difference of the following 
two exterior squares:
\begin{eqnarray}
\label{Extidenity}
\hspace{-0.95in}&&   \, \,  
Ext^2(L_2 \cdot \, M_2 \, + \, \, \lambda) \,\,  - \, Ext^2(L_2 \cdot \, M_2)
 \, \, \, \, =  \\ 
\hspace{-0.95in}&& \quad   
 -4 \cdot \, {{\lambda} \over { \alpha_2(x)\, \beta_2(x)}} \cdot \,
 \Bigl(D_x \, \, 
-\, {{1 } \over {2}} \cdot \, {{ d \ln(\rho_L(x)) } \over {dx}}\Bigr) 
\cdot \,
  \Bigl(D_x \, 
-\, {{1 } \over {2}} \cdot \, {{ d \ln(\rho_R(x)) } \over {dx}}\Bigr),
\quad \quad  \hbox{where:}
\nonumber \\
\hspace{-0.95in}&&   \quad 
 \rho_L(x)
\, \, = \, \, \,
 {{{\cal P}^2 } \over {\alpha_2(x)^2 \cdot \, 
\beta_2(x)^2 \cdot \, w_L(x)^5 \cdot \, w_M(x)^5}}, \quad 
\rho_R(x)\, \, = \, \, \,
 {{ w_L(x)\cdot \, w_M(x) \cdot \, \alpha_2(x) } \over {\beta_2(x)  }},
\nonumber
\end{eqnarray}
where $\, {\cal P}$ is a slightly involved\footnote[2]{Quadratic 
in $\,\alpha_2(x)$ and $\, \beta_2(x)$, 
linear in $\, \alpha_0(x)$ and $\, \beta_0(x)$, and polynomial in
$\,w_L(x)$ and $\,w_M(x)$ and their derivatives up to third derivative.} 
polynomial expression, 
on exterior square of product of the two (not necessarily self-adjoint) 
operators 
\begin{eqnarray}
\hspace{-0.95in}&& \quad \quad  \qquad \quad 
 L_2  \, \, \,  = \, \, \,   \,  \, \, \alpha_2(x) \cdot \, (D_x^2 \,  \,  \,
- \, \, {{d \, \ln(w_L(x)} \over {dx}}  \cdot \, D_x)
\,\,  \, + \, \, \alpha_0(x), \quad \\
 \hspace{-0.95in}&& \quad \quad \qquad  \quad 
  M_2  \, \,  \, = \, \, \, \,   \,  \, \beta_2(x) \cdot \, (D_x^2 \, \,  \,
 - \, \,  {{d\, \, \ln(w_M(x)} \over {dx}} \cdot \, D_x)
\,\, \,  + \, \, \beta_0(x).
\end{eqnarray}
For self-adjoint operators like (\ref{selfop}) and
 the decomposition (\ref{forthcoming2}),
one has $\, \alpha_2(x) \, = \, a(x)/ w_L(x)$ and 
 $\, \beta_2(x) \, = \, a(x)/ w_M(x)$, 
$\, \rho_R \, = \, \, w_M(x)^2$. Furthermore it is simple to see
that the exterior square of the product $\, L_2 \cdot \, M_2$
has the Wronskian of $\, M_2$ as a solution.
Therefore identity (\ref{Extidenity}) means that
the exterior square of 
$\, L_2 \cdot \, M_2 \, + \, \, \lambda$ has $\, w_M(x)$
the Wronskian of $\, M_2$ as a solution.

\vskip .1cm 

\section{Analysis of the order-seven operators associated with the 
exceptional Galois group $\, G_2(C)$}
\label{analysisG2}

\subsection{Solution-series of the order-seven operators }
\label{solser}

The solution-series $\, y_0^{(n)}$, analytic at $\, x\, = \, \, 0$,  
of the order-seven operators $\, E_n$, $\, n\, = \, 1, \, 2 \, \cdots$,
 given in section (\ref{except}),  
are actually series 
with {\em integer coefficients} and read respectively 
\begin{eqnarray}
\hspace{-0.95in}&&  \qquad y_0^{(1)} \, \, = \,\,  \, \, 
 1 \,  \,+2688 \,x  \,+19707264\,{x}^{2}  \,+191647334400\,{x}^{3}
  \,+2133255623587200\,{x}^{4}  
 \nonumber \\
\hspace{-0.95in}&&  \qquad \qquad  \qquad 
\,+25707449648409919488\,{x}^{5}  \, \, + \, \cdots 
 \nonumber 
\end{eqnarray}
\begin{eqnarray}
\hspace{-0.95in}&&  \qquad  y_0^{(2)}  \, \, = \,\,  \, \, 
1 \, \, +384\,x \, +537984\,{x}^{2} \, +1097318400\,{x}^{3} \, 
+2680866518400\,{x}^{4}  \nonumber \\
\hspace{-0.95in}&&  \qquad \qquad  \qquad 
\, +7283382738960384\,{x}^{5} \, \,  + \, \cdots 
 \nonumber 
\end{eqnarray}
\begin{eqnarray}
\hspace{-0.95in}&&  \qquad  y_0^{(3)}  \, \, = \,\,  \,\, 
 1 \, \,  +1512\,x\,+9885240\,{x}^{2}\,
+95782780800\,{x}^{3}\,+1117658718099000\,{x}^{4}
 \nonumber \\
\hspace{-0.95in}&&  \qquad \qquad  \qquad \, 
+14536396497887776512\,{x}^{5}
\, \,  + \,\,  \cdots 
 \nonumber 
\end{eqnarray}
\begin{eqnarray}
\hspace{-0.95in}&&  \qquad  y_0^{(4)}  \, \, = \,\,  \, \, 
 1 \, \, \, +14976\,x \, +1254798720\,{x}^{2} \, 
+159551671910400\,{x}^{3} \nonumber \\
\hspace{-0.95in}&&  \qquad \quad  \qquad 
\,\, +24603126146687088000\,{x}^{4}
 +4241337041632715022974976\,{x}^{5} \, \,   + \,\,  \cdots 
\nonumber 
\end{eqnarray}
\begin{eqnarray}
\label{serieskEn}
\hspace{-0.95in}&&  \qquad  y_0^{(5)}  \, \, = \,\,  \, \, 
 1 \,     \, \, +2678400\,x\, +65172299068800\,{x}^{2}\, 
+2494516941707677286400\,{x}^{3}\, \nonumber \\
\hspace{-0.95in}&&  \qquad \quad  \qquad  \qquad 
 +116986156694543894801624380800\,{x}^{4}\,
 \\
\hspace{-0.95in}&&  \qquad \qquad  \qquad \qquad  \, 
+6160069364202852097613676563979878400\,{x}^{5}
\,  \,   + \,\,  \cdots  \nonumber 
\end{eqnarray}

These order-seven linear differential operators are
 {\em globally nilpotent}~\cite{bo-bo-ha-ma-we-ze-09,Deitweiler}, 
the Jordan reduction of their
 $\, p$-curvature~\cite{Deitweiler,bo-bo-ha-ma-we-ze-09} reading: 
\begin{eqnarray}
\label{pcurvJ7}
\hspace{-0.8in}&&\qquad \qquad \quad \quad 
J_7 \, \, = \, \, \, \,\,
\left[ \begin {array}{ccccccc}
0&1&0&0&0&0&0 \\ \noalign{\medskip}
0&0&1&0&0&0&0 \\ \noalign{\medskip}
0&0&0&1&0&0&0 \\ \noalign{\medskip}
0&0&0&0&1&0&0 \\ \noalign{\medskip}
0&0&0&0&0&1&0 \\ \noalign{\medskip}
0&0&0&0&0&0&1 \\ \noalign{\medskip}
0&0&0&0&0&0&0
\end {array} \right],
\end{eqnarray}
of characteristic polynomial equal to the 
minimal polynomial  $\,P(\lambda)\, = \,  \lambda^7$.

These operators are MUM, so they have the traditional 
``triangular log structure''~\cite{CalabiYauIsing}, 
the formal series solutions with a 
log being of the form  $\, y_1^{(n)} \, \, $
$= \, \, \,  y_0^{(n)}  \cdot \, \ln(x) \, + \, \, \tilde{y}_1^{(n)}$, 
$\, y_2^{(n)} \, \, $
$= \, \, \,  y_0^{(n)}  \cdot \, \ln(x)^2/2 \,
 + \, \, \tilde{y}_1^{(n)} \cdot \, \ln(x) \, 
+ \, \, \tilde{y}_2^{(n)} $, etc. 
The corresponding nomes (called ``special coordinates''
in~\cite{BognerGood}) are defined as
 $\, q^{(n)} \, \, = \, \, \exp(y_1^{(n)}/y_0^{(n)})$
$ \, \, = \, \, x \cdot \, \exp(\tilde{ y}_1^{(n)}/y_0^{(n)})$.
These nomes correspond to series with {\em integer coefficients},
and read respectively:
\begin{eqnarray}
\hspace{-0.95in}&&  \qquad \qquad 
q^{(1)} \, \, = \, \, \,  \,  x \,\,  \,  +7040\,{x}^{2} \, 
+67555904\,{x}^{3} \, +747082784768\,{x}^{4}\,
\nonumber \\
\hspace{-0.95in}&&  \qquad \qquad \qquad  \qquad
  +8968272297124128\,{x}^{5}\,+ \, \, \cdots, 
\nonumber 
\end{eqnarray} 
\begin{eqnarray}
\hspace{-0.95in}&&  \qquad\qquad  
q^{(2)} \, \, = \, \,\,  \,  x\, \,  \, +1152\,{x}^{2}\, 
+2150976\,{x}^{3}\, +4983447552\,{x}^{4}\,
\nonumber \\
\hspace{-0.95in}&&  \qquad\qquad  \qquad  \qquad
  +13054714896672\,{x}^{5}\,  \,+ \, \, \cdots, 
\nonumber 
\end{eqnarray} 
\begin{eqnarray}
\hspace{-0.95in}&&  \qquad\qquad  
q^{(3)} \, \, = \,\,  \,  \, x\,\,  \,  +5562\,{x}^{2}\,+49552317\,{x}^{3}
\,+547802062578\,{x}^{4}\,
\nonumber \\
\hspace{-0.95in}&&  \qquad\qquad  \qquad  \qquad 
 +6855142017357054\,{x}^{5} \,   \,+ \, \, \cdots, 
\nonumber 
\end{eqnarray} 
\begin{eqnarray}
\hspace{-0.95in}&&  \qquad \qquad 
q^{(4)} \, \, = \, \,\,  \,  x \,\,  \,  +72576\,{x}^{2}\, 
+8462979648\,{x}^{3}\, +1230038144557056\,{x}^{4}\,
\nonumber \\
\hspace{-0.95in}&&  \qquad \qquad \qquad  \qquad 
 +203018472128017391904\,{x}^{5}\,   \,+ \, \, \cdots, 
\nonumber
\end{eqnarray} 
\begin{eqnarray}
\hspace{-0.95in}&&  \qquad \qquad 
q^{(5)} \, \, = \, \,\,  \,  x \,\, \,   +20200320\,{x}^{2}\, 
+689499895026240\,{x}^{3}\,
\nonumber \\
\hspace{-0.95in}&&  \qquad \qquad \qquad  \qquad 
 +29916247864887732510720\,{x}^{4} \,
 \\
\hspace{-0.95in}&&  \qquad \qquad \qquad  \qquad  \qquad  
+1488739080271271648779215102240\,{x}^{5} 
\,  \,+ \, \, \cdots \nonumber 
\end{eqnarray}

\subsubsection{Yukawa couplings \newline}
\label{Yuk}

The Yukawa couplings
\begin{eqnarray}
\label{Yukawa}
K(q) \,\,\, = \, \, \,\,\, 
 \Bigl( q \cdot {{d} \over {dq }} \Bigr)^2
 \Bigl(  {{y_2} \over {y_0}}\Bigr),
 \end{eqnarray}
of the five order-seven linear differential operators of section (\ref{except}) 
are series with integer coefficients. Their expansion 
 read respectively: 
\begin{eqnarray}
\hspace{-0.95in}&&  \qquad \qquad  K^{(1)} \, \, = \, \, \,
1  \, \,+768\,q \,-2188032\,{q}^{2} \,+2883403776\,{q}^{3} 
\,-1360234636032\,{q}^{4}
\nonumber \\
\hspace{-0.95in}&&  \qquad \qquad \qquad \qquad
 \,-3787008084959232\,{q}^{5} \,  \, + \, \cdots 
\nonumber \\
\hspace{-0.95in}&&  \qquad \qquad 
K^{(2)} \, \, = \, \, \, 1 \, \, 
+256\,q+728320\,{q}^{2}\,+1640611840\,{q}^{3}\,+3618799525120\,{q}^{4}\,
\nonumber \\
\hspace{-0.95in}&&  \qquad \qquad \qquad  \qquad
+8043817914720256\,{q}^{5} \,  \, + \, \cdots 
\nonumber 
\end{eqnarray}
\begin{eqnarray}
\hspace{-0.95in}&&  \qquad \qquad 
K^{(3)} \, \, = \, \,  \,1 \, \, +1485\,q
\, +19708515\,{q}^{2}\,+206970715890\,{q}^{3}\,
\nonumber \\
\hspace{-0.95in}&&  \qquad \qquad \qquad \qquad 
+2188620549305955\,{q}^{4}\,\,
+23409935555891063985\,{q}^{5} \, \, + \, \cdots 
\nonumber
\nonumber 
\end{eqnarray}
\begin{eqnarray}
\hspace{-0.95in}&&  \qquad \qquad 
K^{(4)} \, \, = \,\, \, 1 \,\, +29440\,q
\, +4438662400\,{q}^{2} \, +621410936504320\,{q}^{3} \,
\nonumber \\
\hspace{-0.95in}&&  \qquad \qquad \qquad 
+88605364227964837120\,{q}^{4} \,
 +12835248124604913684029440\,{q}^{5} \, \, + \, \cdots 
\nonumber
\end{eqnarray}
\begin{eqnarray}
\hspace{-0.95in}&&  \qquad \qquad 
K^{(5)} \, \, = \,\, \, 1 \, \, +17342208\,q\, 
+687629971954944\,{q}^{2}
\nonumber \\
\hspace{-0.95in}&&  \qquad \qquad \qquad \qquad 
 \, +30848876097264182771712\,{q}^{3} \, 
\nonumber \\
\hspace{-0.95in}&&  \qquad \qquad \qquad \qquad \qquad 
+1428770297588004620323742981376\,{q}^{4} \, 
\nonumber \\
\hspace{-0.95in}&&  \qquad \qquad\, \qquad \qquad \qquad \qquad 
+67528440221394152640448454407310942208\,{q}^{5} \,  \, + \, \cdots 
\nonumber
\end{eqnarray}
These Yukawa couplings,  in the $\, x$ variable, read respectively:
\begin{eqnarray}
\hspace{-0.95in}&&  \qquad \qquad  K^{(1)} \, \, = \, \,\,
1 \,\, +768\,x \, +3218688\,{x}^{2}
\, +23958847488\,{x}^{3} 
\nonumber \\
\hspace{-0.95in}&&  \qquad \qquad \qquad \qquad \, 
+229225505561856\,{x}^{4} \,
 +2508123114368335872\,{x}^{5} \,  \,  + \,\cdots 
\nonumber 
\end{eqnarray}
\begin{eqnarray}
\hspace{-0.95in}&&  \qquad \qquad K^{(2)} \, \, = \, \,\,
1 \,\, +256\,x \, 
+1023232\,{x}^{2}+3869310976\,{x}^{3} \,
\nonumber \\
\hspace{-0.95in}&&  \qquad \qquad \, \qquad \qquad 
+14664270683392\,{x}^{4} \, 
+56048323595665408\,{x}^{5} \,\,  + \,\cdots 
\nonumber 
\end{eqnarray}
\begin{eqnarray}
\hspace{-0.95in}&&  \qquad \qquad 
K^{(3)} \, \, = \, \,\,
1 \,\, +1485\,x\, 
+27968085\,{x}^{2}+499793427495\,{x}^{3}
 \nonumber \\
\hspace{-0.95in}&&  \qquad \qquad \, \qquad \qquad 
\,+9018524688844995\,{x}^{4} 
\,+164714785807791646845\,{x}^{5}\,\,  + \,\cdots 
\nonumber \\
\hspace{-0.95in}&&  \qquad \qquad K^{(4)} \, \, = \, \,\,
1 \,\, +29440\,x \, 
+6575299840\,{x}^{2}+1514841782026240\,{x}^{3} \,
\nonumber \\
\hspace{-0.95in}&&  \qquad \qquad \qquad \,\qquad  
 +358624525635384843520\,{x}^{4}  
\nonumber \\
\hspace{-0.95in}&&  \qquad \qquad \qquad \, \qquad \qquad 
\, +86502979031531419474001920\,{x}^{5} \, \,  + \,\cdots 
\nonumber 
\end{eqnarray}
\begin{eqnarray}
\hspace{-0.95in}&&  \qquad \qquad
 K^{(5)} \, \, = \, \,\,
1 \,\, +17342208\,x \, 
+1037948123061504\,{x}^{2}
\nonumber \\
\hspace{-0.95in}&&  \qquad \qquad \,  \qquad 
\, +70587017642949191073792\,{x}^{3}\,
\nonumber \\
\hspace{-0.95in}&&  \qquad \qquad \,  \qquad \qquad 
+5045886607522553002548393221376\,{x}^{4}\,
\nonumber \\
\hspace{-0.95in}&&  \qquad \qquad \qquad \qquad \qquad 
\,+370665145887525483931062348265527902208\,{x}^{5} 
\, \,  + \,\cdots 
\nonumber
\end{eqnarray}

Recalling our results in~\cite{Short,Big},
 one can, for {\em higher order}-operators,
define several Yukawa couplings
from Wronskian-like determinants of the solutions, instead 
of a only one (\ref{Yukawa}) for order-four operators:
\begin{eqnarray}
\label{theKn}
\hspace{-0.95in}&&  \qquad  \qquad 
K_3 \, = \, \, {{W_1^3 \cdot \, W_3 }  \over {W_2^3}}, \quad \qquad 
K_4 \, = \, \, {{W_1^8 \cdot \, W_4 }  \over {W_2^6}}, \quad \qquad 
K_5 \, = \, \, {{W_1^{15} \cdot \, W_5 }  \over {W_2^{10}}}, 
\nonumber \\
\hspace{-0.95in}&&  \qquad \qquad  \qquad \qquad 
K_6 \, = \, \, {{W_1^{24} \cdot \, W_6 }  \over {W_2^{15}}}, \quad \qquad 
K_7 \, = \, \, {{W_1^{35} \cdot \, W_7 }  \over {W_2^{21}}}. 
\end{eqnarray}
The well-known Yukawa coupling (\ref{Yukawa}) is denoted $\, K_3$ 
in the previous set of ``higher orders'' Yukawa couplings
 (\ref{theKn})  (see Appendix C.1 in~\cite{Short}).

One remarks the following non-trivial identities for the five order-seven 
operators $\, E_i$:
\begin{eqnarray}
\label{previousrelKn}
\hspace{-0.95in}&&\qquad \qquad 
K_4^{(i)} \, = \, \, (K_3^{(i)})^2, \qquad 
K_5^{(i)} \, = \, \, (K_3^{(i)})^3, \qquad 
K_6^{(i)} \, = \, \, (K_3^{(i)})^5,\qquad 
\nonumber \\
\hspace{-0.95in}&&  \qquad \qquad \, \qquad \qquad \quad \, 
K_7^{(i)} \, = \, \, (K_3^{(i)})^7, \qquad \qquad i \, = \, 1, \,\cdots, \, 5.
\end{eqnarray}

Therefore, for the five $\, E_i$'s, these various invariants 
{\em just reduce to the unique 
Yukawa coupling} $\, K_3$. These relations correspond 
respectively to the identities on the Wronskian-like determinants $\, W_n$:
\begin{eqnarray}
\hspace{-0.95in}&&\qquad W_1^2 \cdot \,  W_4 \, = \, \, W_3^2, 
\qquad
W_1^6 \cdot \,  W_5 \, = \, \,  W_2 \cdot \, W_3^3, \qquad 
W_1^9 \cdot \,  W_6 \, = \, \, W_3^5, \qquad
\nonumber \\
\hspace{-0.95in}&&  \qquad \qquad \, \qquad \, 
W_1^{14}  \cdot \,  W_7 \, = \, \, W_3^7. 
\nonumber
\end{eqnarray}

It had been seen (see (C.17) in~\cite{Short,Big}), for order-four operators, 
that being self-adjoint up to a conjugation which yields that the Yukawa coupling
$\,  K_3$ is equal to the Yukawa coupling for the adjoint $\,  K_3^{*}$, is 
nothing but relation $\, K_4 \, = \, \, K_3^2$, namely 
$\, W_1^2 \cdot \,  W_4 \, = \, \, W_3^2$.

These relations are no longer valid for non self-adjoint (up to conjugation)
order-seven operators. For order-seven operators taking the adjoint operator 
amounts to performing the following involutive transformation 
on the $\, W_n$'s ($W_0 \, = \, 1$):
\begin{eqnarray}
\hspace{-0.95in}&&\quad  \quad \, \,\,   
 W_n \, \quad  \longleftrightarrow \,\quad  
 W_n^{*} \, = \, \, {{ W_{7-n}} \over {W_{7}}},
 \quad  \quad  \quad  \, 
W_n \, \quad  \longleftrightarrow \,\quad 
  W_7^{*} \, = \, \,  {{1} \over {W_{7}}},
\end{eqnarray}
and, consequently, the Yukawa couplings (\ref{theKn}) 
for the adjoint operator read:
\begin{eqnarray}
\label{theKnstar}
\hspace{-0.95in}&&  \quad  \quad \quad \, 
K_3^{*} \, = \, \, {{W_6^3 \cdot \, W_4 }  \over {W_7 \cdot \, W_5^3}} 
\, = \, \,
 {{K_6^{3} \cdot K_4} \over {K_5^{3} \cdot \,K_7 }}, \quad  \quad  \quad 
K_4^{*} \, = \, \, {{W_6^8 \cdot \, W_3 }  \over {W_7^3 \cdot \, W_5^6}}
 \, = \, \,
 {{K_6^{8} \cdot K_3} \over {K_5^{6} \cdot \,K_7^{3} }}, \quad 
\nonumber \\
\hspace{-0.95in}&&  \quad  \quad \quad  \, 
K_5^{*} \, = \, \, {{W_6^{15} \cdot \, W_2 }  \over { W_7^6 \cdot \, W_5^{10}}}
 \, = \, \,
 {{K_6^{15}} \over {K_5^{10} \cdot \,K_7^{6} }}, \quad  \quad 
K_6^{*} \, = \, \, {{W_6^{24} \cdot \, W_1 }  \over {W_7^{10} \cdot \, W_5^{15}}}
 \, = \, \,
 {{K_6^{24}} \over {K_5^{15} \cdot \,K_7^{10} }},
\nonumber \\
\hspace{-0.95in}&&  \quad \quad  \quad  \, 
K_7^{*} \, = \, \, {{W_6^{35}  }  \over {W_7^{15} \cdot \, W_5^{21}}}
\, = \, \, {{1} \over {K_7}}. 
\end{eqnarray}

Note, for the order-seven operator $\, {\hat E}_1^{(1)}$  
(obtained from $\, {\hat E}_1$ by taking the LCLM with $\, D_x$ 
and rightdividing by $\, D_x$), the Yukawa couplings (\ref{theKn}), 
as well as these adjoint Yukawa couplings (\ref{theKnstar}),
as well as the $\, W_n$'s, as well as the $\, W_n$'s of the adjoint operators,
are still globally bounded (see~\cite{Short,Big}). 

\vskip .1cm 

{\em Do note that the relations} (\ref{previousrelKn}) {\em are such that} 
$\, K_3^{*}\, = \, \, K_3$, $\, K_4^{*}\, = \, \, K_4$,  $\, K_5^{*}\, = \, \, K_5$, 
$\, K_6^{*}\, = \, \, K_6$. 

Taking into account the transformations of the $\, W_n$'s by a pullback $p(x)$
one finds (with $v \, = \, p'$):  
\begin{eqnarray}
\hspace{-0.95in}&&  \qquad 
(W_1, \, W_2, \, W_3, \, W_4, \,W_5, \, W_6, \,W_7) 
\, \quad \quad  \longrightarrow
\\
\hspace{-0.95in}&& \quad  \quad  \qquad     \quad  \quad 
\quad   (W_1, \, v \cdot \, W_2, \, v^3 \cdot \,W_3, \, v^6 \cdot \,W_4, 
\, v^{10}  \cdot \,W_5, \, v^{15} \cdot \,W_6, \,v^{21} \cdot \,W_7), 
\nonumber 
\end{eqnarray}
together with the obvious homogeneity
\begin{eqnarray}
\hspace{-0.95in}&& \qquad 
(W_1, \, W_2, \, W_3, \, W_4, \,W_5, \, W_6, \,W_7)
 \, \quad \quad  \longrightarrow
\\
\hspace{-0.95in}&& \quad  \quad \qquad  \quad \quad \quad 
 (u \cdot W_1, \, u^2 \cdot \, W_2, \, u^3 \cdot \,W_3, \, u^4 \cdot \,W_4, 
\, u^{5}  \cdot \,W_5, \, u^{6} \cdot \,W_6, \, u^{7} \cdot \,W_7), 
\nonumber 
\end{eqnarray}
one finds, if one seeks for invariants of the form $\, W_1^a \cdot W_n /W_2^b/W_3^c$ 
that the only invariants compatible with these symmetries are the (\ref{theKn})
together with\footnote[1]{These ratio being equal to $\, 1$ in our case.} 
\begin{eqnarray}
\hspace{-0.95in}&& \qquad \quad \quad \quad \quad \, \, \quad \quad
{{ K_4} \over {K_3^2}}, \quad \, \quad {{ K_5} \over {K_3^3}}, \quad \,  \quad 
{{ K_6} \over {K_3^5}}, \quad \, \quad 
{{ K_7} \over {K_3^7}}.
\end{eqnarray}

\section{Exceptional Galois groups: two and three parameter operators}
\label{deform2}

\subsection{Exceptional Galois groups: two-parameter deformation of $E_1$}
\label{defor}

Let us introduce the following two-parameter deformation of $\, E_1$ 
(it amounts to changing
 $F(x) \, \rightarrow \, F(x)\cdot \, (x/(1-x))^{1/2}$ 
in the $\, P_2$ of~\cite{DettReit} then changing 
$\, x  \, \rightarrow \,  (x-1)/x$): 
\begin{eqnarray}
\label{twoparam}
 \hspace{-0.95in}&& \quad
\Omega(p,\,q) \,\,  \,  = \, \, \,  \, 
\theta  \cdot \,(\theta^2 \, -p^2) \cdot \,(\theta^2 \, -q^2)
 \cdot \, (\theta^2 \, -(p+q)^2) 
\nonumber \\
\hspace{-0.95in}&& \quad \quad     
 \, -128 \cdot \, x \cdot \, 
\Bigl( (48 \, \theta^4\, +96 \,\theta^3 \, 
+ 124 \, \theta^2 \, +76 \, \theta \, +21) 
\, (2 \, \theta \, +1)^3 \, 
+16 \, \Sigma^2 \cdot \, (2  \, \theta \, +1)^2 
\nonumber \\
\hspace{-0.95in}&& \quad  \quad  \quad  \quad  \quad  
\, -8 \,  \Sigma \cdot \,(8\, \theta^2 \, +8 \, \theta \, +5)
 \cdot \, (2  \, \theta \, +1)^2 
\, -64 \, p^3 \, q^3
\Bigr)  
\nonumber \\ 
 \hspace{-0.95in}&& \quad  \quad  \quad
+4194304 \cdot \, x^2 \cdot \, (\theta \, +1) \cdot \,
 \Bigl(12 \, (\theta \, +1)^2  \,+11 \, -8 \,  \Sigma  \Bigr)
 \cdot \, (2  \, \theta \, +1)^2 \cdot \, (2  \, \theta \, +3)^2
\nonumber \\ 
 \hspace{-0.95in}&& \quad \quad   \quad
-34359738368 \cdot \, x^3 \cdot \, (2  \, \theta \, +5)^2 \cdot \,
 (2  \, \theta \, +1)^2 \cdot \, (2  \, \theta \, +3)^3, 
\end{eqnarray}
where $\, \, \Sigma \, = \, \, p^2 \, +p \, q \, + q^2$.
In the $\, p \, = \, q \, = \, 0$ limit the previous two-parameter 
order-seven operator reduces to $\, {\hat E}_1$. 

For {\em arbitrary values of} $\, p$ and $\, q$ 
the operator $\, x^{-1} \cdot \, \Omega(p,\,q)$
is {\em actually self-adjoint}. Note that $\, \Omega(p,\,0)$ 
and $\, \Omega(p+1,\,0)$
are homomorphic with an order-six intertwiner
for {\em any value of $\, p$}. Also note that
 $\, \Omega(p,\,1/2)$ and $\, \Omega(p+r,\,1/2)$
are homomorphic with an order-six intertwiner
for {\em any value of $\, p$}, for 
$\, r \, = \, 1/2, \, 1/3, \, 1/6, \, 1/11$.  

\vskip .1cm

\subsection{Exceptional Galois groups: families with three parameters}
\label{familly}

Let us consider the following order-seven 
operator\footnote[1]{See operator $\, P_1$ in~\cite{DettReit}.}
depending on {\em three} parameters $\, a$, $\, c$, $\, d$,
  (here $\sigma$ denotes $\,b^2 +b\,c +c^2$):
\begin{eqnarray} 
\label{P1}
\hspace{-0.95in}&& \quad  \quad  \quad  
\Omega_{a,b,c} \,\, \,  = \, \, \,\, \, 
 \theta \cdot \, (\theta^2 \, -b^2) \cdot \,
 (\theta^2 \, -c^2) \cdot \, (\theta^2 \, -(b+c)^2) 
\nonumber \\
\hspace{-0.95in}&& \quad  \quad    \quad    \quad   \quad \, \, 
-x \cdot \,(2\, \theta +1) \cdot \,  (\theta +a) \,\cdot \,  (\theta +1-a) \,  \cdot \,  \Bigl( 
\theta \cdot \, (\theta +1) \, \cdot \, (\theta^2+\theta +1 -\sigma )  \, 
\nonumber \\
\hspace{-0.95in}&& \quad    \quad \quad   \quad   \quad  \quad    \quad  \, 
+2 \, a \cdot \, (1\, -a) \cdot \, 
(\theta^2 +\theta +1 -\sigma \, -a \cdot \, (1\, -a)) 
\Bigr)
\nonumber \\
\hspace{-0.95in}&& \quad      \quad   \quad  \quad  \quad \, 
\,  +\, x^2 \cdot \, (\theta +1)   \cdot \, 
 (\theta \, +a) \cdot \,  (\theta \,+1\, -a ) \cdot \,   (\theta  \,+a \,+1) \cdot \,  
   (\theta  \,+\,  (1 -a) \,+1) \, 
 \nonumber \\
\hspace{-0.95in}&& \quad  \quad 
  \qquad   \quad     \qquad  \quad    \quad   \quad \,\,   \times 
 (\theta  \, +2 \,a) \,  (\theta  \,+2 \cdot \, (1\, -a)).  
\end{eqnarray}
On this explicit expression one sees obviously  that (\ref{P1}) 
is $(b,c)$-symmetric, $\, \, \Omega_{a,b,c} \, = \, \, \Omega_{a,c,b}$
and that it is invariant by the $\, \, a \, \leftrightarrow \, 1\, -a\, $ 
involution, 
$\, \Omega_{a,b,c} \, = \, \, \Omega_{1-a,b,c}$. Less obviously
one notes that $\, \, \Omega_{a,b,c}$  and    $\, \Omega_{a+N,b+M,c+P}$
are homomorphic {\em for any value of the three integers}
 $\, N$, $\, M$, $\, P$. 
This operator can easily be turned into a {\em self-adjoint} operator 
$\,\,   \Omega_{a,b,c}^{s} \, = \, \, 
x^{-1/2} \cdot \,  \Omega_{a,b,c} \cdot \,x^{1/2}\, $ 
(or the self-adjoint operator $\, x^{-1} \cdot \,  \Omega_{a,b,c}$).

The previous order-seven rescaled operators (\ref{rescal}), namely
${\hat  E}_i$ for $\, i \, = \, \, \, 2 \, \cdots 5$ 
can actually be seen as special 
cases of the rescaled (\ref{P1}). For instance
$ \, {\hat  E}_2 \,  = \, \,  \Omega_{1/2,0,0}, \, \, 
{\hat  E}_3 \, = \, \,  \Omega_{1/3,0,0}, $
$\, \, {\hat  E}_4 
\,  = \, \, \Omega_{1/4,0,0}, \, \, {\hat  E}_5 
\,  = \,  \,\Omega_{1/6,0,0}$. Note that $\, \Omega_{0,0,0}$ factors 
into seven products of order-one operators, namely  
$\, \Omega_{0,0,0} \, = \, x^7 \cdot \, (x-1)^2 \cdot \,\omega_{0,0,0}$, 
where $\, \omega_{0,0,0}$ reads:
\begin{eqnarray} 
\hspace{-0.95in}&&   \quad 
\omega_{0,0,0} \,\,  \,  \, = \, \,\, \,\,  \, 
\Bigl(D_x + {{8 \cdot \, x \, -6} \over {(x-1) \cdot \, x}}\Bigr) \, 
\Bigl(D_x + {{7 \cdot \, x \, -5} \over {(x-1) \cdot \, x}}\Bigr) \,
 \Bigl(D_x + {{5 \cdot \, x \, -4} \over {(x-1) \cdot \, x}}\Bigr) 
 \\
\hspace{-0.95in}&&  \quad   \quad   \quad  \qquad  
\times \Bigl(D_x + {{4 \cdot \, x \, -3} \over {(x-1) \cdot \, x}}\Bigr) \,  
\Bigl(D_x + {{3 \cdot \, x \, -2} \over {(x-1) \cdot \, x}}\Bigr) \, 
\Bigl(D_x + {{x \, -1} \over {(x-1) \cdot \, x}}\Bigr) \cdot \, D_x. 
\nonumber 
\end{eqnarray} 

\vskip .1cm 

\vskip .2cm 

\section{Yukawa couplings of the operators (\ref{LNm1})}
\label{Yukawaof}

Let us recall the operators 
\begin{eqnarray}
\label{LNm1bis}
\hspace{-0.95in}&&  \quad \quad  
x \cdot \, L_{N-1} \, \, = \, \,  \,  \,  \,  \,   
N \,x \cdot  \,(N\,\theta +1)\cdot \, (N \,\theta +2)  \, \,  \cdots \, 
  \, (N\,\theta + N-1)\,  \,  \,\,  -\theta^{N-1},  
\end{eqnarray}
annihilating the diagonal of rational functions 
\begin{eqnarray}
\label{simplestbis}
\hspace{-0.95in}&&  \, \quad  \quad  \quad  
S_N \,\, = \, \, \, \,
Diag\Bigl({{1} \over { 1 \,  \, -x_1 \, -x_2 \,  \,  \cdots \, -x_N }}\Bigr) 
\, \,\,\, = \, \, \, \, \, \, 
 \sum_{k=0}^{\infty} \, {{(k\, N)! } \over {(k!)^N }} \cdot x^k. 
\end{eqnarray}
and let us also recall, for these higher orders operators, the 
``higher order'' Yukawa couplings (\ref{theKn}), 
(see Appendix C.1 in~\cite{Short}),
one gets the following series expansions.

For $\, L_6$  one has two independent Yukawa couplings: 
\begin{eqnarray}
\hspace{-0.95in}&&  \,  \quad  
K_3(L_6) \, \, = \, \, \,   1\,+10097920\,x\,
 +381994497763200\,{x}^{2}\,
+16633254043776570088000\,{x}^{3}\,
\nonumber \\
\hspace{-0.95in}&&  \, \qquad \,
 +775506882960998615640344320000\,{x}^{4}
\nonumber \\
\hspace{-0.95in}&&  \, \qquad \, 
+37663047736228445917647206103076800000\,{x}^{5}
\, + \,\,\,\cdots,
\nonumber \\
\hspace{-0.95in}&&  \,  \quad 
 K_4(L_6) \,  = \, \,   1 \, +37273810\,x \,
 +1993144925004100\,{x}^{2} \, 
+110716785445910533561000\,{x}^{3} \,
\nonumber \\
\hspace{-0.95in}&&  \, \qquad \, 
 +6240527867851744863088075810000\,{x}^{4} 
\nonumber \\
\hspace{-0.95in}&&  \, \qquad \, \qquad 
+354307497308094243698303790171562900000\,{x}^{5} \, + \,\,\,\cdots
\nonumber 
\end{eqnarray}
For $\, L_7$  one also has two independent Yukawa couplings: 
\begin{eqnarray}
\hspace{-0.95in}&&  \,  \quad 
 K_3(L_7) \, \, = \, \, \, 1 \, +1998080\,x \, +17805741956352\,{x}^{2} \,
+194576429723517255680\,{x}^{3} \, 
\nonumber \\
\hspace{-0.95in}&&  \, \qquad  \,\qquad \, 
  +2352770839522203863766605056\,{x}^{4}
\nonumber \\
\hspace{-0.95in}&&  \, \qquad \, \qquad \, \qquad \, 
+30251355556001122775209879097376768\,{x}^{5} \, + \,\,\,\cdots,
\nonumber \\
\hspace{-0.95in}&&  \,  \quad 
 K_4(L_7) \, \, = \, \, \,1 \, +8684032\,x \, +117020081027584\,{x}^{2} \, 
+1699286765410547138560\,{x}^{3} \,
\nonumber \\
\hspace{-0.95in}&&  \, \qquad  \,\qquad \,
 +25562078087040978837930064384\,{x}^{4}
\nonumber \\
\hspace{-0.95in}&&  \, \qquad  \,\qquad \,\qquad \,
+392649379685173887316823248478666752\,{x}^{5} \, + \,\,\,\cdots,
\nonumber 
\end{eqnarray}
For $\, L_8$  one  has three independent Yukawa couplings: 
\begin{eqnarray}
\hspace{-0.95in}&&  \,  \quad 
 K_3(L_8) \, \, = \, \, \,1 \, +28165644\,x\, +4505049006911820\,{x}^{2}\, 
+956135990658824836437024\,{x}^{3}\,
\nonumber \\
\hspace{-0.95in}&&  \, \qquad \, \qquad \,
 +233949266493282926229755622721356\,{x}^{4}
\nonumber \\
\hspace{-0.95in}&&  \, \qquad  \, \qquad \,\qquad \,
+62439728262268133355948266259742771574160\,{x}^{5}\,  + \,\,\,\cdots,
\nonumber \\
\hspace{-0.95in}&&  \,  \quad
 K_4(L_8) \, \, = \, \, \, 1 \, +141215076\,x \, +35931446901528372\,{x}^{2}\,
+10427983259188646965239144\,{x}^{3}\,
\nonumber \\
\hspace{-0.95in}&&  \, \qquad \,\qquad \,
 +3234974169704568107122039167181620\,{x}^{4}
\nonumber \\
\hspace{-0.95in}&&  \, \qquad  \, \qquad \,\qquad \,
+1045613138106144593968102802858498597502432\,{x}^{5} 
\,  + \,\,\,\cdots,
\nonumber \\
\hspace{-0.95in}&&  \,  \quad
 K_5(L_8) \, \, = \, \, \, 1 \, +373047525\,x\, +145724830276964841\,{x}^{2}\,
+56876157804695752594934289\,{x}^{3}\, 
\nonumber \\
\hspace{-0.95in}&&  \, \qquad \,\qquad
 \,+22158779200978185414267897869823861\,{x}^{4}
\nonumber \\
\hspace{-0.95in}&&  \, \qquad  \, \qquad \,\qquad \,
+8621411505524839637858169288895426578119277\,{x}^{5}\,  + \,\,\,\cdots,
\nonumber 
\end{eqnarray}
For $\, L_9$  one also has three independent Yukawa couplings: 
\begin{eqnarray}
\hspace{-0.95in}&&  \,  \quad
 K_3(L_9) \, \, = \, \, \,
1 \, +412077600\,x \, +1289641316659740000\,{x}^{2} \, 
+5866947627331695510672000000\,{x}^{3} \,
\nonumber \\
\hspace{-0.95in}&&  \, \qquad \qquad \,
+32243888417489985271109666517337500000\,{x}^{4}
\nonumber \\
\hspace{-0.95in}&&  \, \qquad  \, \qquad \,\qquad \,
\, +198873128042921363581235281228819150221047577600\,{x}^{5} \,  + \,\,\,\cdots,
\nonumber  \\
\hspace{-0.95in}&&  \,  \quad
 K_4(L_9) \, \, = \, \, \,1 \, +2351650400\,x \, +12183365592555300000\,{x}^{2} \, 
+77390125534275385218992000000\,{x}^{3} \,
\nonumber \\
\hspace{-0.95in}&&  \, \qquad  \, \qquad \,
+546304879451395256841297089282462500000\,{x}^{4}
\nonumber \\
\hspace{-0.95in}&&  \, \qquad  \, \qquad \,\qquad \,
+4116981156738304057315870224531007355517746150400\,{x}^{5}
 \,  + \,\,\,\cdots,
\nonumber
\end{eqnarray} 
\begin{eqnarray}
\hspace{-0.95in}&&  \,  \quad
K_5(L_9) \, \, = \, \, \,1 \, +7068028000\,x \,
 +59007489670532260000\,{x}^{2} \, 
\nonumber \\
\hspace{-0.95in}&&  \, \qquad \,
+517997397361455559649200000000\,{x}^{3} \, 
\nonumber \\
\hspace{-0.95in}&&  \, \qquad  \, \qquad \,
+4673650695052899977682502065243662500000\,{x}^{4} \, 
\nonumber \\
\hspace{-0.95in}&&  \, \qquad  \, \qquad \,\qquad \,
+42920054481728604570276319721869711743549574528000\,{x}^{5}
\,  + \,\,\,\cdots,
\nonumber
\end{eqnarray}
For $\, L_{10}$  one  has four independent Yukawa couplings: 
\begin{eqnarray}
\hspace{-0.95in}&&  \,  \quad
K_3(L_{10}) \, \, = \, \, \,
1 \, +6309779256\,x \, +420625737971786884680\,{x}^{2}
\nonumber \\
\hspace{-0.95in}&&  \, \qquad \,
\, +45079261240559432713629269385600\,{x}^{3}\, 
\nonumber \\
\hspace{-0.95in}&&  \, \qquad \, \qquad \,
+6138317582117485921071955303191615926205000\,{x}^{4}
\nonumber \\
\hspace{-0.95in}&&  \, \qquad  \, \qquad \,\qquad \,
+966585253200336527311871686585471862322931324462664256\,{x}^{5}
\,  + \,\,\,\cdots,
\nonumber 
\end{eqnarray}
\begin{eqnarray}
\hspace{-0.95in}&&  \,  \quad
K_4(L_{10}) \, \, = \, \, \,1 \, +40564413164\,x \, 
+4622591813564354886036\,{x}^{2} \, 
\nonumber \\
\hspace{-0.95in}&&  \, \qquad \,
+702661675261046914875321464472560\,{x}^{3} \, 
\nonumber \\
\hspace{-0.95in}&&  \, \qquad  \, \qquad \,
+124030582023288017842696295577576233194974100\,{x}^{4}
\nonumber \\
\hspace{-0.95in}&&  \, \qquad  \, \qquad \,\qquad \,
+24007059648071495889025807725100100128489899260922131664\,{x}^{5}
\,  + \,\,\,\cdots,
\nonumber \\
\hspace{-0.95in}&&  \,  \quad
K_5(L_{10}) \, \, = \, \, \,1 \, +138021017970\,x \, 
+26412005921656543623886\,{x}^{2} \, 
\nonumber \\
\hspace{-0.95in}&&  \, \qquad \,
+5668431522859800858070100130586968\,{x}^{3} \, 
\nonumber \\
\hspace{-0.95in}&&  \, \qquad  \, \qquad \,
+1295519605238134595212195097688057025652309166\,{x}^{4}
\nonumber \\
\hspace{-0.95in}&&  \, \qquad  \, \qquad \,\qquad \,
+308199874468915161097074199781704462799392654807438814820\,{x}^{5}
\,  + \,\,\,\cdots,
\nonumber \\
\hspace{-0.95in}&&  \,  \quad
K_6(L_{10}) \, \, = \, \, \,1 \, +317814173215\,x \, 
+94769164483661457482561\,{x}^{2} \,
\nonumber \\
\hspace{-0.95in}&&  \, \qquad \,
 +27681891546591722088579495934023171\,{x}^{3} \, 
\nonumber \\
\hspace{-0.95in}&&  \, \qquad  \, \qquad \,
+8013406679969563547817150713512983722786791981\,{x}^{4}
\nonumber \\
\hspace{-0.95in}&&  \, \qquad  \, \qquad \,\qquad \,
+2308882670142336707807831384186329396677206570700741275495\,{x}^{5}
\,  + \,\,\,\cdots
\nonumber
\end{eqnarray}

\vskip .2cm

{\bf Remark:} Note that, we have relations between these Yukawa couplings
and the Yukawa couplings of exterior or symmetric powers 
of the same operators. For instance the exterior square 
of $\, L_4$ is an 
irreducible order five operator of nome  
\begin{eqnarray}
\hspace{-0.95in}&&  \,  \quad
q \, \, = \, \, \, x\, +1345\,{x}^{2}\, +2552775\,{x}^{3}
\, +5602757375\,{x}^{4}\, 
+13320846541250\,{x}^{5}
\nonumber \\
\hspace{-0.95in}&&  \, \qquad  \, \qquad
\, +33314508430778394\,{x}^{6}\, 
+86273174430421418330\,{x}^{7}\, 
\nonumber \\
\hspace{-0.95in}&&  \, \qquad  \, \qquad
+229182120170130009397850\,{x}^{8}\,
\nonumber \\
\hspace{-0.95in}&&  \, \qquad \,\qquad \,
 +620754459813846824189800125\,{x}^{9}
\, + \, \, \cdots 
\nonumber
\end{eqnarray}
and of Yukawa couplings:
\begin{eqnarray}
\hspace{-0.95in}&&  \, 
K_3(Ext^2(L_4)) \, \,= \, \, \, 
1\, -575\,x\, -1087500\,{x}^{2}\, 
-2357466250\,{x}^{3}\, -5515348543750\,{x}^{4}\,
\nonumber \\
\hspace{-0.95in}&&  \, \qquad  \, \qquad
 -13549400590159950\,{x}^{5}\,
 -34443162481829737500\,{x}^{6}\, 
\nonumber \\
\hspace{-0.95in}&&  \, \qquad  \, \qquad
-89801360565832417275000\,{x}^{7}\,
 -238760519646901921788093750\,{x}^{8}\,
\nonumber \\
\hspace{-0.95in}&&  \, \qquad  \, \qquad
 -644794600714076957552558593750\,{x}^{9}
\, + \, \, \cdots \nonumber 
\end{eqnarray}
\begin{eqnarray}
\hspace{-0.95in}&&  \, 
K_4(Ext^2(L_4)) \, \,= \, \, \, 
1 \, -1725\,x\, -2270625\,{x}^{2}\, 
-3510633125\,{x}^{3}\,-5943482381250\,{x}^{4}\,
\nonumber \\
\hspace{-0.95in}&&  \, \qquad  \, \qquad
-10616175881261100\,{x}^{5}\, 
-19525058497227228750\,{x}^{6} \nonumber \\
\hspace{-0.95in}&&  \, \qquad  \,    \qquad   
-36235211885062595043750\,{x}^{7}\, 
-66371326425035067092906250\,{x}^{8}\,
\nonumber \\
\hspace{-0.95in}&&  \, \qquad  \, \qquad 
-116122894233894656970457656250\,{x}^{9}
\,\, + \, \, \cdots \nonumber 
\end{eqnarray}
\begin{eqnarray}
\hspace{-0.95in}&&  \, 
K_5(Ext^2(L_4)) \, \,= \, \, \, 
1 \, -2875\,x \, -2131250\,{x}^{2}
\, -1182175000\,{x}^{3}\, +1120605484375\,{x}^{4}
\nonumber \\
\hspace{-0.95in}&&  \, \qquad  \, \qquad
\, +7242481554278375\,{x}^{5}\, 
+23221270993221987500\,{x}^{6}\, 
 \nonumber \\
\hspace{-0.95in}&&  \, \qquad  \,  \qquad 
+64082729395471190031250\,{x}^{7}\,
\nonumber \\
\hspace{-0.95in}&&  \, \qquad  \, \qquad
 +167312972891732043150312500\,{x}^{8}\,
\nonumber \\
\hspace{-0.95in}&&  \, \qquad  \, \qquad
+426590930940677272029245312500\,{x}^{9}
\, \, + \, \, \cdots \nonumber 
\end{eqnarray}
One has the relations
\begin{eqnarray}
\hspace{-0.95in}&&  \, 
K_4(Ext^2(L_4)) \, \,= \, \, \, K_3(Ext^2(L_4))^3, 
\qquad 
K_5(Ext^2(L_4)) \, \,= \, \, \, K_3(Ext^2(L_4))^5, 
\nonumber 
 \qquad 
\end{eqnarray}
and one notes the following relation between the Yukawa coupling 
of $\, L_4$ with the Yukawa coupling of its exterior square: 
\begin{eqnarray}
\hspace{-0.95in}&&  \, \qquad  \, \qquad
K_3(L_4) \, \,= \, \, \, {{1} \over {K_3(Ext^2(L_4))^2}}.  
\end{eqnarray}

\end{document}